\documentclass[useAMS,usenatbib]{mn2e}

\usepackage{longtable,booktabs}
\usepackage[dvips]{graphicx}
\usepackage{epsfig}
\newcommand{\etal}{{\em et al. }}
\usepackage{multicol}
\usepackage[tight]{subfigure}

\title[High Radio Frequency Properties and Variability of Brightest Cluster Galaxies]{High Radio Frequency Properties and Variability of Brightest Cluster Galaxies}
\author[M. T. Hogan \etal ]{M. T. Hogan$^{1,2,3}$\thanks{E-mail:
m4hogan@uwaterloo.ca (MTH)},  A. C.
Edge$^{1}$, J. E. Geach$^{4}$, K. J. B. Grainge$^{5}$, J. Hlavacek-Larrondo$^{6,7,8}$, \newauthor T. Hovatta$^{9,10}$, A. Karim$^{11}$, B. R. McNamara$^{2,3}$, C. Rumsey$^{12}$, H. R. Russell$^{13}$, \newauthor P. Salom\'{e}$^{14}$, H. D. Aller$^{15}$, M. F. Aller$^{15}$, D. J. Benford$^{16}$, A. C. Fabian$^{13}$, \newauthor A. C. S. Readhead$^{9}$, E. M. Sadler$^{17}$ and R. D. E. Saunders$^{12}$ \\
\\
$^{1}$Centre for Extragalactic Astronomy, Department of Physics, Durham University, Durham, DH1 3LE, UK \\
$^{2}$Department of Physics and Astronomy, University of Waterloo, Waterloo, ON, N2L 3G1, Canada \\
$^{3}$Perimeter Institute for Theoretical Physics, Waterloo, ON, N2L 2Y5, Canada \\
$^{4}$Centre for Astrophysics Research, Science $\&$ Technology Research Institute, University of Hertfordshire, Hatfield, AL10 9AB, UK \\
$^{5}$Jodrell Bank Centre for Astrophysics, School of Physics and Astronomy, The University of Manchester, Manchester, M13 9PL, UK \\
$^{6}$D\'{e}partement de Physique, Universit\'{e} de Montr\'{e}al, C.P. 6128, Succ. Centre-Ville, Montreal, Quebec H3C 3J7, Canada \\
$^{7}$Kavli Institute for Particle Astrophysics and Cosmology, Stanford University, 452 Lomita Mall, Stanford, CA 943053 \\
$^{8}$Department of Physics, Stanford University, 382 Via Pueblo Mall, Stanford, CA 94305 \\
$^{9}$Cahill Center for Astronomy $\&$ Astrophysics, California Institute of Technology, 1200 East California Boulevard, Pasadena, CA 91125, USA \\
$^{10}$Aalto University Mets\"{a}hovi Radio Observatory, Mets\"{a}hovintie 114, 02540 Kylm\"{a}l\"{a}, Finland \\
$^{11}$Argelander-Institute of Astronomy, Bonn University, Auf dem Huegel 71, D-53121 Bonn, Germany \\
$^{12}$Astrophysics Group, Cavendish Laboratory, JJ Thomson Avenue, Cambridge, CB3 0HE, UK \\
$^{13}$Institute of Astronomy, Madingley Road, Cambridge CB3 0HA, UK \\
$^{14}$LERMA, Observatoire de Paris, 61 Av. de l'Observatoire, 75014 Paris, France \\
$^{15}$Department of Astronomy, University of Michigan, 311 West Hall, 1085 South University Avenue, MI 48109-1107 USA\\
$^{16}$Observational Cosmology Lab, Code 665, NASA Goddard Space Flight Center, Greenbelt, MD 20771, USA \\
$^{17}$Sydney Institute for Astronomy, School of Physics, The University of Sydney, NSW 2006, Australia}

\begin{document}

\date{Accepted 2015 July 7.  Received 2015 May 23; in original form 2014 September 12}

\pagerange{\pageref{firstpage}--\pageref{lastpage}} \pubyear{2013}

\maketitle

\label{firstpage}

\begin{abstract}
We consider the high radio frequency (15~GHz - 353~GHz) properties and variability of 35 Brightest Cluster Galaxies (BCGs).  These are the most core-dominated sources drawn from a parent sample of more than 700 X-ray selected clusters, thus allowing us to relate our results to the general population.  We find that $\geq$6.0\% of our parent sample ($\geq$15.1\% if only cool-core clusters are considered) contain a radio-source at 150~GHz of at least 3mJy ($\approx$1$\times$10$^{23}$~W~Hz$^{-1}$ at our median redshift of z$\approx$0.13).  Furthermore, $\geq$3.4\% of the BCGs in our parent sample contain a peaked component (Gigahertz Peaked Spectrum, GPS) in their spectra that peaks above 2~GHz, increasing to $\geq$8.5\% if only cool-core clusters are considered.  We see little evidence for strong variability at 15~GHz on short (week-month) timescales although we see variations greater than 20\% at 150~GHz over 6-month timesframes for 4 of the 23 sources with multi-epoch observations.  Much more prevalent is long-term (year-decade timescale) variability, with average annual amplitude variations greater than 1\% at 15~GHz being commonplace.  There is a weak trend towards higher variability as the peak of the GPS-like component occurs at higher frequency.  We demonstrate the complexity that is seen in the radio spectra of BCGs and discuss the potentially significant implications of these high-peaking components for Sunyaev-Zel`dovich cluster searches. 

\end{abstract}


\section{Introduction}
Energetic feedback from accreting supermassive black holes (SMBH) is now widely accepted to play an integral role in the formation and evolution of Universal structure.  Such action is commonly invoked to explain a variety of phenomena such as the high-end curtailment of the galaxy luminosity function \cite[e.g.][]{Benson03, Bower06, Croton06}, the M$_{BH}$ vs M$_{bulge}$ correlation \cite[e.g][]{Magorrian98, Silk98} and the symmetry seen in the cosmic histories of both star formation and active galactic nuclei (AGN) \cite[e.g.][]{Merloni04, Springel05, Hopkins12}.  

In the most massive systems, at the centres of galaxy clusters, it is now well established that the AGN-action of the centrally located Brightest Cluster Galaxy (BCG) prevents runaway cooling \cite[the classical {\it Cooling Flow Problem}: see][for a review]{Fabian94} by imparting energy to its surroundings through mechanical feedback \cite[see the recent reviews by][]{McNamara07, McNamara12, Fabian12}.  Such actions are invoked to explain the deficit of cold gas \cite[e.g.][]{Edge01, Salome03} and star formation \cite[][]{O'Dea08, Rafferty08} compared to what would be expected from cooling dominated systems, as well as the dearth of gas at intermediate cooling temperatures \cite[][]{Peterson03}.  Striking observational evidence for this mechanical feedback is seen through the presence of X-ray cavities in the intra-cluster medium (ICM) \cite[e.g.][]{McNamara00, Fabian00, Hlavacek-Larrondo12a}.  These cavities, which are inflated by the actions of co-spatially observed radio-emitting plasma, subsequently re-distribute accretion energy from the central super-massive black hole (SMBH) to the surroundings as they buoyantly rise.  

There has apparently been a fine balance between heating and cooling within clusters in place for at least half the Hubble time \cite[e.g.][]{Vikhlinin06, Pratt10, McDonald14}. Furthermore, the energy imparted by AGN activity does appear to be sufficient to offset cooling {\it on average} \cite[e.g.][]{McNamara07, Dunn08}.  However, this is not the case at all times suggesting that periods of cooling must be interspersed by periods of AGN energy injection.  Added to this is the growing realisation that in the most settled cool-core (CC) clusters where there is a central peak in the cluster X-ray surface profile indicative of substantial cooling, the BCGs have a radio-loud duty cycle approaching unity \cite[e.g.][]{Burns90, Mittal09, Hogan15}.  This suggests that in these systems there must be cyclic activity whereby the BCG is more active in some periods than others.

Radio observations are integral for tracing this mechanical feedback.  However, most studies have been carried out at $\leq$1.4~GHz \cite[e.g.][]{Best07} and so the radio properties of BCGs in the $\approx$10-300~GHz range are somewhat poorly constrained.  Single-dish surveys of the sky at the higher end of this range are typically shallow with relatively low resolution \cite[e.g.][]{PlanckPCCSXXVIII} whereas even at the lower end, small beam sizes make interferometric surveys of any sizeable area both difficult and expensive.  However, several recent surveys have allowed for huge advances in the understanding of the radio-sky at greater than 10~GHz (e.g. 10C at 15.7~GHz, \citealt{AMI11}; AT20G at 20~GHz, \citealt{Murphy10} and the AT20-deep~(pilot) also at 20~GHz, \citealt{Massardi11, Franzen14}).  Consequently, only a few of the brightest BCGs have well characterised radio-spectra in this crucial spectral range.  

In \cite{Hogan15} (hereforth H15a) we considered the radio properties of BCGs in a parent sample of over 700 X-ray selected clusters comprising the BCS, eBCS and REFLEX cluster catalogues \cite[][respectively]{Ebeling98, Ebeling00, Boehringer04}.  These catalogues are X-ray flux limited, hence our sources are not selected on radio priors and should be representative of the general cluster population.   The clusters in the parent sample were split into CCs and non cool-cores (NCCs) using the presence of optical emission lines around the BCG.  Such lines are only found in systems with central cooling times less than 5$\times$10$^{8}$~years, equivalent to a central entropy less than 30~keV~cm$^{2}$ and hence can be used as a proxy for the cluster state \cite[][]{Cavagnolo08, Rafferty08, Sanderson09}. The radio spectral energy distributions (SEDs) of the BCG in 246 of these were populated, typically between 74~MHz and 20~GHz, and decomposed into active and inactive components attributable to ongoing and historical accretion respectively (see H15a).  Not only was the radio duty cycle of BCGs in CCs seen to be substantially higher than in NCCs, it was found that the majority of CC-hosted BCGs showed evidence for ongoing core activity that manifests itself primarily as a spectral flattening above a few GHz.  Often this emission is missed in low frequency surveys.  Further confounding the lack of information, increased variability is postulated to higher radio frequencies as the emission is expected to originate from increasingly smaller physical scales. 

In this paper we select a sub-sample of the BCGs studied in H15a, believed to contain the most active cores.  These sources all reside in CC clusters where active feedback is prevalent. Furthermore, the pre-eminence of the radio core component in the radio SEDs of these sources indicates that the SMBH is actively accreting at a significant rate.  We have observed them with a variety of facilities to extend their radio coverage up to 353~GHz as well as observing a number of them at different epochs which, alongside historical observations, allows us to study their variability.  We re-iterate that these sources are from an X-ray selected cluster catalogue.  By selecting the sources with the highest expected {\it core} flux we are able to search for variability via short observations.  This permits us to explore the origin of the point-like central radio emission as well as to constrain the amplitude of variation in the accretion rate during active periods of ongoing feedback.  Whilst the feedback powers derived from X-ray cavities trace the AGN energy output averaged over tens of megayears, this shorter term variability provides insights into the more instantanous processes within the core.

One system that has been well-monitored at radio frequencies greater than a few GHz, and indeed constitutes one of the most well-studied examples of active AGN feedback in a cluster core, is NGC1275/3C84 in the Perseus Cluster \cite[e.g.][]{Boehringer93, Conselice01, Abdo09}. Large amplitude variations in the radio spectrum of this source have been known for many years \cite[][]{Pauliny-Toth66}.

Recently \cite{Dutson14} undertook a comprehensive study of the radio and gamma-ray properties of NGC1275, considering its radio variability over five decades in both time and frequency.  The radio spectrum consists of a steep spectrum power-law at frequencies below approximately 1~GHz and an inversion above this leading to a peaked profile. It should be noted that the source is not strongly beamed \cite[][]{Krichbaum92, Nagai10}.  The power-law component, attributed largely to the presence of extended lobes and a 300kpc mini-halo \cite[][]{Burns92}, is constant in its flux.  However the peaked component is found to vary significantly in both flux (more than an order of magnitude) and turnover frequency on few year/decade timescales.  Such variations have been previously linked to individual components in the jet on milliarcsecond scales, as recoverable using Very Long Baseline Interferometry \cite[VLBI, e.g.][]{Suzuki12}.  Interestingly, \cite{Dutson14} find compelling correspondence between this few-year variation of the high radio-frequency peaked component and the high energy gamma-ray emission but no strong connection between the short-term `flaring' seen in the gamma-rays and the 1.3mm flux.  In a study of the core X-ray properties of 57 BCGs, \cite{Russell13} found that roughly half contained an X-ray point source at Chandra resolution.  It is worth noting that three of these (A2052, Hydra-A and M84) were seen to vary over similar 6-month to decade timescales, similar to the radio emission in NGC1275.

One of the aims of the current paper is to investigate whether the high radio frequency properties of NGC1275 mark it out as a peculiar object or whether such periods of high activity in the spectral range above 10~GHz are common amongst the BCG population.

Recently the Sunyaev-Zel`dovich (SZ) effect \cite[][]{Sunyaev72} has been used to compile large catalogues of galaxy clusters \cite[e.g.][]{Vanderlinde10, Reichardt13, Marriage11, Hasselfield13, PlanckERCSCVIII, PlanckINTV, Planck14XXIX}.  Unresolved radio sources present a significant systematic for these searches and may lead to underestimated or completely removed SZ decrement in the 15-200~GHz range \cite[see e.g.][]{Knox04, Coble07, Lin09}.  Furthermore, the single-dish nature of many SZ-observatories means that often the removal of contaminating point sources has to rely on uncertain extrapolation of higher resolution but lower frequency data.  Our results therefore have potentially significant implications for SZ studies of clusters.

This paper is arranged as follows.  In Sections \ref{SAMPLE_SELECTION} and \ref{DATA_SECTION} we describe the sample selection, data collection and reduction.  Our results are presented in Section \ref{RESULTS_SECTION} and discussed in Section \ref{DISCUSSION} before we draw conclusion in Section \ref{CONCLUSIONS_SECTION}.  We have used a standard $\Lambda$CDM cosmology with: $\Omega_{m}$ = 0.3, $\Omega_{\Lambda}$ = 0.7, $H_{0}$ = 70 km s$^{-1}$ Mpc$^{-1}$.  We use the spectral index convention $S~\propto~\nu^{-\alpha}$. Unless other stated, we use the name of the parent cluster to refer to its BCG.

\section[Sample]{Sample}  \label{SAMPLE_SELECTION}

The sample of sources chosen for this study were selected primarily from
H15a as having the brightest {\bf ($>10$mJy at 5~GHz)}, flat-spectrum cores
($\alpha<0.5$) so a detection above 100~GHz was possible. The H15a sample
covers an all-sky, X-ray flux-limited sample of over 700 clusters as
outlined above, but the number of sources matching these flux and index cuts is relatively small ($<30$ or $<4$\%)
. To increase the target list we added seven bright
($>$50mJy at 5~GHz) sources either in fainter clusters and/or clusters
mis-identified until now. Four of these sources are in fainter clusters
(A11, \citealt{Perlman99}; 4C$+$55.16, \citealt{Iwasawa99}; A2270,
\citealt{Healey07}; and RXJ2341+00) and three in clusters above the
eBCS/REFLEX flux limit (RXJ1350+09; RXJ1832+68, \citealt{Boehringer00, 
Gioia03}; and E1821+64, \citealt{Russell10}).  All seven of these sources have been previously identified as AGN, given the association of radio and
X-ray emission.  However, only E1821+64 is actually strongly (more than 50\%) contaminated in the
X-ray by an AGN, in this case a QSO. In the other six cases the cluster has {\it previously} been mis-identified 
as a BLLac as a result of them having seemingly flat radio spectra and so the X-rays had been attributed to the AGN.  However, all six of these clusters have central galaxies with strong, narrow optical line emission, characteristic of cooling flow BCGs \cite[][]{Crawford99} and in all other aspects are similar to the sources selected in H15a. Therefore, we propose a re-identification of each
source such that the X-rays are predominantly from the cluster and not a
central AGN.

The source selection is by no means complete but is representative of the
brightest, core-dominated radio sources in cluster cores. Therefore we
believe that the spectral and variability properties we determine for this
radio-bright sample can be used to constrain the properties of the
complete, X-ray selected sample as a whole.

\section{Data} \label{DATA_SECTION}

\subsection{GISMO}

GISMO is a 150~GHz (2mm) bolometer camera built by the Goddard Space Flight Centre \citep{Staguhn08} for use on the IRAM-30m telescope\footnote{IRAM is supported by INSU/CNRS (France), MPG (Germany) and IGN (Spain)}, which is located at an altitude of 2850m on Pico Veleta, Spain.  GISMO operates as a visitor instrument, being operable for around a two week period every six months.  During its time on the telescope GISMO is operated in a shared risk pool observing mode, during which telescope focus and pointing observations are regularly performed on IRAM calibrator sources.

We obtained data from 3 epochs, using GISMO to observe 29, 24 and 17 sources in the April 2012, November 2012 and April 2013 observing runs respectively, with as many source overlaps between runs as possible (see table \ref{GISMO_FLUXES}).  Sources were typically observed for 5-10 minutes using a standard lissajous scanning script.  The telescope has a FWHM of 16.7'' and the absolute flux calibration of GISMO is found to have a typical 8\% uncertainty\footnote{Performance reports available for each run at http://www.iram.es/IRAMES/mainWiki/Continuum/GISMO/Main}.

Data were reduced using the GISMO specific section of the CRUSH software package \citep{Kovacs08}.  Pointing and flux models specific to each run are updated during each GISMO run and the package itself is being continually updated.  We therefore reduced our data using a version of CRUSH (2.15-2) that post-dates all of our data and hence contains optimised parameters for each of our data-sets.

Most of our sources have signal-to-noise ratios (SNRs) less than 10 for each scan.  We therefore found the {\em -faint} option within CRUSH to deliver the best results in most cases.  For non-detections we re-mapped the data using the {\em -deep} filtering.  Only one source (A1885) was recovered using {\em -deep} but not {\em -faint}.  Since the {\em -deep} option is known to over-filter sources with SNR greater than 5 and will create negative flux haloes, we took our flux measurements from the maps created using the {\em -faint} option.  A1885 was recovered with a SNR of $\approx$4 and hence measuring the flux from the {\em -deep} map for this source is not believed to introduce any additional error.  The specific filtering mode used for each source is shown in table \ref{GISMO_FLUXES}.  Atmospheric extinction is automatically corrected for within CRUSH, using an estimate of the most recent tau225~GHz value as measured by the on-site tau-meter.  However, some of our data (particularly during the Nov. 2012 run) suffered from a faulty tau-meter not inserting the correct measurement into the FITS file.  Time-dependent records of the on-site tau values were recorded manually during each run, which were compared to the values recorded automatically in the log files.  To ensure consistency, we manually inputted the best estimate zenith tau-value using the {\it -tau.225~GHz} option for each of our source reductions.

Maps were produced and fluxes extracted using the CRUSH tool `show'.  Gaussian fits were performed for each source with the resultant flux and FWHM reported along with source peak and map RMS in table \ref{GISMO_FLUXES}.  Only Hydra-A showed significant extent and hence its flux was extracted from a user-defined region.  Fluxes were additionally verified using the kvis data-display tool \citep{Gooch96}.

\subsection{CARMA}
Twenty three sources were observed at 90~GHz using the CARMA interferometer \cite[e.g.][]{Woody04} in D-array between 21st May - 15th June 2012, of which twenty overlap with our GISMO sources.  These observations were performed in queue mode by initiating pre-determined  blocks, which could be started and abandoned part way through,  depending on observing conditions and at the discretion of the telescope operator.  Each block contained an observation of a planet for primary flux calibration.  This observation is run through CARMA pipelines to provide a primary flux scale for the observations.  Additionally, a strong bandpass calibrator was observed for each block (the bandpass calibrator used was either 3C279, 3C345, 3C454.3 or 3C446, depending on observability).  Science targets were visited several times at various hour angles to maximise uv-coverage and sandwiched between 1~minute observations of a nearby phase calibrator.

Data were reduced by the standard methods using the CARMA optimised version of the MIRIAD data-reduction software package (developed and maintained by University of Maryland). Some automatic flagging of bad data is performed, but the visibilities were also inspected and flagged interactively. Files with basic calibration, taken at the time of observation, were provided and used as a calibration `starting point'.  Corrections for antenna positions were performed using the most co-temporal {\it antpos} models provided by the observatory.  Further bandpass,  phase and flux corrections were applied using standard MIRIAD tasks as determined from the observed calibration sources.

Deconvolution and cleaning of the subsequent maps were also performed using MIRIAD and the kvis data-display tool used to extract fluxes, verified by additionally extracting fluxes using the AIPS task \textsc{JMFIT}.

\subsection{AMI}
The Arcminute Microkelvin Imager \cite[AMI,][]{Zwart08} is a versatile instrument located at the Mullard Radio Astronomy Observatory (MRAO). Optimised for study of the SZ effect, AMI consists of two interferometer arrays: the Small Array (SA), with ten 3.7 m antennas with baselines of 5-20~m; and the Large Array (LA) with eight 12.8~m antennas with baselines of 18-110~m.  These arrays are thus optimised for large (3 -- 10 arcmin) and small (30 arcsec -- 3 arcmin) scale observations respectively. This set-up allows study of the large-scale SZ effect with accurate characterisation of the contaminating radio source environment. Both arrays measure the I+Q polarisation at the central frequency of 16~GHz. Data are taken in six channels, each of width 0.72~GHz, over the range 13.9--18.2~GHz, allowing a local spectral index to be determined for strong sources.

We used the AMI-LA to observe seventeen of our sources, with each target visited either two or three times in 2012 (see Table \ref{AMI_FLUXES}).    Observations consisted of 8 minute integrations sandwiched between 1 minute phase calibrator scans.  In the April-June scans sources were themselves used for self-calibration.  For the September follow-ups we additionally observed nearby phase calibrator sources.  3C~48 and 3C~286 were observed for use as amplitude calibrators.

Data reduction was performed using the local in-house software \textsc{Reduce}. Reduction pipelines were used to apply amplitude and phase calibration, flag for telescope errors and Fourier transform to produce frequency channels. Additional flagging for bad data was carried out interactively through visual inspection of the channel data.

Data are written out of \textsc{Reduce} as uv-FITS files which were read into AIPS for deconvolution and cleaning using \textsc{IMAGR}.  Maps were produced for each channel in addition to a single image for the full bandwidth.  None of our sources were found to be resolved and so fluxes were extracted using the AIPS task JMFIT and verified using kvis.  We investigated the flux calibration and find a typical 5\% uncertainty, which we propagate into in our uncertainties.

\subsection{OVRO and UMRAO} \label{OVRO_UMRAO_Intro}
Since 2007 the Owens Valley Radio Observatory (OVRO) has been using its 40m telescope to undertake a 15~GHz monitoring campaign of over 1500 radio sources \cite[mainly blazar Fermi-LAT gamma-ray candidates, see][]{Richards11}.  Five of the sources in our sample have been monitored as part of this OVRO monitoring campaign.  Additionally, since January 2013 eleven extra BCGs with strong high radio-frequency emission selected from this work have been included within the dynamic queue, allowing regular (typically every 10 days) observations for these sources (see Table \ref{OVRO_VI}).  

The OVRO 40~m uses off-axis dual-beam optics and a cryogenic high electron mobility transistor (HEMT) low-noise amplifier with a 15.0~GHz centre frequency and 3~GHz bandwidth. The two sky beams are Dicke switched using the off-source beam as a reference, and the source is alternated between the two beams in an ON-ON fashion to remove atmospheric and ground contamination. A noise level of approximately 3-4~mJy in quadrature with about 2\% additional uncertainty, mostly due to pointing errors, is achieved in a 70 second integration period. Calibration is achieved using a temperature-stable diode noise source to remove receiver gain drifts and the flux density scale is derived from observations of 3C~286 assuming the \cite{Baars77} value of 3.44~Jy at 15.0~GHz. The systematic uncertainty of about 5\% in the flux density scale is not included in the error bars.  Complete details of the reduction and calibration procedure are found in \cite{Richards11}.  We check the pipeline data for, and remove, individual scans whose amplitude is obviously discrepant (of which we find less than 1\%) and remove a small minority of points where uncertainties reach more than 40\%.

Additionally, one of our sources (4C+55.16) that is included within the OVRO monitoring campaign was also monitored at 15~GHz by the 26m telescope of the University of Michigan Radio Astronomy Observatory \cite[UMRAO;][]{Aller85, Aller14}.  The UMRAO monitoring ran between October 1984 and June 2007, hence is not ongoing.  However the addition of the UMRAO data to the OVRO and AMI data means that for this source we have well-sampled lightcurves at 15~GHz extending forwards from 1985 to the present.

\subsection{SCUBA-2} \label{SCUBA2_Intro}

The SCUBA-2 \cite[][]{Holland13} observations were made as part of a poor weather programme (JCMT weather  
Bands 4 and 5, $\tau_{225 \rm GHz}=0.15$--$0.3$) as part of Canadian and UK
projects (M12AC15, M12BC18, M12BU38, M13AC16 and M13AU38) between February
2012 and July 2013. The observations were made in small map (``CV Daisy")
mode with integrations of 30 minutes each.

The observations were reduced using the standard SMURF package \cite[][]{Jenness11, Chapin13}. We used the standard flux calibration of
FCF$_{850}= 527 \pm 26$~Jy~beam$^{-1}$ pW$^{-1}$. The resulting maps
reached noise levels of typically 4--8~mJy~beam$^{-1}$ (depending on
conditions) which was sufficient to detect over half of the sources
observed.

\section{Results} \label{RESULTS_SECTION}

\subsection{Spectral Energy Distributions}

\begin{figure*}    
  \centering
    \subfigure[{\it R0439+05}]{\includegraphics[width=8cm]{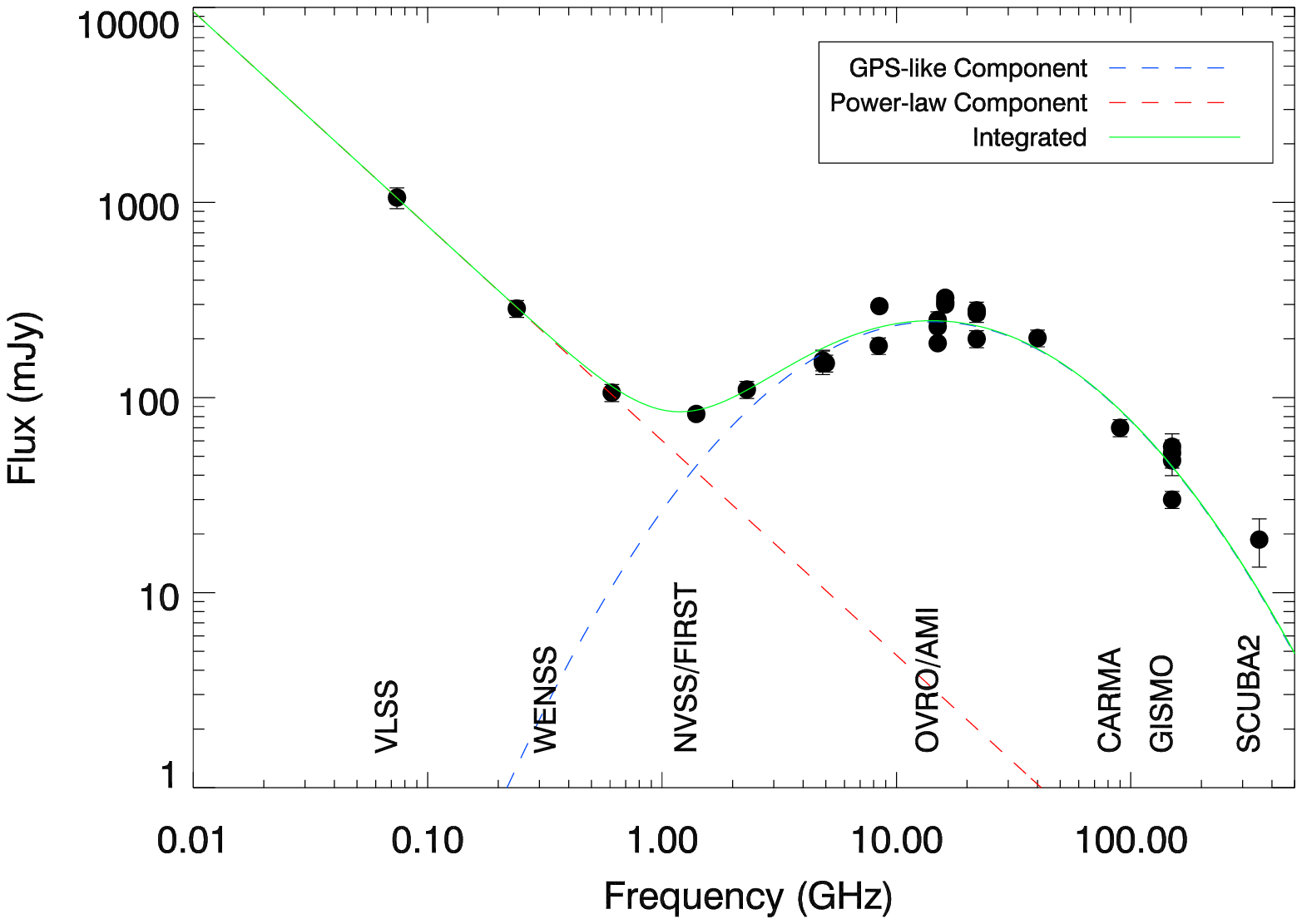}}
    \subfigure[{\it E1821+644}]{\includegraphics[width=8cm]{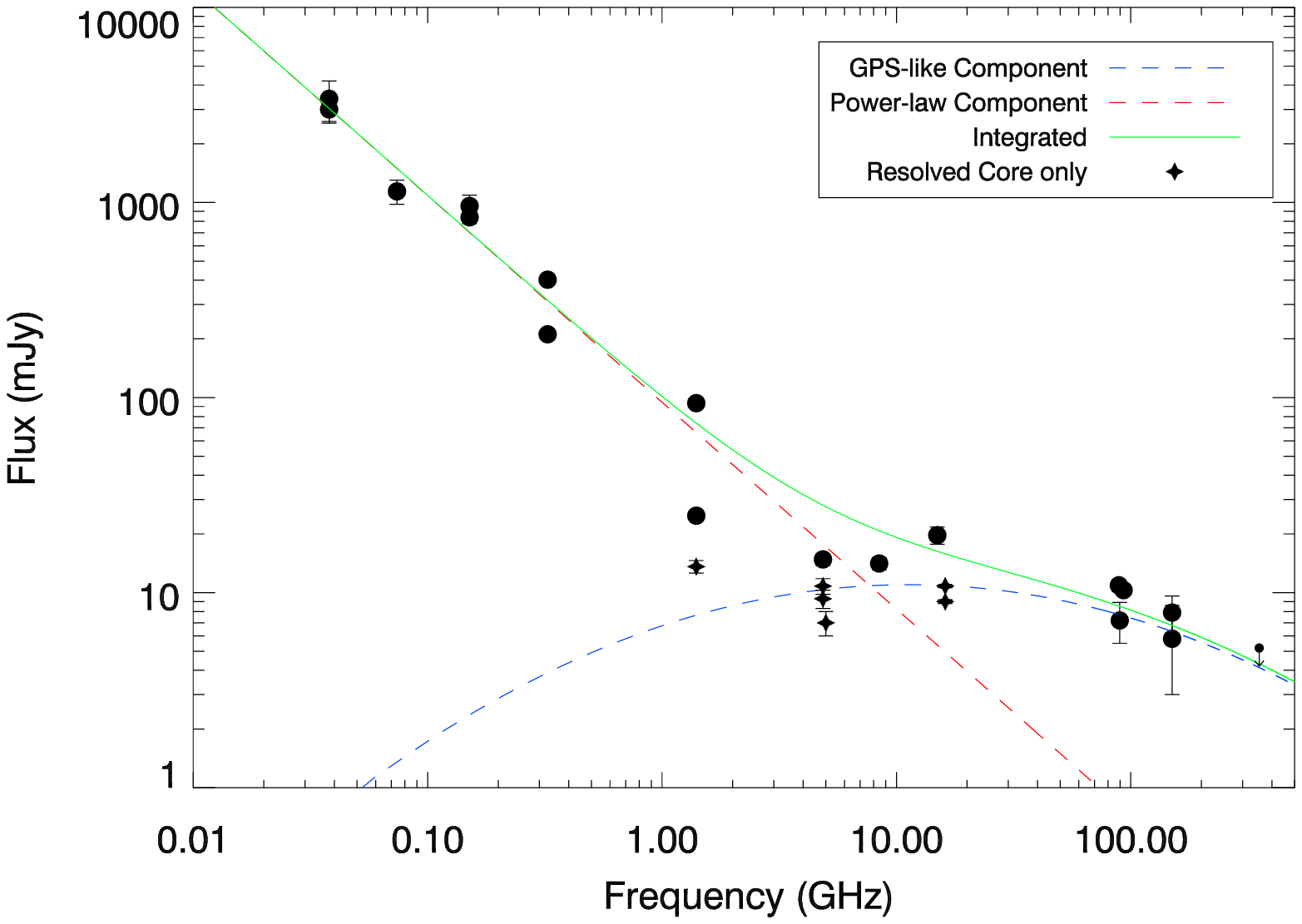}}
    \subfigure[{\it 4C+55.16}]{\includegraphics[width=8cm]{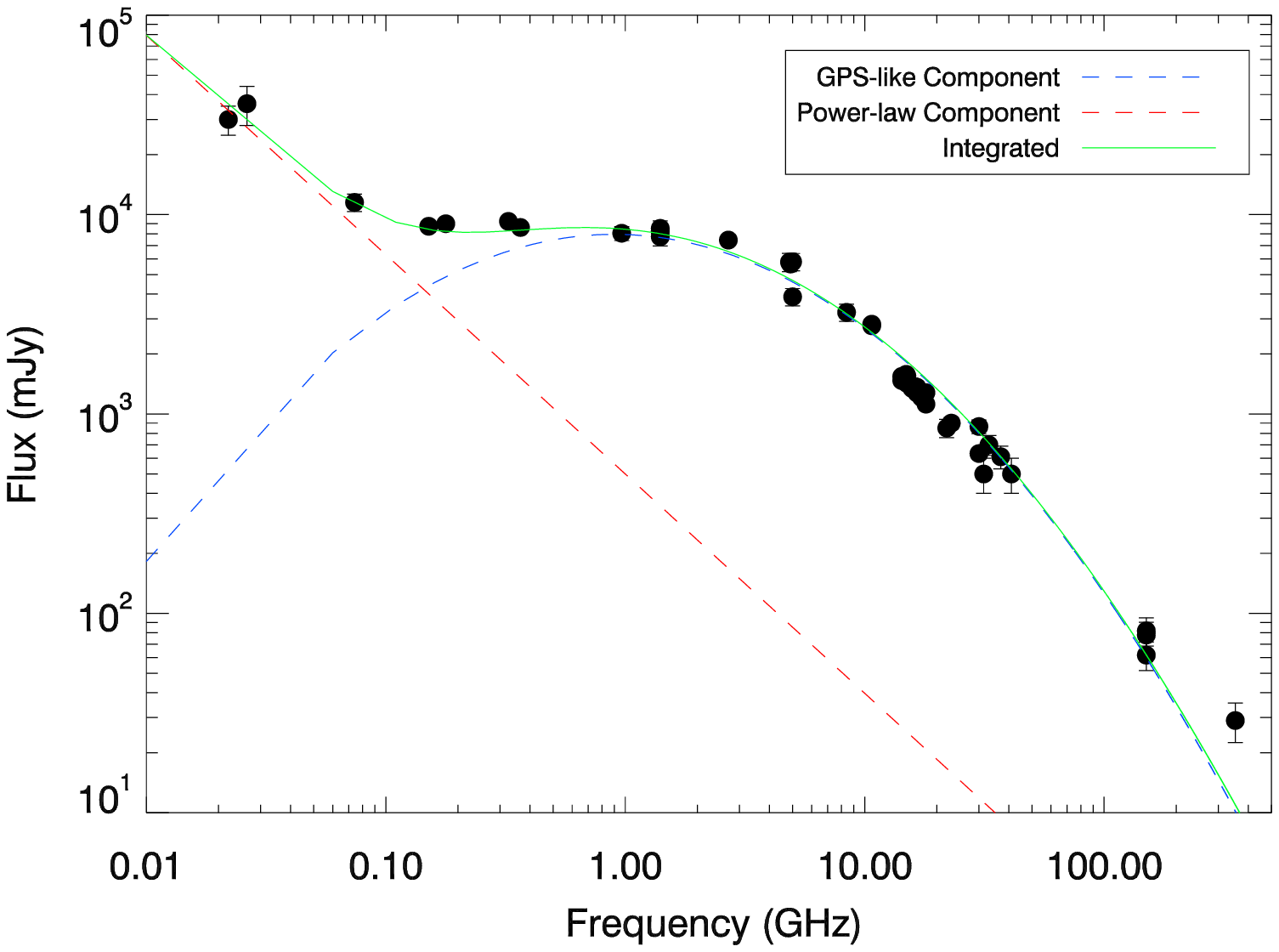}}
    \subfigure[{\it MACS0242-21}]{\includegraphics[width=8cm]{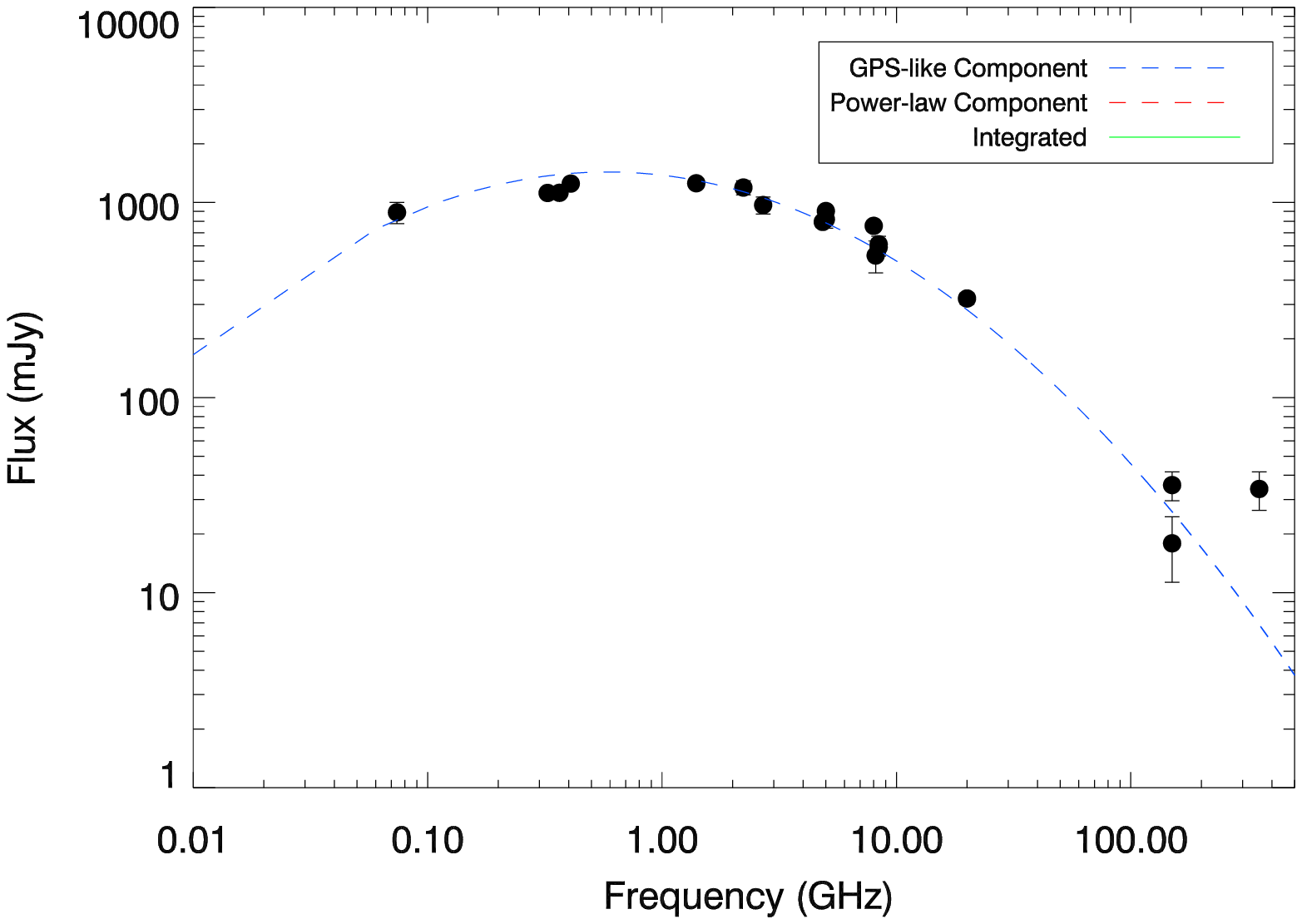}}
  \caption{Example SEDs for four objects.  In panel (a) we highlight the frequencies of the observations presented within this paper, in addition to the widely used radio-surveys NVSS/FIRST, WENSS and VLSS to help contextualise our frequency coverage.  We see a variety of spectral shapes, sometimes well-fit by a single component as in MACS0242-21 whereas in other cases requiring both a power-law and a GPS-like component (see text).  Note that the SCUBA-2 data-points at 353~GHz are included in the SEDs but excluded from the plots as they often appear to be suggesting the presence of an additional, poorly constrained component at the highest frequencies (see Section \ref{ADAF_SCUBA2}).} 
 \label{exampleSEDs}
\end{figure*}

In H15a we performed spectral decompositions for a large number of BCG SEDs, finding that in many cases the spectrum flattens above a few GHz, indicative of these sources containing an underlying strong core component.  In the current paper we expand the spectral coverage for our subsample of 35 of these sources into the mm/sub-mm regime.  We indeed see that in many instances these active components extend to high frequency and typically rollover at the highest radio-frequencies, consistent with recently accelerated synchrotron populations.

We performed fits to each of our extended SEDs, using the \textsc{CURVEFIT} program of IDL.  Generally our sources could be well fitted with either a power-law (equation \ref{Power_law_equation}), Giga-hertz Peaked Source (GPS) \cite[equation \ref{GPS_equation}, also see][]{Orienti14} like component or a combination of these.  Individual SEDs, fitting notes and parameters are presented in Appendix \ref{Fitting_Notes}.  Our SCUBA-2 data were included on the SEDs although excluded from the spectral fits as in several cases it appears to be indicative of an additional albeit poorly constrained component becoming prominent in the mm/sub-mm regime (see Section \ref{ADAF_SCUBA2}).  Four example SEDs are shown in Figure \ref{exampleSEDs}.

In some instances, as in RXJ0439+05 (panel a of Figure \ref{exampleSEDs}) the GPS-like component is distinct and inverts the spectrum above a break frequency below which we see a steep spectrum power-law tail to lower frequency.  These sources are likely to be undergoing powerful recent activity in their cores.  At frequencies below the self-absorption turnover of the core component the spectrum becomes dominated by an underlying steep spectrum power-law component, that may be suggestive of the presence of an amorphous halo of confused emission \cite[][]{Kempner04, Hogan15}.  In other instances, as shown in E1821+644 and 4C+55.16 (panels b and c of Figure \ref{exampleSEDs}) the integrated spectrum shows only a flattening in the GHz range which may stay flat to high frequencies as in E1821+644 or tail off as in the case of 4C+55.16.  That almost all CC-hosted BCGs contain radio-AGN suggests these radio-sources are long lived.  The variety that we see in their radio spectra suggests that the emitted radio flux is not a constant for any given source.  The timing of the most recent period of increased radio emission may then dictate the ratio of high radio-frequency core emission to steeper low-frequency emission from more physically extended regions.  Yet other sources, such as MACS0242-21 (panel d, Figure \ref{exampleSEDs}) appear well fit by a GPS model with a turnover frequency below 1~GHz. Such sources may be classified as compact steep spectrum objects \cite[CSS,][]{O'Dea98}. These may be dominated by a single period of older activity where the ejected emission has propagated outwards and expanded, permitting the spectral peak to move towards lower frequency.

\subsubsection{Notes on the General BCG Population}

\begin{figure}
\begin{center}
\includegraphics[width=9cm]{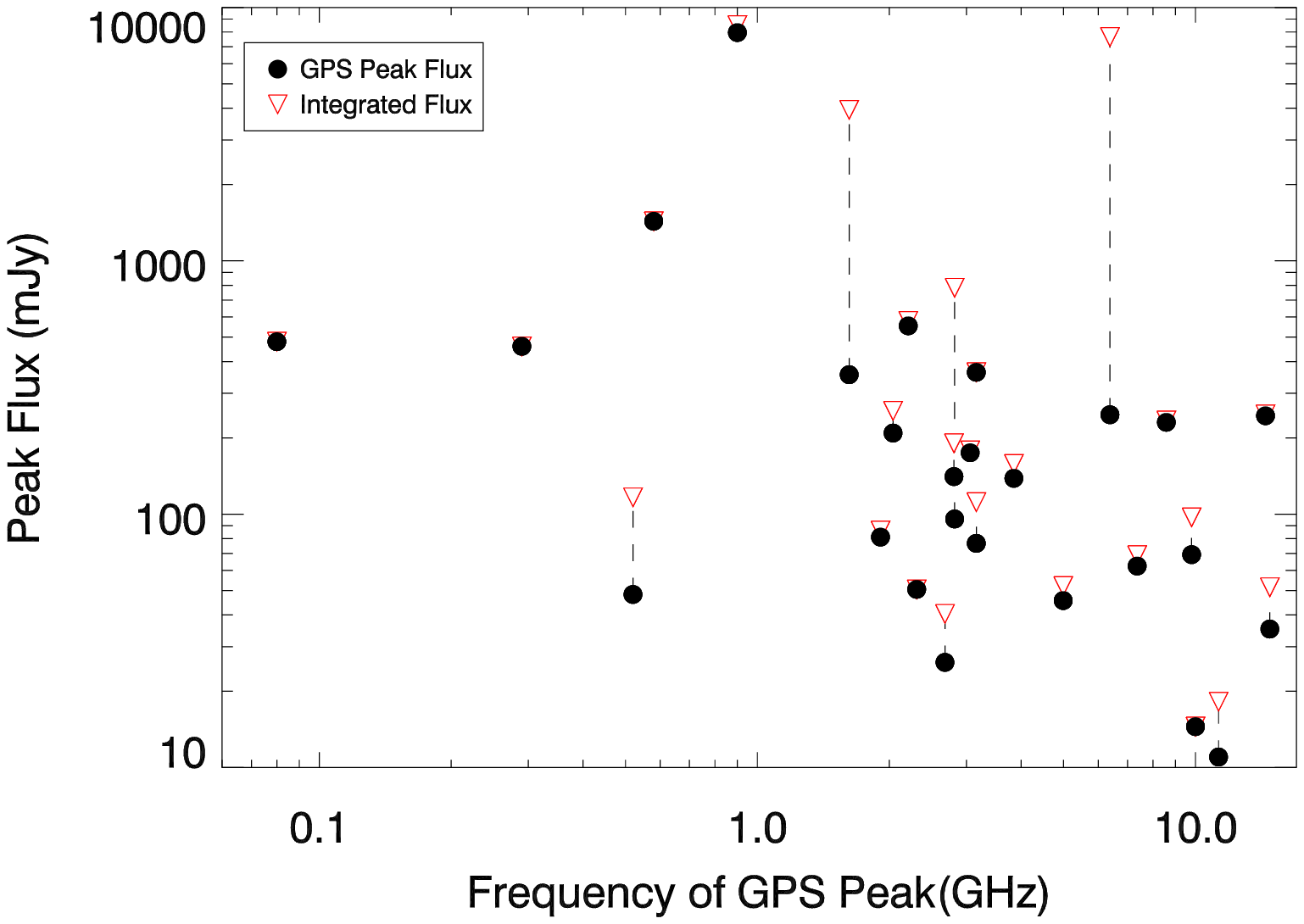}
\caption{Flux of the GPS-like component as a function of the frequency at the peak.  Note that in the classification scheme of O'Dea \etal (1998), the lowest peaking of these components, if an isolated source would likely be identified as CSS rather than GPS.  However, we note that in the evolution scenario for this family of sources, CSS sources are just slightly older GPS sources and hence we refer to all of these peaked components as `GPS-like' for brevity.  Each source is represented twice (connected by dashed lines).  Circles show the peak flux of the GPS component whereas downfacing triangles show the combined flux of both the GPS and any power-law component measured at the GPS turnover frequency.  In the majority of instances we see that at the turnover frequency the GPS component is dominating the flux.  It should be noted that {\it virtually all} radio-loud BCGs {\it may} contain a peaked active component, although in most systems these would be expected to peak far below the integrated flux and hence be undetectable in all but the brightest systems.}
\label{GPS_peak_vs_flux}
\end{center}
\end{figure}

Twenty-six of the thirty-five sources are found to contain a GPS-like peaked component.  Physically, GPS sources are believed to be young radio sources whose spectra are peaked due to synchrotron self-absorption \cite[][]{O'Dea98}.  That we find peaked components within our BCG spectra, in addition to steep components at low-frequency, is indicative of recently enhanced radio emission in the central regions of these objects.  This could be associated with a long-lived source having a varying accretion rate, or with the launching of distinct knots in the radio jets.  Either scenario is consistent with the high duty cycle expected for these cool-core hosted AGN. We plot in Figure \ref{GPS_peak_vs_flux} both the peak flux of the fitted GPS-component and the combined power-law + GPS-component flux of each of our GPS-containing sources as a function of the GPS turnover frequency.  These are mainly found to be GPS dominated at the frequency of the peak, with only a minority of sources found where the combined flux is significantly greater than the GPS component alone.

We find that the spectral turnover of the GPS component in nineteen of these twenty-six sources lies above 2~GHz. Only Hydra-A has a sub-dominant core at the frequency of the core's spectral peak.  The presence of the core here therefore does not overtly affect the integrated spectral shape in Hydra-A at frequencies greater than a few GHz.  In the simplest case we can use this to put a limit on the number of BCGs containing powerful peaked cores that greatly affect the spectrum above the observing frequency of most wide-sky surveys \cite[e.g. NVSS/FIRST at 1.4~GHz, WENSS/WISH at 325~MHz, VLSS at 74~MHz, see Appendix \ref{Fitting_Notes},][]{Condon98, White97, Rengelink97, DeBreuck02, Cohen07}.  Our sample was drawn from the 726-source Parent Sample of H15a.  If we subtract from this the 196 sources that fall below Declination -30$^{\circ}$ and so were not considered by our initial GISMO selection criteria (see Section \ref{SAMPLE_SELECTION}) then we find that $\geq$3.4\% of BCGs contain a synchrotron self-absorbed active core component peaking above 2~GHz that is brighter than 10mJy at C-band ($\approx$ 5~GHz).  This fraction increases to $\geq$8.5\% if we consider only the cool-core clusters. Our incomplete spectral coverage of the full sample means that this number is only a lower limit and that the true fraction of BCGs with GPS-cores may be much higher.  Indeed this simple analysis neglects systems whose spectra appear to be persistently flat out to high radio frequencies where multiple superposed flux components may be present.

That strong spectral deviations at frequencies higher than a few GHz are not uncommon in BCGs has important implications.  This includes, but is not limited to, the activity of the population as a whole and also the expected contamination rate of high-frequency peaking radio components in BCGs on SZ signals.  We consider this further in Section \ref{DISCUSSION}.

\subsection{Nature of Variability}

Understanding the radio variability of AGN is important as it informs us on several physical processes.  Short-term variation allows {\bf us} to place size constraints on the region from where any observed radio-flux propagates (i.e. a source cannot vary on less than its crossing-time) whereas longer term (few year) variation can inform us as to the likely fueling mechanisms of AGN.  At radio frequencies, the variation itself can be due to differing energy densities within the jet, which may be due to a change in the accretion structure and fuelling rate at the jet-base.  Alternatively, changes in flux may be due to individual components within a jet becoming brighter as they interact with other material.  However, fully characterising the radio variability of AGN is inherently difficult as the measured variability can be attributed to different mechanisms and will be most prevalent on different timescales depending on the observing frequency.  

As an example, consider a synchrotron self-absorbed source sampled at a frequency just below the turnover which shows a flux increase.  Without contemporary observations at and above the peak it would be  difficult to say whether the additional received flux is due to an increase in total  power-input to the jet by the AGN (and hence the normalisation of the entire SED  should increase) or whether the overall power-output is falling but that the lower intensity and expanded physical scales of the emitting regions lead to the turnover moving to lower frequency and so giving a flux increase below the peak while the overall normalisation remains constant. Ideally the entire SED would be sampled simultaneously and fitted, accounting for self-absorption at a variety of epochs in  order to recover the true variability in the underlying power output.  Unfortunately, such a campaign is prohibitively expensive.  Nevertheless, limits can be placed on the variability of BCGs at different frequencies and timescales.

\subsection{Monochromatic Variability}

\subsubsection{Variability Index}
We use the debiased variability equation \cite[e.g.][]{Akritas96, Barvainis05, Sadler06} to attempt to place quantitative limits on the monochromatic flux variability of our sources.  This index accounts for the  uncertainties in flux measurements and hence should be robust against artificial  positives.  The Variability Index (V.I.) is defined as
\begin{equation} \label{VI_equation}
V_{RMS} = \frac{100}{<S>}\sqrt{\frac{{\sum{(S_{i} - <S>)^2} - \sum{\sigma_{i}^2}}}{N}}
\end{equation} 
where $S_{i}$ is an individual flux measurement, $\sigma_{i}$ the associated  uncertainty, $<S>$ the mean flux and N the number of observations of a given source.
%Updated
For lightcurves with a large number of points this statistic gives a measure of the typical variability of a source about its mean flux.  This analysis begins to falter for lightcurves with a small number of points where the mean is not necessarily representative of all measured fluxes.  However, it is useful even in the limit of N=2 as a positive returned V.I. is indicative of significant variation whereas an imaginary returned V.I. suggests any observed flux difference is most likely attributable to measurement uncertainty.  A limitation of this method is that over-estimation of the error on flux measurements can mask real variability.  However, this variability would be below any limits found for variation below the given errors.  Given limits on variability would therefore be correct, albeit less restrictive than if less conservative flux uncertainties were used.  Although it should be noted that such a limit does not of course restrict the potential for the source to vary below this level. Similarly, under-estimation of errors can falsely indicate variability where there is none.

\begin{figure}
\begin{center}
\includegraphics[width=9cm]{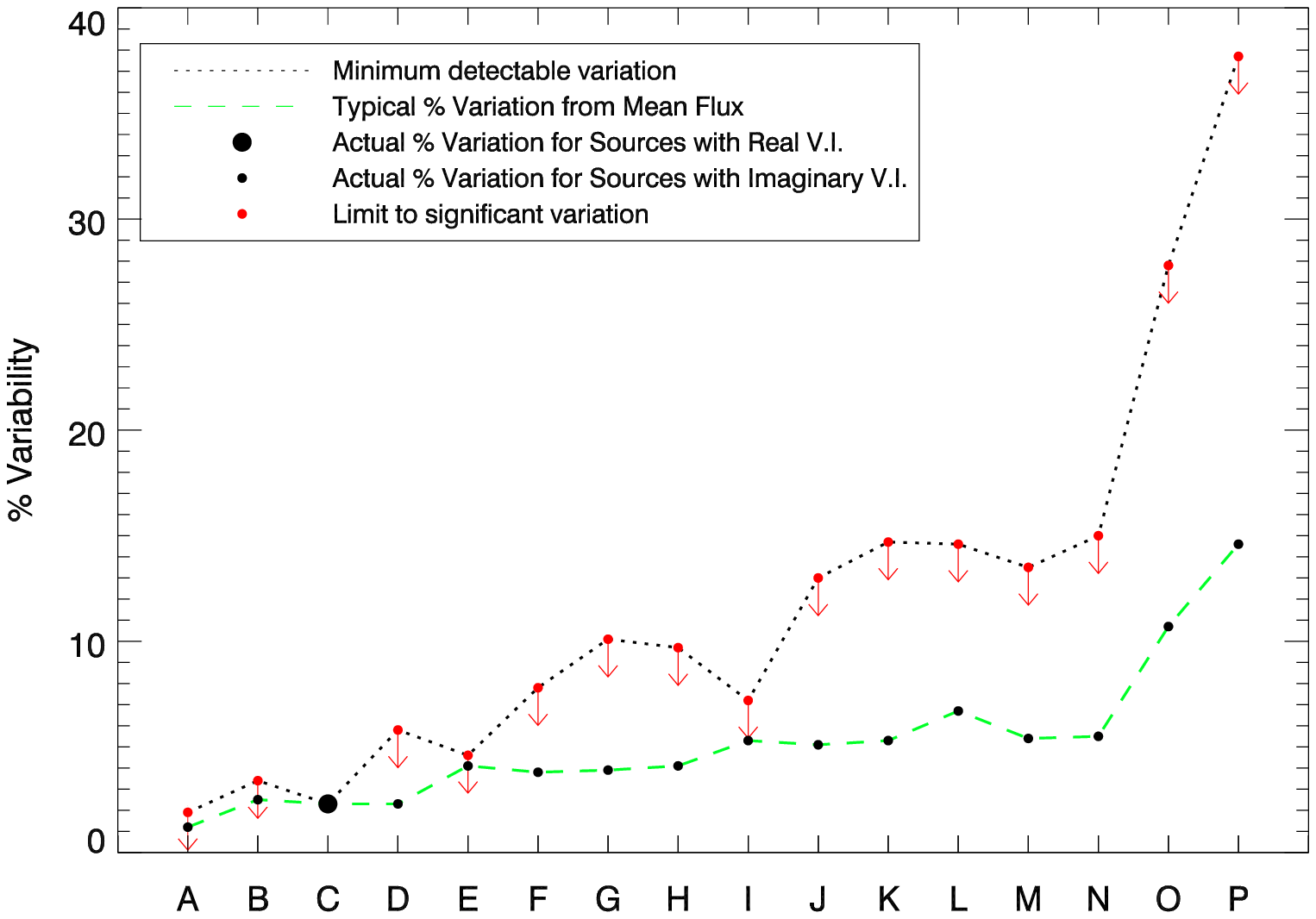}
\caption{Dotted line shows the minimum percentage variabilities that would have corresponded to a positive Variability Index (typical deviation from mean flux greater than mean fractional error) for the OVRO sources, sampled at 15~GHz on typically one-week timescales.  The green dashed line shows the actual typical deviation from the mean flux for each source respectively.  Note that only A2270 shows significant variation (dashed line above dotted line).  Sources are arranged left-to-right by their maximal measured flux at 15~GHz.  The detectable percentage variation is typically much lower for brighter sources, with some deviation from a one-to-one correlation due to uncertainties on flux measurements varying by source.  Sources are: A) 4C+55.16, B) RXJ0439*, C) A2270*, D) A2052, E) RXJ1558*, F) RXJ2341, G) RXJ0132, H) AS780*, I) RXJ1350*, J) A2055, K) A2415, L) Z8276, M) A2627, N) A2390, O) A646 and P) A1348. Note that five of these (indicated with *) were in the initial OVRO monitoring list and hence longer time-baseline data are available, where plotted values are those calculated when the time-frame was matched to the other eleven.  See also Table \ref{OVRO_VI}.}
\label{OVRO_var}
\end{center}
\end{figure}

\begin{figure}
\begin{center}
\includegraphics[width=9cm]{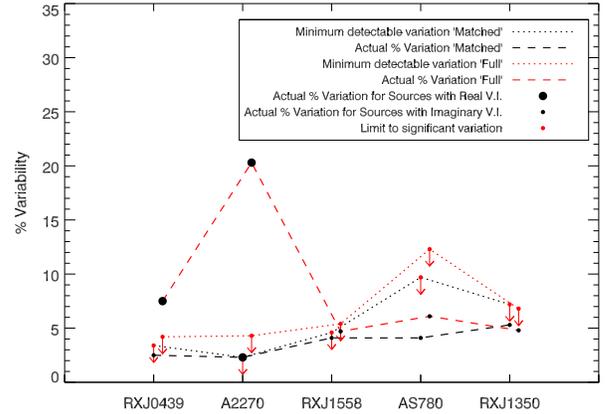}
\caption{OVRO variabilities for only the five sources in the original lists, measured for the full (from January 2008 onwards, red lines and offset for clarity) and restricted (January 2013 onwards, black lines) time-frames. Dotted lines show the minimum percentage variabilities that would have corresponded to a positive Variability Index (typical deviation from mean flux greater than mean fractional error), whereas dashed lines show the actual typical deviation from the mean flux for each source respectively.  Note that only A2270 shows low-level significant variation (dashed line above dotted line) when only data during the matched time period whereas when using the full data, both A2270 and RXJ0439 show significant variability, as would be expected by considering the lightcurves in Figure \ref{SUPER_5_LIGHTCURVES}. Although not a perfect measure, this does show evidence that variability at 15~GHz is most prominent over longer time-frames, hence considering variation of BCGs on longer than single year timescales is important.  As in Figure \ref{OVRO_var} we order the sources left to right by their maximal measured flux at 15~GHz.  See also Table \ref{OVRO_VI}. }
\label{OVRO_var_extra5}
\end{center}
\end{figure}

\begin{figure*}    
  \centering
    \subfigure{\includegraphics[width=14cm]{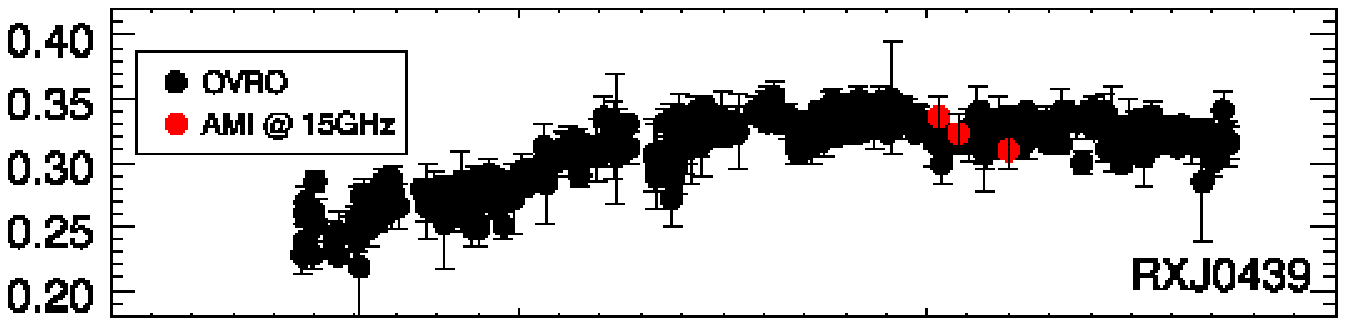}}
    \subfigure{\includegraphics[width=14cm]{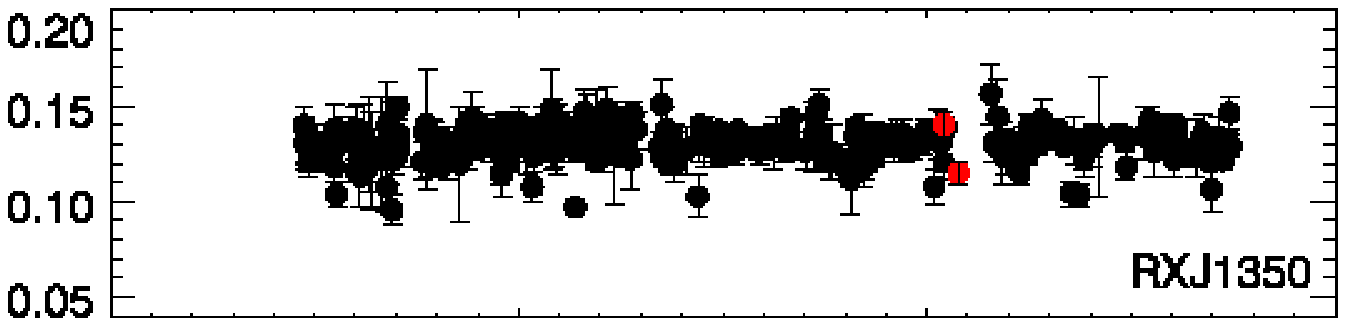}}
    \subfigure{\includegraphics[width=14cm]{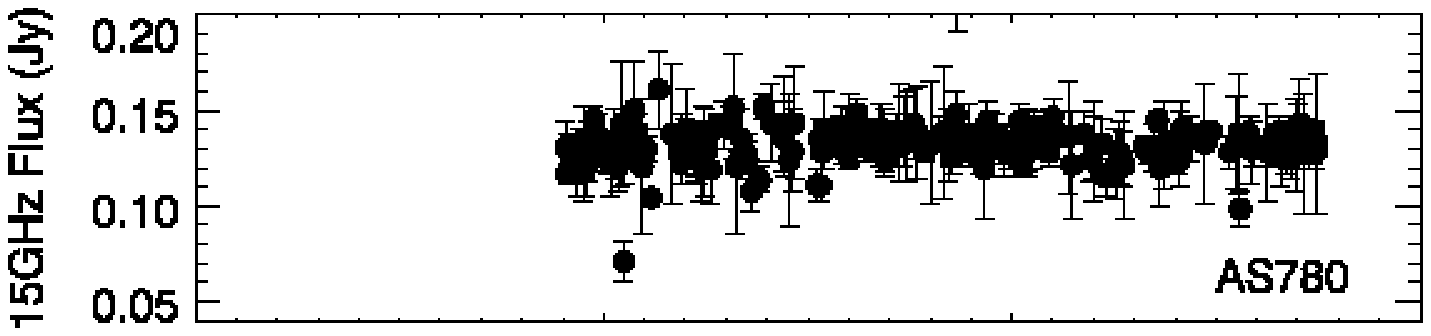}}
    \subfigure{\includegraphics[width=14cm]{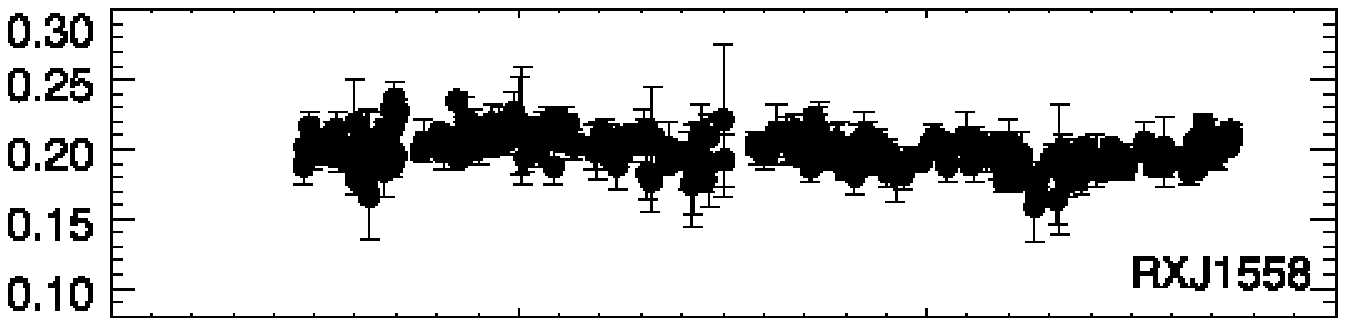}}
    \subfigure{\includegraphics[width=14cm]{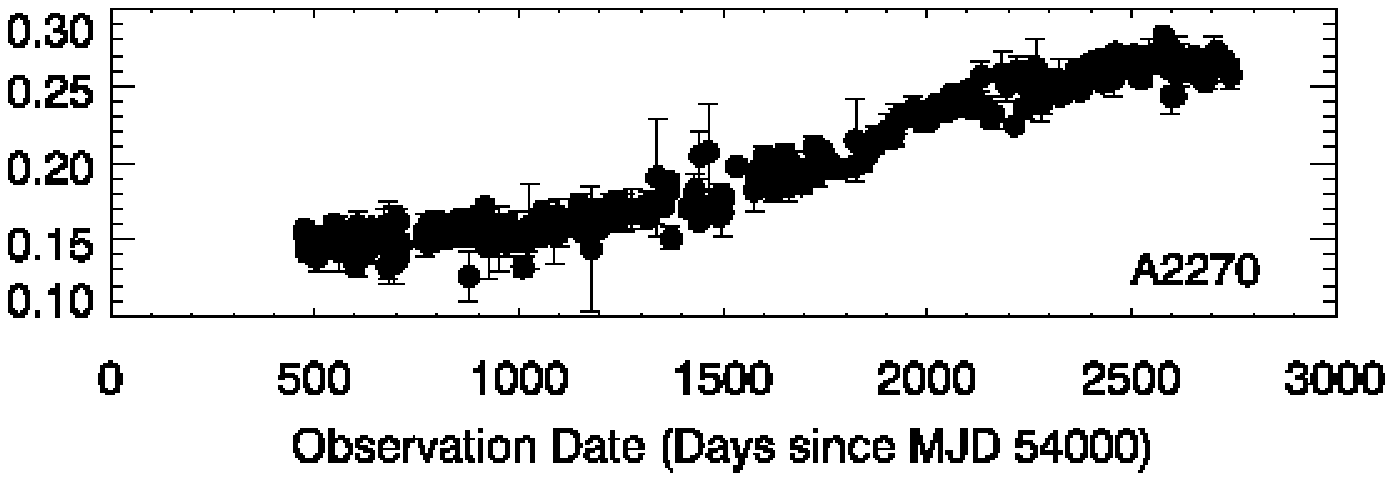}}
  \caption{Longer term lightcurves for the five sources that were in the original list of OVRO targets.  Where AMI observations were available, the flux is corrected to 15~GHz using an index fitted to the AMI data split over the six channels for that observation specifically.  We see strong variation over typically year timescales in both RXJ0439 and A2270 at 15~GHz. Note that MJD 54000 corresponds to September 22$^{nd}$ 2006.}
 \label{SUPER_5_LIGHTCURVES}
\end{figure*}

\subsubsection{Variability at 15GHz: OVRO}
For the eleven sources added to the OVRO monitoring campaign circa MJD 56320 (January 2013; see bottom panel, Table \ref{OVRO_VI}) we find that no sources have a positive Variability Index.  This suggests that for all eleven sources the difference in measured flux is within the 1-$\sigma$ uncertainties of the individual measurements such that this leads to an imaginary value for the square root.  This is somewhat surprising, since for a non-varying population we might expect around half to have slightly positive V.I.s and half slightly negative V.I.s.  The lack of any positive V.I.s indicates that our flux uncertainties are over-estimated.  A positive V.I. requires that the numerator within the root of Equation \ref{VI_equation} be positive, which at the most basic level just requires that the average fractional flux deviation from the mean flux be larger than the mean fractional error.

In Figure \ref{OVRO_var} we plot the typical percentage deviation from the mean flux for each source, as well as the mean fractional error on the flux (displayed as a percentage).  Additional data are provided in Table \ref{OVRO_VI}.  We see that in all instances of the eleven sources monitored since January 2013 the typical flux deviation is less than the typical error, hence showing why no real V.I.s were returned.  For a non intrinsically-varying population of sources we might expect $\approx$68\% of flux measurements to lie within one standard error of the mean.  That our typical percentage uncertainties are around twice our typical flux deviation from the mean suggests that our errors are over-estimated by a factor of 2-3.  We cannot steadfastly claim that our sources are intrinsically non-varying and should therefore adhere to Gaussian statistics, hence we merely claim that they typically are restricted to varying by $\leq$10\% on roughly one-week timescales.  However, we note that this limit may be a factor of 2-3 lower.

Two of the five sources that were in the OVRO monitoring campaign over the full period (top and middle panels, Table \ref{OVRO_VI}) return real Variability Indices when the V.I. is calculated over the full timerange (A2270 and RXJ0439) but only one (A2270) is recovered when the V.I. is calculated over a timerange restricted to match that of the other eleven sources (see Figure \ref{OVRO_var_extra5}).  In Figure \ref{SUPER_5_LIGHTCURVES} we show the 15~GHz lightcurves for these five sources.  Clearly both A2270 and RXJ0439 display large flux increases over the full timescales considered.  It is interesting to then note that the variability of RXJ0439 as shown by the increase from MJD 54500 - 55500 would be missed if considering only the restricted timescale. This highlights a shortcoming of the V.I. technique. If a source varies significantly (e.g. RXJ0439, Figure \ref{SUPER_5_LIGHTCURVES}, MJD ~54500-55500) then its average flux difference from the mean will be much larger than the mean flux and so a positive V.I. will be returned.  If the source then continues to be monitored but does not subsequently vary much (e.g. RXJ0439, MJD ~55500-57000) then the {\it average} flux difference from the mean will drop significantly.  If data from monitoring over a substantially long time is combined, this will eventually result in the average difference from the mean dropping below the average error and so the measured V.I. is no longer truly reflective of the variability seen early on.  The inverse of this effect can be seen by considering A2270 in Figures \ref{OVRO_var_extra5} \& \ref{SUPER_5_LIGHTCURVES}.  A2270's lightcurve shows that it varies significantly.  However, when only the matched timeframe data is used, A2270 returns a barely significant variability (Figure \ref{OVRO_var}).  Conversely, when the longer-term monitoring is included then a very high level of variability is seen (Figure \ref{OVRO_var_extra5}), which better reflects what is evident from its lightcurve..  These effects suggest that at 15~GHz, variability of sources is most likely to be seen in campaigns with high cadence over several year monitoring timescales.

The measureable percentage variation that would have returned a real V.I. (and therefore a significant measure of variability above the noise) varies from source to source, typically being lower for brighter objects.  However, by inspection of Figure \ref{OVRO_var}, we can claim that the 15~GHz variability of these BCGs is typically restricted to $\leq$10\% on roughly one-week timescales (see Table \ref{OVRO_VI}).  Additionally, we see evidence that much larger magnitude variation may be common in BCGs on few-year timescales.  This mirrors the finding of \cite{Hovatta07} who found variability on few-year timescales in a sample of blazars from a greater than 25~year monitoring campaign with the UMRAO and Mets\"{a}hovi telescopes. The OVRO monitoring campaign is ongoing hence the statistics on this longer-term variability will naturally improve with time.

\begin{table*}
\begin{minipage}[b]{\linewidth}\centering
\caption{Variability Indices for the OVRO data.  The mean and median timescales between observations are given in days, with N being the total number of times OVRO observed each source during the considered monitoring period (MJD 54470 to MJD 56750, 5$^{th}$ Jan 2008 -- 3$^{rd}$ Apr 2014).  Note that a real V.I. is recovered for RXJ0439 when the full monitoring period is considered but not when only using the restricted timescale (see also Figure \ref{SUPER_5_LIGHTCURVES}.  A much better understanding of the longer-term variability of BCGs is expected to become apparent with continued monitoring.} \label{OVRO_VI}
\begin{tabular}{|c|c|c|c|c|c|c|c|}
\hline\hline
Source & V.I. & Corresponding \% & Min. detectable \% &    Mean    &  Median   & Mean       & N \\
       &      &                  &                    & Timescale  & Timescale & Flux (mJy) &   \\
\hline
   &   & \multicolumn{3}{c}{\it Extended Time Baseline (from MJD 54470)}     &     &  &  \\
RXJ1558  & NaN  & NaN  & 5.9  & 7.5  &  4.0  & 195.8  &  303  \\
RXJ0439  & 8.0  & 4.0  & 4.0  & 5.8  &  4.0  & 131.9  &  393  \\
AS780    & NaN  & NaN  & 10.3 & 9.0  &  6.0  & 130.0  &  205  \\
A2270    & 22.0 & 4.4  & 4.2  & 7.0  &  4.0  & 308.4  &  326  \\
RXJ1350  & NaN  & NaN  & 6.8  & 7.1  &  4.0  & 201.2  &  321  \\
\hline
   &   & \multicolumn{3}{c}{\it Matched Time Baseline (from MJD 56320)}     &     &   & \\
RXJ1558  & NaN  & NaN & 5.5  & 8.3  & 5.0   & 261.1 & 52 \\
RXJ0439  & NaN  & NaN & 3.8  & 6.3  & 5.0   & 131.5 & 67 \\
AS780    & NaN  & NaN & 10.5 & 9.6  & 5.9   & 128.9 & 45 \\
A2270    & 1.7  & 2.5 & 2.5  & 7.1  & 5.0   & 321.9 & 60 \\
RXJ1350  & NaN  & NaN & 6.3  & 8.4  & 5.0   & 195.5 & 50 \\
\hline
   &   & \multicolumn{3}{c}{\it Added to Monitoring List MJD 56320}     &     &   & \\
4C+55.16 & 0.48 & 2.2 & 2.2  & 6.8  & 5.0   & 1539.2 & 63 \\
A1348    & NaN  & NaN & 15.9 & 11.4 & 7.0   & 55.2   & 37 \\
A2052    & NaN  & NaN & 6.4  & 6.1  & 4.2   & 272.5  & 71 \\
A2055    & NaN  & NaN & 11.8 & 6.2  & 4.9   & 120.1  & 70 \\
A2390    & NaN  & NaN & 11.9 & 5.6  & 4.8   & 88.5   & 76 \\
A2415    & NaN  & NaN & 10.6 & 7.3  & 5.0   & 123.1  & 59 \\
A2627    & NaN  & NaN & 11.3 & 5.5  & 4.2   & 99.4   & 77 \\
A646     & NaN  & NaN & 14.8 & 9.1  & 5.0   & 58.2   & 47 \\
RXJ0132  & NaN  & NaN & 9.3  & 7.2  & 5.0   & 155.5  & 60 \\
RXJ2341  & NaN  & NaN & 7.2  & 6.6  & 4.1   & 156.9  & 65 \\
Z8276    & NaN  & NaN & 11.6 & 6.8  & 4.3   & 95.3   & 62 \\
\hline\hline
\end{tabular}
\end{minipage}
\end{table*}

\subsubsection{Variability at 150~GHz: GISMO}
Of the thirty-five individual sources observed with GISMO, twenty-three had at least one repeat observation hence allowing variability to be measured.  Of these, and considering only detections, seven sources were observed at all three epochs with the remaining sixteen observed twice.  Only four sources returned real two-point Variability Indices, namely Z8276 (V.I.=43.5 : 2 observations), A2270 (8.2 : 3), MACS0242 (23.2 : 2) and RXJ1558 (10.2 : 3).

\begin{figure}
\begin{center}
\includegraphics[width=9cm]{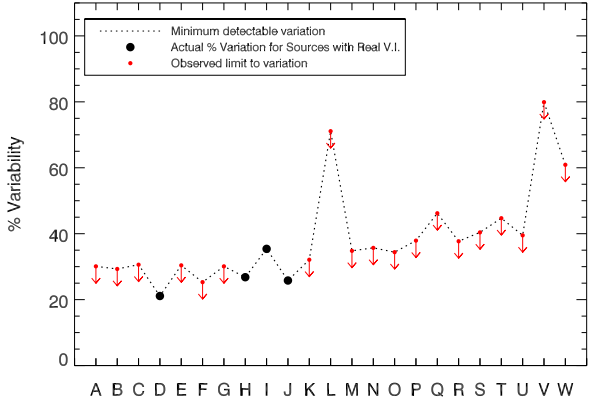}
\caption{Detectable variabilities that would have corresponded to a positive Variability Index for the GISMO sources, sampled at 150~GHz on approximately 6 month timescales.  Sources are arranged left-to-right by their maximal measured flux at 150~GHz.  The detectable percentage variation is typically much lower for brighter sources, with some deviation from a one-to-one correlation due to uncertainties on flux measurements varying by source.  Sources are: A) Hydra-A, B) 4C+55.16, C) A3581, D) A2270, E) RXJ0439, F) RXJ2341, G) A2052, H) RXJ1558, I) MACS0242, J) Z8276, K) A2415, L) RXJ0132, M) AS780, N) A2627, O) RXJ1350, P) A2055, Q) Z8193, R) A646, S) RXJ1832, T) A2390, U) A496, V) E1821, W) PKS0745. }
\label{GISMO_var}
\end{center}
\end{figure}

However it should be noted that working out the V.I. for our GISMO observed sources is less informative than for those observed with OVRO since the V.I. works primarily by determining the difference from the mean flux.  Clearly, when only two (or three) observations of a source are available then such an average is less meaningful due to the large uncertainties on the flux.  Nonetheless, a real V.I. still indicates which sources are really variable at the investigated cadence.  Therefore instead of calculating the minimum detectable typical percentage variation about the mean (as was done for the OVRO data, see Figure \ref{OVRO_var}), for GISMO we instead calculate: \\
i) For sources with two observations - the {\it actual} minimum required percentage change between the first and second fluxes to return a real V.I.  (assuming the reported uncertainties)\\
ii) For sources with three observations - the {\it total} minimum required percentage change across three epochs, from the first measured flux assuming a simple model where the observations are symmetrically distributed around the flux of the central observation (i.e. S$_{1}$ - S$_{2}$ = S$_{2}$ - S$_{3}$).  This is a simplistic approach and would require a source that does not vary between epochs 1 and 2 to vary by twice the expected mean between epochs 2 and 3 to be detected, however it does allow an {\it estimate} of the minimum average variation per six-month window that would be detectable.  \\
The results of this are plotted in Figure \ref{GISMO_var}.  We see that we are sensitive only to variation of above typically 30\%.  The {\it actual} percentage variations of the four sources with real V.I. values are also shown.

\begin{figure}
\begin{center}
\includegraphics[width=9cm]{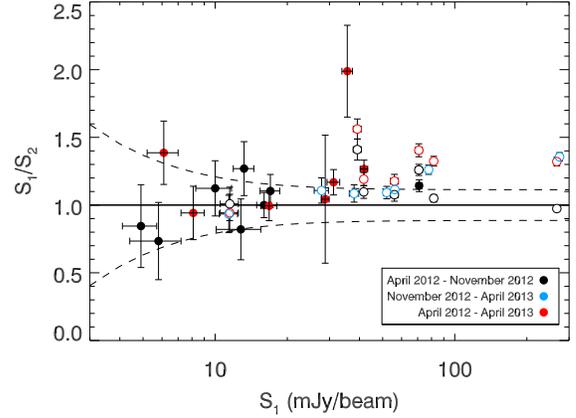}
\caption{Ratio of the measured GISMO fluxes at 150~GHz between each pair of epochs.  The dashed lines indicate the error envelope due to the 8\% typical flux uncertainty.  Open symbols are for sources that appear more than once due to being observed during all three epochs; consult also table \ref{GISMO_FLUXES}.  Sources typically appear to be fading, as expected for a sample chosen as the brightest at the selection epoch.  The source that appears to show very strong variation is Z8276, which is seen to fade drastically at 150~GHz between April 2012 and April 2013.}
\label{GISMO_Variability}
\end{center}
\end{figure}

In Figure \ref{GISMO_Variability} we plot the fractional flux changes between each pair of GISMO observations for a source.  As in Figure \ref{GISMO_var} we see that the majority of observed variability is within the error envelope.  The large variations of some sources as seen in Figure \ref{GISMO_var} are again apparent. One point that becomes obvious in this particular representation is that the majority of our observed variation is of sources fading over the 1-2 years covered.  This may hint towards systematic flux offsets between GISMO runs.  However, these are carefully checked for\footnote{Performance reports available for each run at http://www.iram.es/IRAMES/mainWiki/Continuum/GISMO/Main} and should not dominate.  A natural alternative explanation arises simply because we selected the sources to be (likely) the brightest amongst our parent sample at 150~GHz as a result of their lower frequency spectral shapes.  These most active sources are therefore likely to have been selected at their peak brightnesses and subsequent follow-ups will naturally be expected to see the flux falling.    

The April 2013 observation of M0242-21 was observed with a high optical extinction (tau = 0.70), which may contribute to the GISMO flux drop of about 50\%.  However, we note that this source is a Submillimeter Array (SMA) calibrator and shows a similar percentage flux decrease over this period (at 1.3mm\footnote{http://sma1.sma.hawaii.edu/callist/callist.html}) suggesting that the variability is real.

Given the high error envelopes, that we detect variability amongst our sample of repeat observations at 150~GHz suggests that significant variation of such sources at high radio frequencies may be relatively commonplace over 6-month to year timescales.  Such variability has important implications particularly for Sunyaev-Zel`dovich observations (see Section \ref{DISCUSSION}).

\subsection{Measuring Percentage Variation} \label{VAR_PERCENT}

\begin{figure*}
  \begin{minipage}{\linewidth}
  \centering
  \includegraphics[width=16cm]{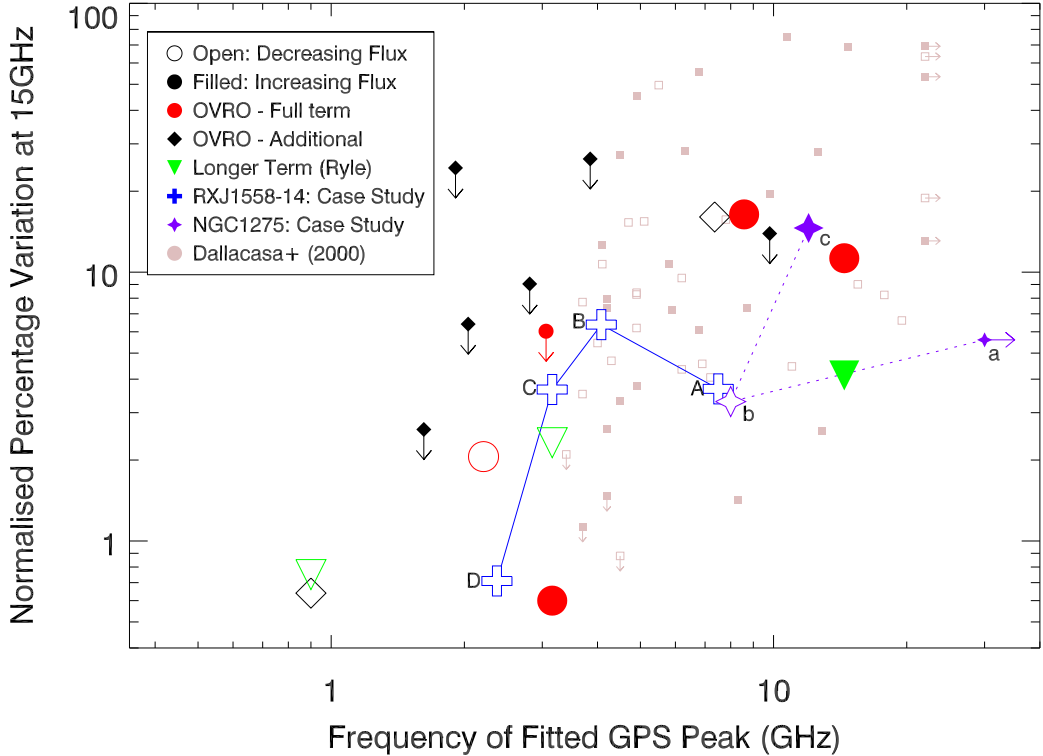}
  \end{minipage}
  \caption{Percentage variation at 15~GHz as a function of the position of the fitted GPS peak.  Note that the percentage variation has been normalised to the implied {\it annual} variation.  This allows comparison of sources monitored over different timescales and with a variety of sampling timescales.  Values for OVRO sources correspond to those calculated in Section \ref{VAR_PERCENT}.  Three sources have historical observations from the 1970/80s at 15~GHz with the Ryle telescope (see text Section \ref{VAR_PERCENT}) and we include the percentage variations over these longer timeframes, calculated in a similar way for comparison.  Open symbols denote flux decreases over the monitored period whereas filled symbols denote flux increases.  We see a general tendency for higher peaking sources to be more variable although this is not a strong trend and the ratio is highly dynamic, as highlighted by the case studies of RXJ1558-14 (shown as large crosses) and NGC1275 (shown as large stars - also see Section \ref{DISCUSSION}).  For comparison we plot as faded points the normalised percentage variations in the OVRO monitoring for the HFP sample of Dallacasa \etal (2000). These inhabit a similar distribution to the peaked components within our BCGs showing that their behaviours is not dissimilar to young radio sources elsewhere.}
  \label{VAR_PERCENT_PLOT}
\end{figure*}

In addition to a V.I., we also determined a measure of the typical annual absolute percentage variability of the sources monitored by OVRO at 15~GHz. Measuring the maximal percentage variation of a source over any given time period is highly susceptible to measurement uncertainty, since one bad unidentified outlying flux can greatly skew the result.  Instead we effectively determine a robust measure of the maximal gradient of the lightcurve during the monitored period. To do this, we measure the mean of the six percentage differences between the 95$\&$5th, 94$\&$6th, 93$\&$7th, 92$\&$8th, 91$\&$9th and 90$\&$10th percentiles (n.b. for sources only in the monitoring list since January 2013 with less than 100 observations, we used the 5th to 95th percentile range and the 5 unique ranges below this to a minimum separation of the 85$\&$15th percentiles - beyond this we automatically classified any resulting measured variability as an upper-limit). 

A mean over six ranges was found, by trial and error, to be a reasonable compromise between too few measurements being susceptible to random fluctuations but too many requiring the inclusion of less-separated percentile differences, over which less variation is expected even for varying sources.  We found that taking the 5th to 95th percentile range as our maximum ensured that our measurements were robust against outlying data-points.  Additionally, averaging over an even number of percentile ranges allows non-varying sources to have variations averaging to effectively zero (note that in practice the probability of it {\it actually} averaging to precisely zero is minimal, however the probability of it approaching zero is increased with an even number).

To determine whether these measured percentage variations most reliably constitute genuine variability or an upper-limit on variability, we took a measure of the two-point V.I. values for each of these six percentile ranges.  We required that a minimum of four of these return real V.I. values for a value to be assigned to the percentage variation, otherwise the measured percentage variation was taken to be an upper-limit.  We normalise our mean measured percentage variation values by the mean difference in years between the measurements used, thus recovering a measure of typical annual variation.  Where only a limit on variation is recovered, we normalise by the total monitored period in years (roughly 1~year for the 11 sources added to OVRO list in January 2013).

Three sources (4C+55.16, RXJ0439+05 and Z8193) have historical 15~GHz data preceding OVRO monitoring from either the UMRAO campaign or pointed observations with the Ryle Telescope.  We combine these with the OVRO data and find the percentage variations over longer timescales.

In Figure \ref{VAR_PERCENT_PLOT} we show our calculated absolute percentage variations at 15~GHz as a function of the peak position of the fitted GPS.  A mixture of sources increasing and decreasing in brightness is seen.  A weak general trend for the highest peaking sources to show most variability is seen.  Such a trend is expected, as a higher turnover frequency suggests self-absorption closer to the jet-base and hence emission from smaller scales, which can more easily translate to faster emitted variability.  However, it must be noted that the GPS-peak frequency is expected to move and also the variability is non-constant, hence both parameters are expected to undergo linked evolution.  We illustrate this dynamic relationship by including on Figure \ref{VAR_PERCENT_PLOT} evolutionary tracks for two sources with long-term monitoring; RXJ1558-14 and NGC1275.  Both are seen to move extensively across Figure \ref{VAR_PERCENT_PLOT} but always remain within the region occupied by the other points.  We consider these sources as case studies in sections \ref{R1558_section} and \ref{1275_CASE_STUDY}.

For comparision with the more general galaxy population we additionally include on Figure \ref{VAR_PERCENT_PLOT} the normalised percentage variations at 15~GHz for the fifty sources (of fifty six) in the High Frequency Peaker (HFP) sample of \cite{Dallacasa00} that are in the OVRO monitoring list.  These show a similar weak trend to the peaked components of our BCG spectra, suggesting that our peaked components show similar behaviour to young recently activated sources. This indicates that they are associated with recent episodes of enhanced radio emission.

\subsubsection{A GPS link to variability?}
The behaviour of NGC1275 and RXJ1558-14 suggest that a direct link between GPS-peak frequency and variability is not present.   Instead a dynamic ratio is seen between these parameters whereby the position (and indeed presence) of a self-absorption peak in relation to the observed frequency of a source can have a large effect on the amount of variability seen.  

The position of the peak appears to weakly correspond to how quickly a source can vary at a given frequency although not necessarily how quickly it does vary over any given epoch.  Indeed, whilst the presence of a flat or peaked component in a BCG's SED can be indicative of it being more likely to vary at high radio frequencies there does not appear to be a single observable proxy for the level of variation.  The only way to robustly remove high frequency contaminants remains contemporaneous observation.

It should be noted that GPS sources actually constitute a range of types and that many sources classified in the literature as GPS and HFPs are actually flaring flat spectrum quasars or blazars \cite[see Discussion by][]{Torniainen07}.  The variability of these sources is naturally expected to be different to genuine, long-term GPS objects.  Amongst genuine GPS sources the peaked spectral shape is maintained over decade timescales and long-term variation of the absolute flux level of this is seen \cite[][]{Aller02}.  This longer term variation (years to decades) is similar to what we see in the peaked components of NGC1275 and RXJ1558-14 and hence further supports that the variability herein is likely due to similar physical processes to that in more typical GPS objects, which may be related to opacities within the jet flow.

\section{Discussion} \label{DISCUSSION}

\subsection{Case Study: 4C+55.16}

\begin{figure}
\begin{center}
\includegraphics[width=9cm]{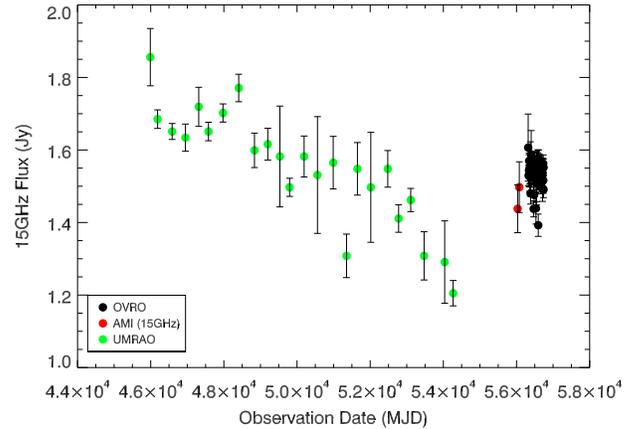}
\caption{Lightcurve for 4C+55.16.  UMRAO points are the yearly average.  Note that AMI observations have a central useable frequency of 16.05~GHz, which was corrected to the value shown here using the in-band spectral index.  Note that MJD 46000 corresponds to 27th October 1984.}
\label{4C55p16_lightcurve}
\end{center}
\end{figure}

As mentioned in Section \ref{OVRO_UMRAO_Intro} one of our sources, 4C+55.16, has near-continuous 15~GHz lightcurves available for almost two decades, allowing us to consider its longer term activity.  In Figure \ref{4C55p16_lightcurve} we show the combined lightcurve consisting of UMRAO, AMI and OVRO observations.  Note that our AMI observations have a central frequency of 16~GHz.  We fit a single spectral index to the SED of 4C+55.16 above 10~GHz, recovering an index of 1.29\footnote{S=A$\nu$$^{-\alpha}$} and use this to correct our AMI fluxes to 15~GHz.

A steady decline in total flux density is seen towards the end of the UMRAO lightcurve for this source (2007, MJD~$\approx$~54000).  Our AMI fluxes are consistent with those recovered with OVRO, within the errors.  However, if we remove the absolute flux calibration of the AMI observations then they are perhaps self-consistent with having caught the source as it brightens, before it is then reasonably steady throughout the OVRO monitoring.  Variability of this at 15~GHz is slow, varying by approximately 20\% on decade timescales.  This compares to significantly less variability on typically one-week timescales.  However, from the GISMO observations of this source we do get faster variation at higher frequencies showing that although the overall SED may only vary slowly over several years, short-term `flickering' of the flux at frequencies above 100~GHz is still evident over much shorter periods.

\subsection{Case Study: RXJ1558-14} \label{R1558_section}

\begin{figure}
  \begin{minipage}[b]{0.5\linewidth}
  \centering  
  \includegraphics[width=9cm]{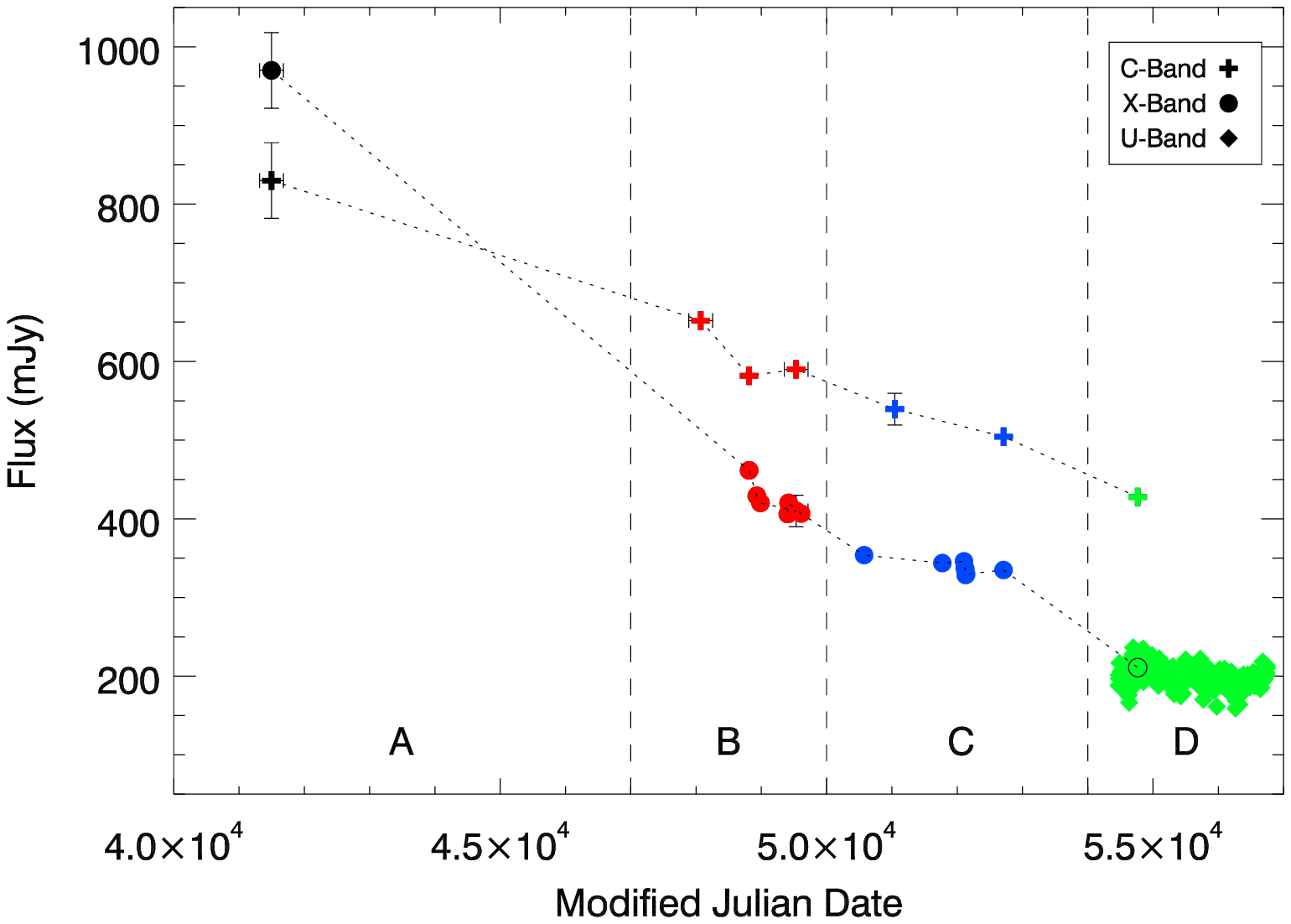}  
  \end{minipage}
  \begin{minipage}[b]{0.5\linewidth}
    \includegraphics[width=9cm]{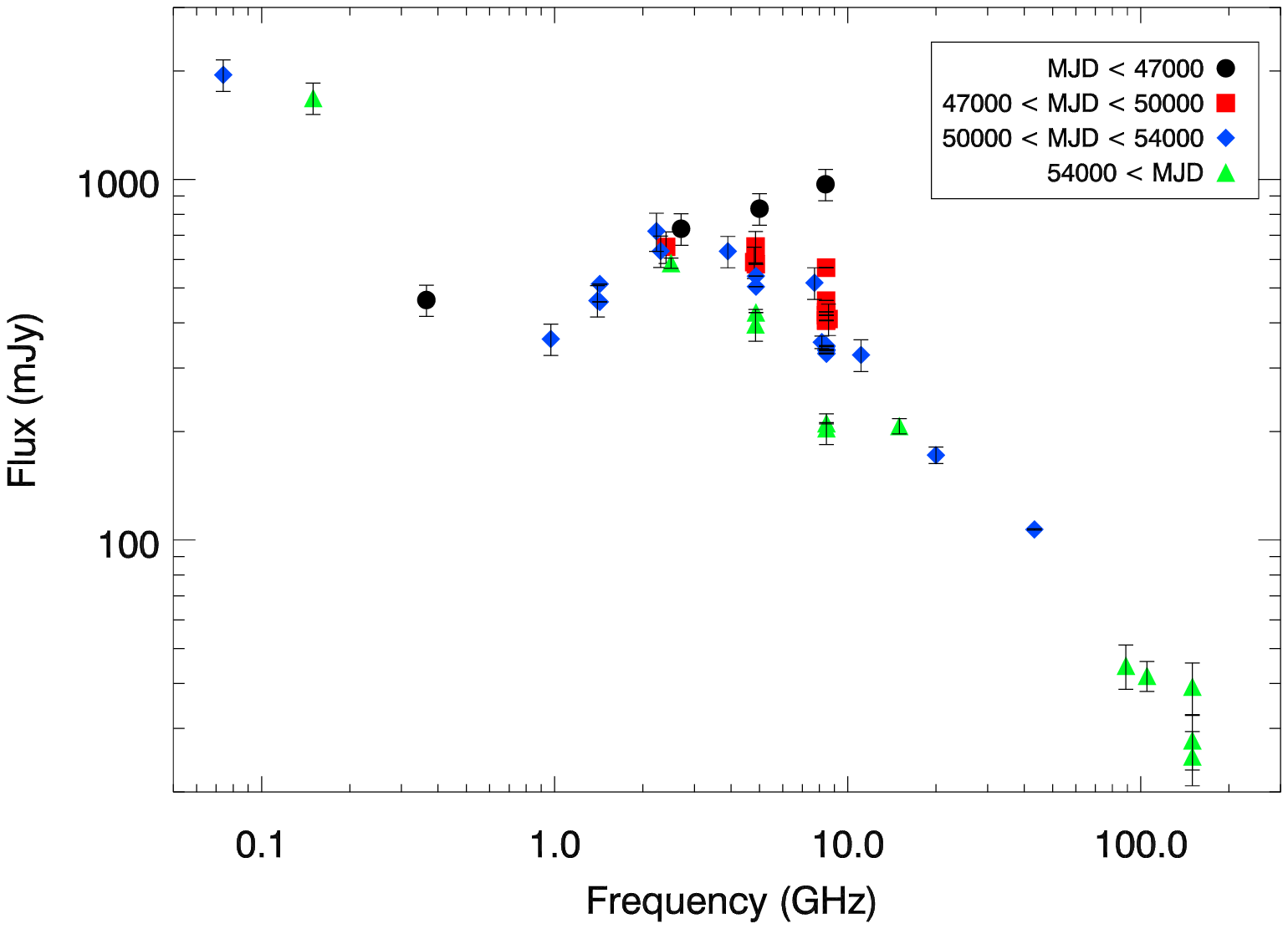}
  \end{minipage}
  \begin{minipage}[b]{0.5\linewidth}
    \includegraphics[width=9cm]{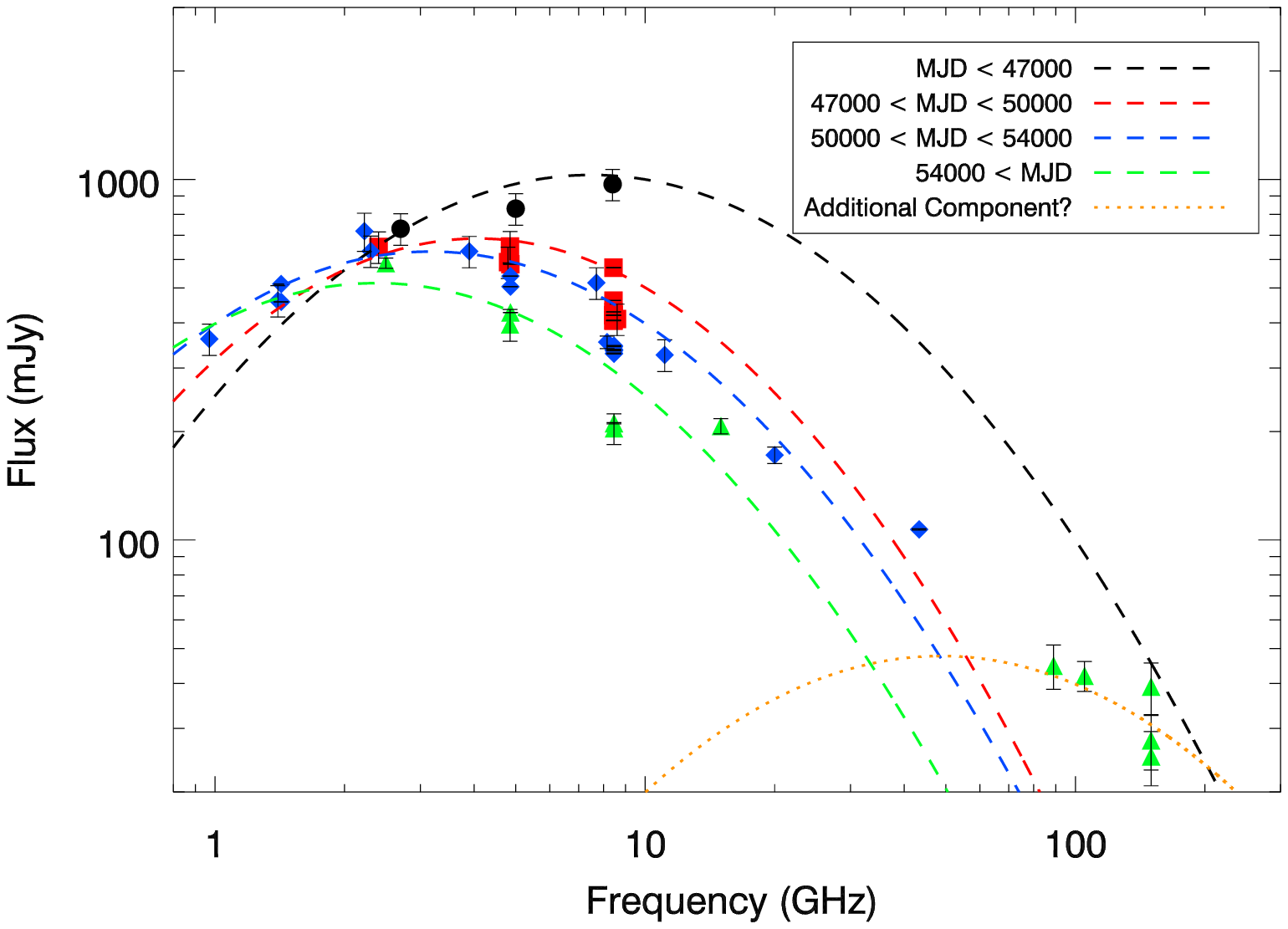}  
  \end{minipage}
  \caption {Top panel: C- ($\approx$5~GHz), X- ($\approx$8) and U-Band ($\approx$15) light-curves for RXJ1558-14, showing steady decline since the first data point, in 1972. Note that the X-band datum point in time window D has a border included to differentiate it from the overlapping U-band data. Middle panel: SED where points coloured corresponding to the time-windows indicated in top-panel.  Bottom panel: Illustrative fits to the GPS component of the SED for each of the time windows.  As time goes on the GPS-peak appears to move both downwards and to the lower frequency.  Note that the highest frequency points in time-window D (and to a lesser degree window C) appear discrepantly high.  This may indicate a new period of activity, illustrated as a dotted orange line in the lower panel.  Such indication suggests that a continued VLBI monitoring of this source may see new parsec-scale knots form and propagate outwards within the next few years.} 
  \label{R1558_SEDs}
\end{figure}

RXJ1558-14 shows remarkably similar behaviour to NGC1275, albeit displaying around an order of magnitude less flux.  NGC1275 is a well-studied source which constitutes the prototypical example of a variable BCG with a GPS-like component showing steady mm-variability over few year timescales \cite[see][for a thorough discussion of the variability properties of NGC1275]{Dutson14}.  

RXJ1558-14 has been monitored by OVRO since 2008 and shows little variation at 15~GHz over this timeframe.  However, the source has historically been used as both a VLA and VLBA phase calibrator and has a wealth of observations available in the NRAO archive.  We can therefore study its behaviour over long timelines. 

In the top panel of Figure \ref{R1558_SEDs} we show long-term lightcurves at both X-band (8.4~GHz) and C-band (4.8~GHz) for RXJ1558-14, as well as the more recent U-band (14.9~GHz) monitoring.  Data were compiled from the literature and by downloading FITS images for all available observations in the NRAO-archive and extracting flux measurements using \textsc{IMFIT}.  We split the lightcurves into four time-windows, indicated on Figure \ref{R1558_SEDs} as `A', `B', `C' and `D'.  In panel A \cite[see][]{Wright90} we see that the 5-to-8~GHz spectrum was inverted\footnote{We use `inverted' within this paper to refer to radio spectra with increasing flux to higher frequency, which is the opposite to the usual scenario for a typical, unobscured synchrotron spectrum.}, although this is not the case at later times.  

In the middle panel we plot the SED for RXJ1558-14 with data-points coded to correspond to these four time-windows.  In the bottom panel we focus only on the GPS-like part of the spectrum, again coding the data-points although here we additionally show illustrative GPS-models to the data in each of the time windows.  As we move through time windows A to D we see the GPS-peak appear to move both downwards in flux and also in turnover frequency.  

RXJ1558-14 displays double-lobed structure on parsec scales with an unresolved central core, as seen at 2.224 and 8.154~GHz in 1997 by \cite{Fey00} and more recently at 4.86~GHz in 2012 (see Hogan \etal 2015c).  Although these observations are at different frequencies and thus cannot be used to get any secure measure on the expansion, it is worth noting that no new features are seen to arise between these observations.  The lobes seen in the 4.86~GHz map of Hogan \etal 2015c lie approximately equidistantly 7~milli-arcseconds either side of the central component.  If we associate the high core fluxes observed in 1972 with the launching of these knots then in the intervening 40~years the knots travel with an apparent transverse velocity of approximately 1.02~c, permitting viewing angles between 45$^{\circ}$ and 90$^{\circ}$ that are entirely consistent with the symmetry seen in the source.  It is therefore consistent, and indeed highly likely, that the very high core fluxes observed in the 1970s were attributable to the emission of the features which are now observed as expanding lobes on milliarcsecond scales and the subsequent fall in flux is attributable to these lobes propagating away from the core.  Such a scenario would suggest that variability of RXJ1558-14 (and indeed other sources) above a few GHz may be associated with the launching of individual jet components and may precede the time at which these are observable given current angular resolutions.  VLBI monitoring of self-absorbed sources that are varying above a few GHZ may therefore allow such flux changes to be directly linked to individual parsec-scale jet components, as has been observed elsewhere \cite[e.g. NGC1275:][]{Suzuki12}.

The highest frequency data in window D (and a lesser degree, window C) appear discrepantly high and may be indicative of a new period of activity that will become apparent at lower frequencies only as the emitting knots move down-jet and the self-absorption turnover peaks to lower frequency.  VLBI monitoring of the source may therefore detect new knots forming and propagating outwards in the coming decade.

In further analogy to NGC1275 we note that RXJ1558-14 has a steep spectrum power-law tail to low frequency.  In NGC1275 this tail is associated with a 300kpc radio mini-halo.  A similar structure may be present around RXJ1558-14.  However, we note that in the TIFR GMRT Sky Survey (TGSS) imaging of this source there appears to be reasonably symmetric structure extending almost 200~kpc from the core (PA $\approx$ 91.8) and hence the steep spectrum emission may be indicative of large scale lobes from a previous large outburst.  Interestingly, X-ray cavities are detected by Chandra in this source \cite[][]{Hlavacek-Larrondo13, Russell13} with total physical extent of 17.1$\pm$2.3kpc.

Due to its relatively good long-term SED monitoring we consider RXJ1558-14 as a case study in Figure \ref{VAR_PERCENT_PLOT}.  We calculate the GPS-peak frequency and measured variability in RXJ1558-14 for each of the four epochs highlighted in Figure \ref{R1558_SEDs}. We can directly measure the 15~GHz percentage variability in epoch D only, during which OVRO monitoring data are available.  For each of the other three epochs, we measure the C band (4.86~GHz) and X band (8.44~GHz) values at the epoch boundaries, using these to to extrapolate an estimate of the 15~GHz flux.  We plot the position of RXJ1558-14 on Figure \ref{VAR_PERCENT_PLOT} during each of these four epochs.  Initially, in epoch A we find typically 4\% variation annually at 15~GHz.  During this epoch the peak is just below 15~GHz and hence the spectrum is still relatively self-absorbed at this frequency.  In epoch B we are further in time from the event that caused the peaked component of RXJ1558-14's spectrum. The peak has moved to lower frequency meaning that a steeper part of the spectrum crosses 15~GHz and hence contrary to the general trend we see higher variability during this epoch.  In epochs C and D the variability continues to lessen as the turnover frequency drops.  It may be possible that a second, sub-dominant component peaked at much higher frequency (see Figure \ref{R1558_SEDs}, bottom panel) is also present in the spectrum. If this component varied differently to a lower peaked component then it could show increasing flux as the other was decreasing at 15~GHz and hence could lessen observed monochromatic variability.  

\subsection{Case Study: NGC1275} \label{1275_CASE_STUDY}

As presented and discussed in \cite{Dutson14}, NGC1275 has been regularly monitored for over four decades and shows long term variability both in terms of its GPS-peak frequency and spectral normalisation.  

In order to place NGC1275 on our Figure \ref{VAR_PERCENT_PLOT} we take estimates of the GPS-peak frequency and 15~GHz variability in three epochs: 1979-83 (epoch a, Figure \ref{VAR_PERCENT_PLOT}), 1983-2005 (epoch b) and 2005-2013 (epoch c) \cite[see also Figure 5,][]{Abdo09}.  Initially, during epoch a, the peak is well above the 15~GHz monitoring frequency and we see a sharp increase of around 5.5\% annually. During epoch b, the self-absorption peak `rolls up' the spectrum to lower frequency and is accompanied by a relatively steady decline of about 3\% annually over an extended period that may be coinciding with expanding features on parsec scales. Finally in epoch c there is another period of sharply increasing flux (almost 12\% per year) which coincides with the peak moving to higher frequencies.  Overall the trend is for long-term climbs followed by troughs, with the variability at 15~GHz dependent on both the position of the peak relative to this as well as the underlying variability in the total normalisation of the spectrum.

\subsection{Comparisons to the general GPS population}  

An important consideration to make is whether the variability and wide variety of spectral shapes that we are seeing in BCGs, is exclusive only to this special class of objects or whether they are applicable to the wider population of radio sources.  

There are many examples of high peaking sources in the literature \cite[e.g.][]{Rodriguez14} but large samples are required to fully determine how common these are amongst the overall radio-source population.  Classifying peaked radio sources is difficult.  Often only non-contemporaneous observations are available meaning that variability of inherently flat spectrum sources can lead them to be mis-classified \cite[e.g. see discussion in][and references therein]{Tornikoski09}.  Alternatively, even when a spectrum is sampled at multiple frequencies and a peak observed, without follow-up over several years it is very difficult to determine whether this is a true, slowly varying GPS source or a usually flat-spectrum source undergoing a rapid flare.  Further complicating the issue, it appears as if the contamination of GPS catalogues by variable sources and BL-LACs is dependent upon the GPS host galaxy; quasar-type GPS sources are much more commonly mis-identified than GPS sources in more typical galaxies \cite[][]{Torniainen05,Torniainen07}.

In the AT20G survey of the southern sky at 20~GHz \cite[][]{Murphy10}, there are 3763 sources (detection limit of 40mJy at 20~GHz) that have simultaneous observations at 5, 8 and 20~GHz. Of these, 21\% are found to have peaked spectra, 14\% show a spectral upturn over this range and 8\% are inverted, suggesting a peak above 20~GHz.  This fraction of inverted and peaked sources is significantly higher than the $\geq$3.4\% of our parent sample peaking above 2~GHz.  There is however a clear selection bias towards a higher fraction of inverted and peaked sources in the 20~GHz selected sample than in our sample of BCGs detected at high radio frequencies but selected from a parent sample unbiased by radio priors.  Indeed, in the 5 deg$^{2}$ sampled down to 2.5mJy for the AT20{\em -deep} pilot survey \cite[][]{Franzen14} 83 sources are detected and have near-simultaneous spectra from 1.4-20~GHz.  Of these only 2.4\% show a spectral upturn, 15.7\% are peaked and 6.0\% are peaked above 20~GHz.  These reduced fractions of `exotic' spectral types in comparison to the full AT20G suggest that the area-limited nature of the survey coupled with the deeper detection limit means that more of the `typical' single-spectrum sources that constitute the bulk of extragalactic radio sources are detected. \cite{Franzen14} find that about 12\% of their sources vary by more than 15\% over 3 years at 20~GHz, which is comparable to the levels of variability we detect for our sources at 15~GHz (see Figure \ref{VAR_PERCENT_PLOT}).

\cite{Dallacasa00} matched 1740 sources with S$_{4.9~GHz}$~$>$~300mJy in the GB6 catalogue \cite[][]{Condon94} to the NVSS catalogue \cite[][]{Condon98}, finding 102 candidate inverted spectra.  Simultaneous follow-up of these candidates at 1.365 - 22.46~GHz with the VLA revealed 55 to be genuinely inverted sources whereas the remaining 47 were flat spectrum sources whose variability meant that non-contemporaneous observations had caused their spectra to appear peaked.  These 55 sources (their HFP `bright sample') thus mean that 3.2\% of their initial matched sample have spectral peaks $\geq$3.4~GHz.  Taking the same frequency cut we find more than 1.9\% of our parent sample peak at similarly high frequencies.  Whilst still lower than the detected fraction of \cite{Dallacasa00}, if we consider that our sample contains BCGs detected irrespective of their radio-loudness and hence contains a portion of radio-quiet objects then our detection samples are in reasonable agreement.  We note that in a follow-up paper,  \cite{Stanghellini09} define an HFP `faint sample' of sources with S$_{4.9~GHz}$ between 50 and 300~mJy in the GB6 sample.  Sixty-one HFPs with peaks at frequencies above 1.3~GHz are identified in this sample however they do not state the size of the parent sample, meaning a comparison to their detection fraction cannot be made.

Overall it appears as if the peaked components of our BCG spectra are similar in their properties to the general GPS/HFP populations.  However, a worthwhile point to note is that of the 26 sources in which we fit a GPS-like component, 20 of these (representing 76.9\%) also contain a steep spectrum power-law to low frequencies.  The usual interpretation of peaked radio sources is that they are young \cite[][]{O'Dea98}.  The presence of an accompanying steep component in the majority of our BCGs rules out these being truly young radio sources.  Instead, the peaked components could be interpreted as due to either enhanced radio emission at recent times in a long-lived source, or that the radio source in BCGs can be restarted on short timescales.  The low level variation that we see in our BCGs, coupled with the high duty cycle of detection, points to these being very long-lived radio sources whose radio output can show quite large variations in magnitude over time. \cite{Hancock10} followed up 21 sources found to be inverted in the AT20G between 8 and 20~GHz with the Australia Telescope Compact Array (ATCA) at 40 and 95~GHz.  Of these 21 targets, 12 were found to be genuine peaked sources, of which 3 (25\%) showed evidence of being restarted rather than truly young radio-sources, indicating them to be longer-lived sources that have undergone a recent episode of high activity.  Although caution must be employed for the small numbers that we are considering, this tentatively supports the idea that peaked components in BCGs are much more commonly attributable to enhanced activity in long-lived sources than in the GPS population as a whole, which is dominated by truly `young' sources.  In other words, BCGs do not just have a single period of strong jet-emitting activity and then fade.  They instead live for a long time, but remain active and effectively young at heart.

We point out that continuous radio core activity over a long time is known in galaxies not selected as BCGs; for example, FRII radio sources, which have measured ages of up to 10$^{8}$ years {\bf \cite[see e.g.][]{Machalski07, Mathews12}} -- and are {\em still} being powered -- can have cores with radio luminosities similar to those in this paper, with many having flat or rising spectra.  However, if the AGN activity of BCGs has been regulating cluster cores for around half the Hubble time \cite[e.g.][]{Vikhlinin06, Pratt10, McDonald14} then this implies activity timescales over an order of magnitude longer than even these long-lived sources.

\subsection{Implications as an SZ contaminant}

Considering the SED shapes of our sampled sources (see Appendix \ref{SED_APPENDIX}), it is clear that extrapolation of the specta from below 10~GHz towards the interesting range for SZ decrements (between roughly 15 and 200~GHz) will in many cases underestimate or completely overlook the contribution of an active self-absorbed component.  Added to this is the further complication that variability brings, requiring contemporaneous high-resolution observations to ideally account for contaminants (as is of course possible with interferometric SZ instruments such as AMI and CARMA).

For a sample of 45 galaxy clusters observed at 140~GHz with BOLOCAM, \cite{Sayers13} used the 1-30~GHz spectral index in addition to limits from their 140~GHz maps to constrain the contamination by point sources.  Although they concluded that typically only about 25\% of the clusters showed a greater than 1\% fractional change of the SZ signal, they noted that this level of contamination was much more prevalent in cool-core clusters (11/17, roughly 65\%).  We note that {\it all} of the BCGs in our current sample are believed to lie in cool-core clusters.  \cite{Sayers13} detect no clear point sources at 140~GHz from their sample of 17 cool cores, whereas from our significantly larger parent sample \cite[40\% of our 530 cluster parent sample are tagged as cool core:][]{Hogan15} we detect 32 at 150~GHz.  This suggests $\geq$6.0\% of BCGs in all clusters exhibit bright BCG emission in the mm-range rising to $\geq$15.1\% if only cool-core clusters are considered. These values are lower-limits since our 150~GHz follow-up is flux limited and also incomplete in that some clusters are not observed at 150~GHz. The true level of contamination in cool-core clusters could therefore be even higher. We note that the prevalence of flattened or inverted spectral components in non cool-core clusters at frequencies below 20~GHz is much reduced \cite[see][]{Hogan15}, hence the level of contamination at 150~GHz is expected to be significantly lower in these systems.  

Interestingly, of the $\geq$15.1\% of the CC-hosted BCGs in our sample, more than half ($\geq$8.5\% of the total CCs) have a peaked component with a turnover above 2~GHz.  Underestimating the point-source contamination will be particularly severe for extrapolation of the lower frequency spectrum in this type of object. SZ catalogues are therefore potentially biased against the inclusion of cool-core clusters. The true magnitude of line-of-sight integrated comptonization (Y$_{SZ}$) in these systems may be commonly underestimated, due to BCG radio emission canceling the SZ decrement.  Interestingly this could suggest that the mass bias between X-ray and SZ derived cluster masses is more prevalent than previously assumed.

During their observations, the Planck Consortium compiled extensive catalogues of high radio frequency sources \cite[e.g.][]{PlanckPCCSXXVIII, PlanckERCSCXIII}.  In \cite{PlanckERCSCXIII} the extragalactic source counts agree with those of the South Pole Telescope \cite[SPT,][]{Mocanu13}, Atacama Cosmology Telescope \cite[ACT,][]{Marriage11b} and the Wilkinson Microwave Anisotropy Probe \cite[WMAP,][]{Komatsu11}, at the lower frequency range of Planck.  However, \cite{PlanckERCSCXIII} show that there may be a steepening of the typical spectral index above 70~GHz which could mean that contamination of the Cosmic Microwave Background (CMB) power spectrum by radio sources below the Planck detection limit may be less than currently thought.  However, this may only apply to the bright end of the luminosity function.  A population of GPS/HFP sources (in addition to a potentially ubiquitious mm/sub-mm component, see Section \ref{ADAF_SCUBA2}) in low-luminosity AGN (LLAGN) and BCGs could present a low level contaminant to the CMB power spectrum.  Deep, high radio frequency radio surveys are required to shed light on the luminosity function of faint sources in this range.

\subsection{Potential Additional sub-mm Component} \label{ADAF_SCUBA2}

Examination of our SEDs (see Section \ref{SED_APPENDIX}) shows that a number of sources have inverted spectra between our GISMO observations at 150~GHz and SCUBA-2 observations at 353~GHz.  That is, the spectral slope switches from being falling as frequency increases at frequencies below 150~GHz to rising just afterwards.  This is most notable in A646, MACS1931-26 and RXJ1504-02.  The observed indices are too shallow ($\alpha~\approx~1.0$, in all three cases) to be solely attributable to dust-emission \cite[expected $\alpha$$>$$\approx$2.5, ][]{Edge99}.  However, combining the extrapolated flat spectrum core component with the expected dust contribution determined from Herschel-SPIRE observations at 250, 350 and 500$\mu$m \cite[Herschel Lensing Survey (HLS),][]{Egami10, Rawle12} can explain the spectral upturn in MACS1931-26.  However, there is still unaccounted for flux of $\approx$20 and 10 mJy at 353~GHz in A646 and R1504-02 respectively after accounting for both these contributions, roughly equivalent to P$_{353}$~GHz of 1 $\times$ $10^{24}$ W~Hz$^{-1}$ for both sources.

An intriguing explanation to account for this `missing flux' is that we may be seeing an additional component in the radio/sub-mm spectrum caused by an Advection Dominated Accretion flow \cite[ADAF: e.g.][]{Mahadevan97, Narayan98, Ulvestad01, Doi05, Narayan08}.  These are believed to exist, commonly in conjunction with jets, in low radiative efficiency accretion sources.  Multiple models exist that aim to explain the bolometric spectra of low luminosity AGN, often invoking a combination ADAF~+~jet \cite[e.g.][]{Falcke96, Wu07, Nemmen14}.  A common feature of ADAF models is the existence of a synchrotron self-absorbed component that typically peaks at mm-wavelengths.  To investigate whether such a component could explain our anomalous SCUBA-2 fluxes we take the base model of \cite{Mahadevan97} for a black hole mass of 5~$\times$10$^{9}$M$_{\odot}$, with a viscosity parameter of 0.3 and a ratio of gas to total pressure of 0.5.  For a typical BCG accretion rate in the ADAF range \cite[e.g. see][]{Russell13} of 3~$\times$10$^{-4}$~--~3~$\times$10$^{-5}$~M$_{Edd}$ this predicts an ADAF-power of $\sim$ 1 $\times$ $10^{22-23}$ W~Hz$^{-1}$ at 353~GHZ.  We note that this an order of magnitude lower than would be inferred if all of the missing flux at 353~GHz in A646 and R1504-02 were from an ADAF component.  The modeled ADAF-powers scale with black hole mass and accretion rate.  However, to align these with our missing flux would require unfeasibly high black hole masses or accretion rates that would no longer be compatible with an ADAF-like accretion structure \cite[e.g.][]{Done07}.  Furthermore, the upturned spectral index is expected to be close to $\alpha$~$\propto$~0.4 for an ADAF \cite[][]{Mahadevan97, Anderson04}.  Whilst our fitted core components in A646 and RXJ1504-02 underpredict the measured 150~GHz fluxes in both instances (by 1.0 and 0.3mJy respectively), a simple interpolation between the `missing fluxes' at 150- \& 353~GHz would give indices far too steep to be attributable to simplest case self-absorption.  Alternatively, extrapolation with an index $\propto$~0.4 from the missing flux at 353~GHz would overpredict our measured fluxes at 150~GHz.  

For the reasons given above we rule out the presence of a strong ADAF component to explain our missing flux.  We note that our findings do not rule out the presence of an ADAF component hidden well below our observed (or missing) flux.  The most likely explanation for the `missing flux' then becomes a combination of low level dust emission and variability in the jetted core component.

\section{Conclusions} \label{CONCLUSIONS_SECTION}

We have expanded the radio SEDs of a sub-sample of 35 BCGs selected from the sample of over 700 clusters considered by H15a.  These sources are all hosted by cool-core clusters where active feedback is believed to occur, and show the brightest flat or inverted components in their spectra at high radio frequencies.  These bright cores suggest that the central SMBH is currently accreting at an appreciable rate and so determining the spectra of these cores at higher frequency allows us to investigate possible physical origins for the radio emission. We considered the variability of these sources at both 15 $\&$ 150~GHz, enabling us to better understand the amplitudes of variation in SMBHs during their more active phases.  We find a wide variety of spectral shapes and in many cases see that these core components exhibit a spectral peak above 2~GHz, similar to the GPS/HFP population of young radio sources.  Our results can be summarised as follows:

\begin{itemize}
\item Cool-core hosted brightest cluster galaxies contain distinguishable active radio core components in over 60\% of cases (H15a).  This core can become dominant at frequencies higher than a few GHz.  We see in our current sub-sample that $\geq$15.1\% of cool-core hosted BCGs contain a radio source greater than 3mJy at 150~GHz (equivalent to a radio power of $\approx$ 1.2$\times$10$^{23}$W~Hz$^{-1}$ at our median redshift of 0.126) and that more than half of these ($\geq$8.5\% of all CCs) contain a distinguishable peaked component to their spectra with a turnover frequency above 2GHz that would be missed by lower frequency observations.  This shows that the majority of CC-hosted BCGs show recent activity, of which a significant fraction show strong active accretion at any given time.

\item These core components appear to be similar to the GPS/HFP populations that are usually interpreted as young radio sources.  That these peaked components are usually accompanied by steeper spectrum power-law emission at lower frequencies suggests that BCGs show near-continuous activity - these sources essentially enjoy a repeated youth.

\item Sources do not commonly show strong variability at 15~GHz on short (week to month) timescales, with a typical limit of $<$10\% indicating relatively steady accretion over these timescales.  We note however that there could be variability that we are missing due to it being hidden within error.  There is an increased incidence of variation on 6 month timescales at 150~GHz although, accounting for measurement uncertainty, we find that in most cases this variability can only be restricted to an upper limit of $<$35\%.  That we do see variability on less than 6 months at 150~GHz suggests this emission originates within the inner 0.01~pc of the and hence is tracing the inner regions of the jet or accretion flow.

\item We find that sources can show steady variation at 15~GHz over 1-5 year timescales, although typically with less than 10\% per annum.  Variability of up to 20\%~pa is observed in a small number of sources during the most active periods of their lightcurves (although we note that our selection of the currently brigthest systems may more favourably select highly varying objects).

\item This year-scale variability is found to be weakly related to the position of the peak in the GPS component of the spectrum.  Whilst for any individual source the position of the peak at any given time {\it does not} serve as a proxy for the definite amount that source will vary, a higher peak {\it does} indicate an increased likelihood of large scale fluctuations.

\item We find that $\geq$3.4\% of BCGs in our parent sample of 530 sources at Declinations greater than 30$^{\circ}$ contain peaked components peaking above 2~GHz.  

\item The fraction of BCGs with a peaked components peaking above 2~GHz increases to $\geq$8.5\% if only cool-core clusters are considered.  Overall, we find $\geq$15.1\% of cool-core clusters contain a 150~GHz point source greater than 3mJy.  This suggests that much more than half of all cool-core clusters with bright 150~GHz central point sources have spectra whereby even well-determined spectral indices below a few GHz would give very large underestimates of the flux at 150~GHz if extrapolated and hence constitute a potential contaminant for SZ surveys as these sources can wipe out any SZ decrements if not fully accounted for.

\end{itemize}

\section*{Acknowledgments}
We thank the anonymous referee for useful comments and suggestions that have greatly improved this work.  MTH acknowledges the support of the Science and Technologies Funding Council (STFC) through studentship number ST/I505656/1. ACE acknowledges support from STFC grant ST/I001573/1.  CR acknowledges the support of STFC. ACF and HRR acknowledge support from ERC Advanced Grant Feedback.  We wish to thank the staff of the OVRO for allowing us access to data from their monitoring campaign and additional data for the sources they kindly added to the observing schedule. We would also like to thank the observatory staff at the MRAO (AMI), IRAM-30m (GISMO), CARMA, the JCMT (SCUBA-2) and the GISMO team for their help in preparing and obtaining data. The OVRO 40-m monitoring program is supported in part by NASA grants NNX08AW31G and NNX11A043G, and NSF grants AST-0808050 and AST-1109911.  This research has made use of data from the University of Michigan Radio Astronomy Observatory which has been supported by the University of Michigan and by a series of grants from the National Science Foundation, most recently AST-0607523. The James Clerk Maxwell Telescope is operated by the Joint Astronomy Centre on behalf of the Science and Technology Facilities Council of the United Kingdom, the National Research Council of Canada, and (until 31 March 2013) the Netherlands Organisation for Scientific Research. Additional funds for the construction of SCUBA-2 were provided by the Canada Foundation for Innovation. Support for CARMA construction was derived from the states of California, Illinois, and Maryland, the James S. McDonnell Foundation, the Gordon and Betty Moore Foundation, the Kenneth T. and Eileen L. Norris Foundation, the University of Chicago, the Associates of the California Institute of Technology, and the National Science Foundation. Ongoing CARMA development and operations are supported by the National Science Foundation under a cooperative agreement, and by the CARMA partner universities.  The research leading to the IRAM-30m results has received funding from the European Commission Seventh Framework Programme (FP/2007-2013) under grant agreement No 283393 (RadioNet3). This research work has used the TIFR GMRT Sky Survey (http://tgss.ncra.tifr.res.in) data products.

%%%%%%%%%%%%%%%%%%%%%% GISMO TABLE %%%%%%%%%%%%%%%%%%%%%%%%%%%%%%%%%%%%%%%%%%%%%%%%%%%%%%%%%%%%%%%%%%%%%%%%%%%%%%%%%%%%%%%%%%%%%%%%
\onecolumn
\begin{longtable}[l]{@{}llllccccccc@{}}
\caption{ GISMO (150~GHz) fluxes of the observed clusters.} \label{GISMO_FLUXES} \\
%%This is the header for the first page of the table...
  \hline\hline
   Cluster & Obs. Date & RA & DEC & Tau & Exposure & CRUSH  & Flux  & FWHM & Peak  & Map \\
           &           &    &     &     &    (min) & Option & (mJy) & (``) & (mJy) & RMS \\
  \hline\hline
\endfirsthead
%This is the header for the remaining page(s) of the table...
\multicolumn{3}{l}{{\tablename} \thetable{} -- Continued} \\
  \hline\hline
   Cluster & Obs. Date & RA & DEC & Tau & Exposure & CRUSH  & Flux  & FWHM & Peak  & Map \\
           &           &    &     &     &    (min) & Option & (mJy) & (``) & (mJy) & RMS \\
  \hline\hline
  \\
\endhead
%%This is the footer for the last page of the table...
  \\ \hline \hline
\endlastfoot
%%Now the data...
  \\
\multicolumn{3}{l}{\bf{\emph{Epoch 1 : April 2012 }}}   \\
\hline
Z235        &  20/04/12  &  00:43:52.30  & 24:24:24.5  & 0.32 & 10  & -faint & 4.7  $\pm$ 1.2  & 18.0 & 4.4  & 0.9 \\ 
RXJ0132-08  &  20/04/12  &  01:32:41.16  & -08:04:07.6 & 0.17 & 7   & -faint & 28.8 $\pm$ 4.7  & 17.7 & 30.9 & 1.1 \\
MACS0242-21 &  21/04/12  &  02:42:36.11  & -21:32:27.3 & 0.20 & 2   & -faint & 35.6 $\pm$ 6.0  & 18.1 & 37.5 & 1.8 \\
A496        &  21/04/12  &  04:33:37.82  & -13:15:46.5 & 0.23 & 10  & -faint & 6.1  $\pm$ 1.3  & 16.4 & 5.9  & 0.9 \\
RXJ0439+05  &  18/04/12  &  04:39:02.32  &  05:20:37.0 & 0.35 & 5   & -faint & 56.0 $\pm$ 9.2  & 16.4 & 56.5 & 2.1 \\
PKS0745-191 &  18/04/12  &  07:47:31.34  & -19:17:42.7 & 0.19 & 25  & -faint &  4.9 $\pm$ 1.1  & 21.9 & 4.7  & 0.8 \\ 
A646        &  17/04/12  &  08:22:08.86  &  47:05:54.1 & 0.49 & 20  & -faint & 11.5 $\pm$ 2.1  & 13.4 & 10.9 & 1.0 \\
4C+55.16    &  17/04/12  &  08:34:55.21  &  55:34:20.6 & 0.10 & 2   & -faint & 81.7 $\pm$ 13.2 & 15.6 & 79.4 & 1.7 \\
Hydra-A     &  17/04/12  &  09:18:05.84  & -12:05:41.1 & 0.14 & 4  & -faint & 266.7 $\pm$ 42.7\footnote{Resolved source, integrated flux} & 24.8 & 179.9 & 2.3 \\
A1348       &  18/04/12  &  11:41:24.26  & -12:16:37.7 & 0.15 & 2   & -faint &     $<$6.3   &   -  &   -  & 2.1 \\  
RXJ1347-11  &  21/04/12  &  13:47:31.84  & -11:45:10.9 & 0.20 & 20  & -deep  &     $<$3.3   &   -  &   -  & 1.1 \\
RXJ1350+09  &  11/04/12  &  13:50:22.01  &  09:40:16.9 & 0.10 & 10  & -faint & 16.0 $\pm$ 2.8  & 12.7 & 14.9 & 1.1 \\
A3581       &  21/04/12  &  14:07:30.19  & -27:01:08.6 & 0.18 & 2   & -faint & 71.0 $\pm$ 11.6 & 16.4 & 72.4 & 2.1 \\
AS780       &  18/04/12  &  14:59:29.98  & -18:10:47.8 & 0.10 & 5   & -faint & 17.0 $\pm$ 3.2  & 16.1 & 16.0 & 1.6 \\
A2052       &  11/04/12  &  15:16:44.50  &  07:01:16.8 & 0.10 & 30  & -faint & 41.7 $\pm$ 6.7  & 17.3 & 42.5 & 0.7 \\
A2055       &  22/04/12  &  15:18:46.13  &  06:13:49.3 & 0.26 & 5   & -faint & 13.2 $\pm$ 2.5  & 17.2 & 12.9 & 1.3 \\
RXJ1558-14  &  18/04/12  &  15:58:22.19  & -14:10:02.7 & 0.10 & 5   & -faint & 39.2 $\pm$ 6.4  & 14.4 & 38.4 & 1.5 \\
Z8193       &  18/04/12  &  17:17:19.06  &  42:26:59.2 & 0.13 & 2   & -faint & 12.8 $\pm$ 3.1  & 10.5 & 13.8 & 2.7 \\
A2270       &  18/04/12  &  17:27:23.71  &  55:10:52.3 & 0.11 & 2   & -faint & 70.8 $\pm$ 4.1  & 14.9 & 70.0 & 2.4 \\
Z8276       &  23/04/12  &  17:44:14.88  &  32:59:30.8 & 0.29 & 5   & -faint & 35.1 $\pm$ 5.8  & 17.4 & 36.8 & 1.3 \\
E1821+644   &  18/04/12  &  18:21:57.49  &  64:20:35.6 & 0.13 & 10  & -faint &  5.8 $\pm$ 2.8  & 13.3 & 5.4  & 1.4 \\
RXJ1832+68  &  18/04/12  &  18:32:35.80  &  68:48:06.3 & 0.12 & 10  & -faint & 10.0 $\pm$ 2.3  & 11.5 & 9.3  & 1.7 \\
MACS1931-26 &  21/04/12  &  19:31:49.74  & -26:34:33.2 & 0.12 & 20  & -faint &  7.2 $\pm$ 2.1  & 18.9 & 7.6  & 0.9 \\
A2390       &  22/04/12  &  21:53:36.74  &  17:41:40.2 & 0.09 & 10  & -faint &  8.1 $\pm$ 1.6  & 13.3 & 7.7  & 0.9 \\
A2415       &  18/04/12  &  22:05:38.62  & -05:35:35.1 & 0.12 & 5   & -faint & 31.2 $\pm$ 5.3  & 19.1 & 33.0 & 1.9 \\
A2597       &  18/04/12  &  23:25:19.93  & -12:07:26.8 & 0.11 & 5   & -faint & 15.0 $\pm$ 2.8  & 16.3 & 15.2 & 1.4 \\
A2627       &  20/04/12  &  23:36:42.29  &  23:55:31.4 & 0.40 & 5   & -faint & 16.8 $\pm$ 3.0  & 17.9 & 16.8 & 1.3 \\
RXJ2341+00  &  18/04/12  &  23:41:07.10  &  00:18:30.9 & 0.10 & 5   & -faint & 41.8 $\pm$ 6.9  & 16.1 & 42.2 & 1.5 \\
\\
\multicolumn{3}{l}{\bf{\emph{Epoch 2 : November 2012 }}}   \\
\hline
RXJ0352+19  &  31/10/12  &  03:52:58.97  &  19:40:59.6 & 0.13 & 10  & -deep  &     $<$3.6   &   -  &   -  & 1.2 \\
RXJ0439+05  &  31/10/12  &  04:39:02.11  &  05:20:42.9 & 0.15 & 5   & -faint & 52.0 $\pm$ 8.5  & 16.6 & 52.5 & 1.6 \\ 
PKS0745-191 &  31/10/12  &  07:47:31.30  & -19:17:45.9 & 0.13 & 5   & -faint &  5.8 $\pm$ 1.8  & 17.8 &  5.3 & 1.5 \\
A646        &  31/10/12  &  08:22:10.05  &  47:05:54.9 & 0.10 & 6   & -faint & 11.4 $\pm$ 2.1  & 18.4 & 12.0 & 1.0 \\
4C+55.16    &  31/10/12  &  08:34:55.26  &  55:34:22.9 & 0.13 & 5   & -faint & 77.8 $\pm$ 12.5 & 18.0 & 80.1 & 1.4 \\  
Hydra-A     &  31/10/12  &  09:18:05.83  & -12:05:46.5 & 0.11 & 5   & -faint & 273.8 $\pm$ 43.8$^{7}$ & 24.5 & 167.8 & 1.8 \\
A1084       &  31/10/12  &  10:44:33.11  & -07:04:11.1 & 0.11 & 10  & -faint &  4.5 $\pm$ 1.2  & 19.0 &  4.1 & 1.0 \\
A1348       &  31/10/12  &  11:41:24.37  & -12:16:40.0 & 0.11 & 5   & -faint &  6.9 $\pm$ 1.7  & 10.7 &  6.5 & 1.3 \\
A1774       &  31/10/12  &  13:41:05.19  &  39:59:47.1 & 0.15 & 10  & -faint &  6.9 $\pm$ 1.5  & 19.7 &  6.9 & 1.0 \\
RXJ1347-11  &  31/10/12  &  13:47:31.84  & -11:45:10.9 & 0.12 & 10  & -deep  &     $<$3.0   &   -  &   -  & 1.0 \\
A1795       &  31/10/12  &  13:48:52.25  &  26:35:32.2 & 0.19 & 10  & -faint &  5.1 $\pm$ 1.4  & 17.7 &  5.4 & 1.1 \\
RXJ1350+09  &  31/10/12  &  13:50:22.04  &  09:40:07.9 & 0.19 & 10  & -faint & 16.0 $\pm$ 2.7  & 17.0 & 15.7 & 1.0 \\
A3581       &  31/10/12  &  14:07:30.00  & -27:01:07.8 & 0.12 & 5   & -faint & 62.1 $\pm$ 10.1 & 19.0 & 63.2 & 2.0 \\
A1885       &  31/10/12  &  14:13:43.86  &  43:39:40.0 & 0.10 & 10  & -deep  &  4.0 $\pm$ 1.0  & 12.4 &  4.2 & 0.8 \\
AS780       &  31/10/12  &  14:59:29.14  & -18:10:51.3 & 0.11 & 10  & -faint & 15.4 $\pm$ 2.7  & 17.9 & 15.3 & 1.2 \\
A2052       &  31/10/12  &  15:16:44.64  &  07:01:16.3 & 0.11 & 5   & -faint & 38.0 $\pm$ 6.3  & 16.3 & 38.4 & 1.7 \\
A2055       &  02/11/12  &  15:18:46.19  &  06:13:56.5 & 0.50 & 10  & -faint & 10.4 $\pm$ 2.5  & 18.9 & 10.2 & 1.8 \\
RXJ1558-14  &  31/10/12  &  15:58:22.22  & -14:10:03.1 & 0.11 & 5   & -faint & 27.8 $\pm$ 4.8  & 15.1 & 27.6 & 1.9 \\
NGC6338     &  31/10/12  &  17:15:23.25  &  57:24:41.5 & 0.12 & 10  & -faint &  7.3 $\pm$ 1.5  & 11.0 &  6.8 & 1.0 \\
Z8193       &  31/10/12  &  17:17:19.11  &  42:26:59.2 & 0.11 & 10  & -faint & 15.6 $\pm$ 2.8  & 15.9 & 15.0 & 1.2 \\
A2270       &  31/10/12  &  17:27:23.52  &  55:10:52.5 & 0.11 & 5   & -faint & 56.2 $\pm$ 9.1  & 18.0 & 58.3 & 1.3 \\
Z8276       &  02/11/12  &  17:44:14.82  &  32:59:27.4 & 0.53 & 10  & -faint & 12.5 $\pm$ 2.7  & 16.1 & 13.2 & 1.8 \\
E1821+644   &  31/10/12  &  18:21:57.00  &  64:20:32.9 & 0.11 & 10  & -faint &  7.9 $\pm$ 1.7  & 11.3 &  7.8 & 1.2 \\
RXJ1832+68  &  31/10/12  &  18:32:35.08  &  68:48:02.2 & 0.12 & 7   & -faint &  8.9 $\pm$ 1.7  & 17.2 &  9.0 & 1.0 \\
\\
\multicolumn{3}{l}{\bf{\emph{Epoch 3 : April 2013 }}}   \\
\hline
RXJ0132-08  &  11/04/13  &  01:32:41.05  & -08:04:07.8 & 0.83 & 5   & -faint & 27.6 $\pm$ 13.7 & 17.4 &  24.0 & 13.0 \\
M0242-21    &  11/04/13  &  02:42:35.99  & -21:32:24.9 & 0.70 & 10  & default & 17.9 $\pm$ 6.6 & 15.4 &  22.9 & 6.0 \\
A496        &  10/04/13  &  04:33:37.75  & -13:15:40.7 & 0.57 & 20  & -faint &  4.4 $\pm$ 1.1  & 13.1 &  4.3  & 0.8 \\
RXJ0439+05  &  10/04/13  &  04:39:02.31  &  05:20:42.7 & 0.69 & 5   & -faint & 47.6 $\pm$ 7.8  & 17.4 &  48.6 & 1.6 \\
A646        &  10/04/13  &  08:22:09.50  &  47:05:53.4 & 0.59 & 10  & -faint & 12.2 $\pm$ 2.3  & 10.7 &  11.4 & 1.3 \\
4C+55.16    &  10/04/13  &  08:34:54.79  &  55:34:19.8 & 0.63 & 5   & -faint & 61.7 $\pm$ 10.0 & 16.9 &  62.7 & 1.8 \\
Hydra-A     &  10/04/13  &  09:18:05.79  & -12:05:39.6 & 0.61 & 5   & -extended & 201.9 $\pm$ 32.9$^{7}$ & 25.0 & 153.5 & 6.2 \\
RXJ1504-02  &  11/04/13  &  15:04:07.43  & -02:48:15.8 & 0.59 & 10  & -faint &  5.5 $\pm$ 1.4  & 11.2 &  5.1  & 1.1 \\
A2052       &  11/04/13  &  15:16:44.48  &  07:01:18.9 & 0.58 & 5   & -faint & 35.0 $\pm$ 5.8  & 17.0 &  36.6 & 1.6 \\
RXJ1558-14  &  11/04/13  &  15:58:22.01  & -14:09:58.2 & 0.63 & 5   & -faint & 25.1 $\pm$ 4.3  & 14.9 &  23.8 & 1.6 \\
A2270       &  11/04/13  &  17:27:23.58  &  55:10:52.8 & 0.64 & 5   & -faint & 50.4 $\pm$ 8.2  & 14.5 &  51.2 & 1.7 \\
Z8276       &  11/04/13  &  17:44:14.47  &  32:59:29.4 & 0.68 & 5   & default &     $<$44.7 &   -  &   -   & 14.9 \\
A2390       &  11/04/13  &  21:53:36.96  &  17:41:40.4 & 0.65 & 10  & -faint &  8.6 $\pm$ 2.0  & 14.1 &  8.6  & 1.4 \\
A2415       &  11/04/13  &  22:05:38.64  & -05:35:31.7 & 0.65 & 5   & -faint & 26.7 $\pm$ 4.7  & 16.3 &  25.8 & 1.9 \\
A2627       &  11/04/13  &  23:36:41.84  &  23:55:28.2 & 0.66 & 10  & -faint & 16.9 $\pm$ 3.0 & 14.9 &  15.9 & 1.3 \\
RXJ2341+00  &  15/04/13  &  23:41:07.01  &  00:18:33.0 & 0.39 & 5   & -faint & 33.0 $\pm$ 2.9  & 15.2 &  32.5 & 1.8 \\
\end{longtable}
\twocolumn

%%%%%%%%%%%%%%%%%%%%%%%%%%%%%%%%%%%%%%%%%%%%%%%%%%%%%%   AMI TABLE %%%%%%%%%%%%%%%%%%%%%%%%%%%%%%%%%%%%%%%%%%%%%%%%%%%%%%%%%%%%%%%%%%%%%%%
\onecolumn
\begin{longtable}[l]{@{}lccccccccc@{}}
\caption{AMI (16~GHz) fluxes of the observed BCGs.} \label{AMI_FLUXES} \\
%%This is the header for the first page of the table...
  \hline\hline
   Cluster & Obs. Date & Flux & RMS & Obs. Date & Flux & RMS & Obs. Date  & Flux & RMS  \\
  \hline\hline
\endfirsthead
\endhead
%%This is the footer for the last page of the table...
  \\ \hline \hline
\endlastfoot
%%Now the data...
Z235       & 11/04/12  & 17.7   $\pm$ 0.9  & 0.2 &  04/06/12  & 16.7   $\pm$ 0.9  & 0.3  & -         &  -               & -   \\ 
RXJ0132-08 & 17/04/12  & 139.5  $\pm$ 8.3  & 4.5 &  02/06/12  & 126.1  $\pm$ 6.9  & 2.7  & -         &  -               & -   \\ 
RXJ0439+05 & 12/04/12  & 324.8  $\pm$ 16.3 & 1.6 &  30/05/12  & 307.5  $\pm$ 15.5 & 2.1  & 29/09/12  & 299.9 $\pm$ 15.1 & 2.0 \\ 
A646       & 10/04/12  & 45.6   $\pm$ 2.3  & 0.5 &  30/05/12  & 45.7   $\pm$ 2.7  & 1.4  & 27/09/12  & 47.3  $\pm$ 2.5  & 0.7 \\ 
4C+55.16   & 14/04/12  & 1317.7 $\pm$ 66.0 & 3.5 &  30/05/12  & 1372.3 $\pm$ 70.6 & 16.8 & -         &  -               & -   \\ 
RXJ1350+09 & 24/04/12  & 133.7  $\pm$ 6.7  & 0.9 &  14/06/12  & 107.4  $\pm$ 5.6  & 1.5  & -         &  -               & -   \\ 
A2052      & 24/04/12  & 231.8  $\pm$ 11.8 & 2.2 &  14/06/12  & 220.8  $\pm$ 11.2 & 1.9  & -         &  -               & -   \\ 
A2055      & 24/04/12  & 66.7   $\pm$ 3.5  & 1.2 &  14/06/12  & 68.6   $\pm$ 3.6  & 1.2  & -         &  -               & -   \\ 
Z8193      & 24/04/12  & 79.8   $\pm$ 4.1  & 1.1 &  26/06/12  & 79.3   $\pm$ 4.1  & 1.2  & -         &  -               & -   \\ 
A2270      & 24/04/12  & 185.0  $\pm$ 9.4  & 1.6 &  26/06/12  & 253.8  $\pm$ 12.8 & 1.4  & 26/09/12  & 226.9 $\pm$ 11.6 & 2.6 \\ 
Z8276      & 14/04/12  & 88.6   $\pm$ 4.5  & 0.7 &  31/05/12  & 95.9   $\pm$ 4.9  & 0.8  & 26/09/12  & 87.6  $\pm$ 4.6  & 1.3 \\ 
E1821+644  & 25/04/12  & 9.0    $\pm$ 0.5  & 0.2 &  30/06/12  & 10.8   $\pm$ 0.8  & 0.2  & -         &  -               & -   \\ 
RXJ1832+68 & 19/05/11  & 45.6   $\pm$ 2.3  & 0.4 &  25/04/12  & 42.2   $\pm$ 2.1  & 0.1  & 05/07/12  & 42.3  $\pm$ 2.1  & 0.3 \\ 
A2390      & 14/04/12  & 68.5   $\pm$ 12.4 & 0.8 &  31/05/12  & 66.4   $\pm$ 3.4  & 0.9  & -         &  -               & -   \\ 
A2415      & 11/04/12  & 77.4   $\pm$ 4.2  & 1.7 &  02/06/12  & 77.2   $\pm$ 4.5  & 2.3  & -         &  -               & -   \\ 
A2627      & 14/04/12  & 83.1   $\pm$ 4.2  & 0.7 &  31/05/12  & 73.3   $\pm$ 4.0  & 1.5  & -         &  -               & -   \\ 
RXJ2341+00 & 11/04/12  & 147.0  $\pm$ 7.4  & 1.2 &  31/05/12  & 140.9  $\pm$ 7.2  & 1.5  & 28/09/12  & 209.2 $\pm$ 11.8 & 5.4 \\ 
\end{longtable}
\twocolumn

%%%%%%%%%%%%%%%%%%%%%%%%%%%%%%%%%%%%%%%%%%%%%%%%%%%%%%   CARMA TABLE %%%%%%%%%%%%%%%%%%%%%%%%%%%%%%%%%%%%%%%%%%%%%%%%%%%%%%%%%%%%%%%%%%%%%
\onecolumn
\begin{longtable}[c]{@{}lccccc@{}}
\caption{CARMA (90~GHz) fluxes of the observed BCGs.} \label{CARMA_FLUXES} \\
%%This is the header for the first page of the table...
  \hline\hline
   Cluster & Obs. Date & RA & DEC & Flux & RMS  \\
  \hline\hline
\endfirsthead
\endhead
  \\ \hline \hline
\endlastfoot
%%Now the data...
Z235        & 11/06/12 & 00:43:52.21 & 24:24:21.5  & 9.9   $\pm$ 1.8  & 4.5 \\
RXJ0132-08  & 21/05/12 & 01:32:41.11 & -08:04:05.7 & 98.8  $\pm$ 11.9 & 6.7 \\
MACS0429-02 & 15/06/12 & 04:29:36.00 & -02:53:07.0 & $<$14.1          & 4.7 \\
A496        & 15/06/12 & 04:33:37.89 & -13:15:42.3 & 11.4  $\pm$ 3.6  & 3.4 \\
RXJ1350+09  & 21/05/12 & 13:50:22.12 & 09:40:10.7  & 22.8  $\pm$ 2.8  & 1.7 \\
A3581       & 21/05/12 & 14:07:29.83 & -27:01:04.2 & 82.5  $\pm$ 9.1  & 2.8 \\
AS780       & 21/05/12 & 14:59:28.78 & -18:10:44.9 & 29.5  $\pm$ 4.4  & 2.3 \\
A2052       & 21/05/12 & 15:16:44.58 & 07:01:18.1  & 126.7 $\pm$ 14.3 & 9.1 \\
A2055       & 21/05/12 & 15:18:46.45 & 06:13:57.9  & 53.6  $\pm$ 9.0  & 21.6 \\
RXJ1504-02  & 21/05/12 & 15:04:07.50 & -02:48:16.9 & 8.3   $\pm$ 1.6  & 1.4 \\
RXJ1558-14  & 21/05/12 & 15:58:21.91 & -14:09:58.9 & 44.8  $\pm$ 4.9  & 6.3 \\
Z8193       & 22/05/12 & 17:17:19.21 & 42:26:57.8  & 22.4  $\pm$ 2.9  & 1.8 \\
A2270       & 22/05/12 & 17:27:23.49 & 55:10:53.9  & 164.7 $\pm$ 17.2 & 14.4 \\
Z8276       & 22/05/12 & 17:44:14.47 & 32:59:27.4  & 35.4  $\pm$ 6.0  & 4.9 \\
E1821+644   & 24/05/12 & 18:21:56.24 & 64:20:58.0  & 7.2   $\pm$ 1.8  & 1.7 \\
RXJ1832+68  & 22/05/12 & 18:32:35.52 & 68:48:07.2  & 19.6  $\pm$ 4.0  & 3.5 \\
MACS1931-26 & 11/06/12 & 19:31:49.64 & -26:34:32.4 & 9.6   $\pm$ 2.9  & 2.7 \\
A2390       & 04/06/12 & 21:53:36.81 & 17:41:44.8  & 22.3  $\pm$ 3.2  & 2.3 \\
A2415       & 04/06/12 & 22:05:38.53 & -05:35:33.7 & 25.7  $\pm$ 3.4  & 2.3 \\
A2597       & 21/05/12 & 23:25:19.82 & -12:07:28.6 & 19.7  $\pm$ 5.8  & 2.0 \\
A2627       & 04/06/12 & 23:36:42.82 & 23:55:23.8  & 19.8  $\pm$ 4.5  & 4.0 \\
RXJ2341+00  & 21/05/12 & 23:41:06.94 & 00:18:33.3  & 72.8  $\pm$ 8.0  & 3.2 \\
\end{longtable}
\twocolumn

%%%%%%%%%%%%%%%%%%%%%%%%%%%%%%%%%%%%%%%%%%%%%%%%%%%%%%  SCUBA2 TABLE %%%%%%%%%%%%%%%%%%%%%%%%%%%%%%%%%%%%%%%%%%%%%%%%%%%%%%%%%%%%%%%%%%%%%
\onecolumn
\begin{longtable}[c]{@{}lccc@{}}
\caption{SCUBA-2 (353~GHz) fluxes of the observed BCGs.} \label{SCUBA2_FLUXES} \\
%%This is the header for the first page of the table...
  \hline\hline
   Cluster & Obs. Date & Flux (mJy) &  RMS  \\
  \hline\hline
\endfirsthead
\endhead
  \\ \hline \hline
\endlastfoot
%%Now the data...
RXJ0132-08  &   23/10/12  &     -  &  13.1  \\
MACS0242-21 &   01/10/12  &   34.0 &  7.6   \\
RXJ0439+05  &   01/10/12  &   18.7 &  5.2   \\
A646        &   27/02/12  &   25.9 &  6.6   \\
4C+55.16    &   27/02/12  &   29.0 &  6.5   \\
Hydra-A     &   08/10/12  &   76.3 &  8.8   \\
A1348       &   27/12/12  &    -   &  7.9   \\
RXJ1347-11  &   29/12/12  &     -  &  5.5   \\
A1795       &   27/12/12  &    -   &  7.4   \\
RXJ1350+09  &   23/02/12  &     -  &  5.7   \\
A3581       &   28/01/13  &   59.2 &  9.1   \\
AS780       &   15/04/12  &    -   &  5.0   \\
RXJ1504-02  &   12/01/13  &   13.7 &  4.0   \\
A2052       &   28/01/13  &    -   &  7.8   \\
RXJ1558-14  &   15/04/12  &   40.3 &  4.7   \\
Z8193       &   14/05/12  &     -  &  8.5   \\
A2270       &   30/09/12  &   34.9 &  4.7   \\
Z8276       &   30/09/12  &   27.7 &  5.0   \\
RXJ1832+68  &   05/09/12  &     -  &  5.8   \\
MACS1931-26 &   16/09/12  &   16.8 &  4.8   \\
A2390       &   24/04/13  &    -   &  4.9   \\
A2415       &   01/10/12  &   18.2 &  6.7   \\
A2597       &   24/10/12  &    -   &  7.2   \\
A2627       &   24/10/12  &  20.1  &  7.3   \\
RXJ2341+00  &   31/08/12  &   33.2 &  3.9   \\
\end{longtable}  
\twocolumn

\appendix

\section{Notes on Individual Sources} \label{Fitting_Notes}
Individual source SEDs were populated using the main radio catalogues (e.g. Australia Telescope 20~GHz Survey (AT20G), \citealt{Murphy10}; NVSS and Faint Images of the Radio Sky at Twenty-cm (FIRST) at 1.4~GHz, \citealt{Condon98, White97}; SUMSS at 843~MHz, \citealt{Mauch03}; Texas Survey of Radio Sources (TEXAS) at 365~MHz, \citealt{Douglas96}; Westerbork Northern Sky Survey (WENSS) and Westerbork In the Southern Hemisphere (WISH) at 325~MHz, \citealt{Rengelink97, DeBreuck02}; VLA Low-Frequency Sky Survey (VLSS) at 74~MHz, \citealt{Cohen07}), the NASA/IPAC Extragalactic Database (NED) and the High Energy Astrophysics Science Archive Research Center (HEASARC) database, the National Radio Astronomy Archive (NRAO), literature searches, the data presented herein and supplemented by data from H15a.

Due to the lack of self-similarity between these spectra each source was fit on an individual basis.  Where possible, active components were isolated and fit using the GPS model of \cite[][:]{Orienti14}
\begin{equation} \label{GPS_equation}
Log(S) = A_{0} + Log(\nu)(A_{1} + A_{2}Log(\nu))
\end{equation}
Steep spectrum components and active flat spectrum components where no peak was apparent were both fit using a simple power-law of the form:
\begin{equation} \label{Power_law_equation}
{\bf S = A \nu^{-\alpha} }
\end{equation}
Where both an active and inactive component were found to be present, the core contribution is removed from the total emission prior to fitting the more extended emission.  In a couple of cases clear spectral curvature attributable to synchrotron ageing is present in the steeper spectrum components and is modelled as described in the notes below.  Additionally, flattening of a steep spectrum component to low frequencies (below 100MHz) is seen in a minority of sources.  Where morphologically it is clear that this is extended emission, and hence the flattening is most likely due to free-free absorption then these data are excised from the fits as described below.\\
\\
{\bf 4C+55.16} \\
2MFGC 06756, z=0.2411.  This is a powerful radio source that is clearly extended on milliarcsecond scales as observed using VLBI \cite[][see also Hogan \etal 2015c, in prep.]{Helmboldt07}.  The SED has a flat index in the GHz range that appears to be caused by the superposition of a broad peaked GPS-like component with a steep spectrum power-law component to lower frequency (below about 100~MHz).  The SED is fit using a GPS-like model with parameters A$_{0}$=3.90$\pm$0.03, A$_{1}$=-0.04$\pm$0.02 and A$_{2}$=-0.43$\pm$0.01 and a steep power-law to lower frequency with $\alpha$=1.1$\pm$0.1 and A=500.0$\pm$10.0. \\
{\bf A1084} \\
2MASX J10443287-0704074, z=0.1326.  The SED is sparsely sampled, although the non-detection in VLSS \cite[][]{Cohen07} coupled to its flat returned index shows the source to be core-dominated.  The SED is fit with a power-law of $\alpha$=0.36$\pm$0.04 and A=36.7$\pm$2.9. \\
{\bf A1348} \\ 
LCRS B113851.7-115959, z= 0.1195. The source is best fit using a GPS-like model, with parameters A$_{0}$=1.96$\pm$0.20, A$_{1}$=0.62$\pm$0.41 and A$_{2}$=-0.53$\pm$0.20 and then a steeper spectrum component that becomes apparent below 1~GHz that is fit using a power-law with $\alpha$=1.05$\pm$0.07 and A=80.2$\pm$1.5. \\
{\bf A1774} \\ 
2MASX J13410515+3959456, z=0.1715. The SED of this source is consistent with a single power-law fit with parameters of $\alpha$=0.61$\pm$0.03 and A=104.9$\pm$5.0, that persists to high frequency.  Note that the FIRST \cite[][]{White97} datum-point was removed from this fit as it appears to have resolved out some emission.  The source appears one-sided in FIRST but the core is not isolated and the lack of higher-resolution data therefore means that a separate core component cannot be fit.  The intermediate spectral index combined with the resolved morphology suggests that an active core is likely to be present in this system, which may be distinguishable with higher resolution observations. \\
{\bf A1795} \\
CGCG 162-010, z=0.0632. Extended source with several observations of high enough resolution to create a separate SED for the isolated core only \cite[see][]{Laurent-Muehleisen97, Lin09, Liuzzo09}.  This core spectrum is flat to high frequency, and fit using a power-law with $\alpha$=0.65$\pm$0.06 and A=120.2$\pm$14.2.  The SED of the more extended emission appears to exhibit spectral curvature to high frequency, hence to account for this we allow an exponential rollover to the steep power-law, of the form: 
\begin{equation}
S = A\nu^{-\alpha}(1 - Be^{-\frac{\nu_{0}}{\nu}}) 
\end{equation}
\cite[][]{Hogan15}. The best fit returns parameters A=1262.8, $\alpha$=0.86, B=1.07 and $\nu_{0}$=7.51 for the unknown parameters in this model. \\
{\bf A1885} \\
2MASX J14134379+4339450, z=0.090. The SED is core dominated with a hint of variability seen from historical C-band observations \cite[][]{Laurent-Muehleisen97, Gregory91}.  Additionally, the WENSS \cite[][]{Rengelink97} detection alongside the VLSS \cite[][]{Cohen07} non-detection shows that the spectrum must peak in the GHz range.  The SED can be fit with a GPS model, returning parameters of 1.66$\pm$0.03, 0.24$\pm$0.04 and -0.33$\pm$0.03. \\
{\bf A2052} \\
UGC 09799/3C~317, z=0.0355.  A separate core only SED can be created for the resolved isolated core.  This can be fit with a GPS model (returning parameters; 2.54$\pm$0.02, 0.10$\pm$0.02 and -0.24$\pm$0.01).  Accounting for this core contribution, fit a power-law (with $\alpha$=1.20$\pm$0.08 and A=6404.5$\pm$1142.8) to the lower frequency emission.  Note that there are a number of very low frequency (less than 50MHz) data-points that are flatter than this steep spectrum would imply.  These are possibly affected by free-free absorption and hence were excised from the power-law fit. \\
{\bf A2055} \\
2MASX J15184574+0613554, z=0.1019. High resolution data reveals an underlying core component with a GPS-like spectrum.  Note that the CARMA datum is tagged as potentially unreliable and indeed appears inconsistent and has therefore been excluded from the fit.  A GPS fit to the core-only SED returns parameters of 1.28$\pm$0.04, 1.13$\pm$0.02 and -0.57$\pm$0.01. The integrated SED is increasingly dominated by a power-law component at frequencies less than 1~GHz.  Fit this separately, accounting for the expected core contribution, to recover a power-law fit with $\alpha$=0.70$\pm$0.04 and A=519.2$\pm$122.7. \\
{\bf A2270} \\
2MASX J17272346+5510538, z=0.2473. Source appears highly variable.  There is a limit only in the VLSS-redux catalogue \cite[][]{Lane14} but an overlay of the map shows there to be a 3$\sigma$ map detection at the optical position of the BCG.  Along with other low frequency detections from 7C \cite[][]{Hales07} and WENSS \cite[][]{Rengelink97}, this hints at there being a low-powered low frequency power-law component within this source.  The data are widely spaced in observation date, making separate fits by epoch for the GPS component unfeasible.  Fitting a GPS to fit to all the available data at higher frequencies than L-band, recovers an average GPS with parameters A$_{0}$=2.11$\pm$0.06, A$_{1}$=0.56$\pm$0.06 and A$_{2}$=-0.31$\pm$0.04. Account for the contribution of this to the lower frequency data and find a resultant power-law component with $\alpha$=0.94$\pm$0.07 and A=28.0$\pm$0.7\\
{\bf A2390} \\
ABELL 2390:[YEA96] 101084, z=0.2328. The spectrum appears to have a split power-law like shape below 10~GHz although the higher frequency observations show that the active component curves.  Hence the SED is best fit using a combination of a GPS component (that returns parameters; 2.28$\pm$0.03, 0.26$\pm$0.04 and -0.42$\pm$0.05) and then a power-law component with $\alpha$=1.34$\pm$0.25 and A=121.5$\pm$60.6.  \citet[][]{Smail02} claim a 6.7mJy SCUBA detection at 330~GHz for this source.  This suggests that there may be additional intermittent activity at these higher frequencies or alternatively there may be a dust component coming in (although non-detection ($<$60mJy) at 660~GHz opposes this).  Note that we do not include this claimed detection on our SED, including only our more contemporaneous upper-limit at 330~GHz.   \\
{\bf A2415} \\
2MASX J22053865-0535330, z=0.0573. The SED has a flat index from around 1~GHz up to high frequency although a subtle drop-off means that this is best fit with a broad-peaked GPS model (returned parameters: 2.09$\pm$0.08, 0.26$\pm$0.02 and -0.29$\pm$0.01) in addition to a steep power-law component at lower frequencies with $\alpha$=1.31$\pm$0.02 and A=193.2$\pm$9.4. \\
{\bf A2597} \\
PKS 2322-12, z=0.0830.  Source appears very flat to higher frequencies.  Archival VLA observations at C, X and U-bands from the NRAO Image Archive allow isolated core measurements which in conjunction with higher frequency data suggest a broad GPS-like core that is well fit with parameters A$_{0}$=1.92$\pm$0.05, A$_{1}$=0.27$\pm$0.09 and A$_{2}$=-0.30$\pm$0.04, although there may be further flattening to yet higher frequencies as suggested by SCUBA observations \cite[][]{Zemcov07, Knudsen08}.  Lower frequency data are fit with a power-law of $\alpha$=1.18$\pm$0.06 and A=2336.4$\pm$13.4, where the data below 100~MHz have been excluded as here the spectrum flattens, possibly due to free-free affected.  Note that a simple extrapolation of this power-law in conjunction with the GPS core gives a wholly inaccurate integrated spectrum and hence the older, extended component must roll-off at some frequency.  This has been simulated here by subtracting the GPS from the simple power-law to give the presented curved spectrum. \\
{\bf A2627} \\
This cluster has two bright galaxies in close proximity, both of which are radio-sources.  The more dominant source - A2627(a)~/~B2 2334+23 `the BCG' - is at position 23:36:42.12 23:55:29.0 and z=0.126.  The other galaxy - A2627(b)~/~2MASX J23364245+2354442 is at position 23:36:42.47 23:54:44.6 and has a redshift given in NED as z=0.122.  VLA C-band data (H15a) separates the sources and overlaying other maps onto this we see that at NVSS \cite[1.4~GHz,][]{Condon98} and lower frequencies the sources are confused.  They are similarly confused in our AMI maps.  The BCG can be isolated in the CARMA map and a VLA-A array image from the NRAO Image Archive.  These show it to be flat spectum.  Along with the GISMO data, fit a flat spectrum power-law to this ($\alpha$=0.37$\pm$0.06 and A=109.0$\pm$22.9).  Morphologically, A2627(a) is core dominated although it does show some wispy emission that is likely to contribute at lower frequencies.  A2627(b) is fainter (by roughly a factor of 4 at C-band) but appears to be steep spectrum and is likely to become increasingly dominant at lower frequency.  Fit the integrated emission (accounting for the flat component) with a power-law of $\alpha$=0.65$\pm$0.06 and A=499.8$\pm$6.4 noting that this is highly confused.  We currently do not have sufficient coverage to fully deblend the sources. \\
{\bf A3581} \\
IC 4374, z=0.0218.  Flat spectrum core dominated system with hints of variation in repeat observations around C- and X-bands.  Unresolved at VLA-A in a map from the NRAO Image Archive but the source is resolved on milli-arcsecond scales with the VLBA (Hogan \etal 2015c, in prep.).  This small-scale resolved emission is linked to ongoing activity and the integrated flux recovered with the VLBA is consistent with lower resolution fluxes showing that source is core-only.  Fit a power-law to the integrated SED (returning a fit of $\alpha$=0.49$\pm$0.04 and A=758.2$\pm$6.9) and then highlight the VLBA peaks on the SED. \\
{\bf A496} \\
MCG -02-12-039, z=0.0328.  This source has a GPS-like core with a steeper power-law component to low frequency.  The core is fit with the GPS-like model with parameters A$_{0}$=1.45$\pm$0.02, A$_{1}$=0.60$\pm$0.02 and A$_{2}$=-0.43$\pm$0.03 and then accounting for this component, the remaining extended emission is fit with a power-law of $\alpha$=1.71$\pm$0.03 and A=99.8$\pm$6.4. \\
{\bf A646} \\
2MASX J08220955+4705529, z=0.1268. The SED of this source can be fit using a GPS model for the core component, which returns parameters 1.89$\pm$0.02, 0.14$\pm$0.01 and -0.25$\pm$0.08 and a power-law component of $\alpha$=1.54$\pm$0.09 and A=13.5$\pm$0.3 to account for the lower frequency, extended emission.  Note that the combination of these two components at 1.4~GHz over-estimates the NVSS flux \cite[][]{Condon98} which may be attributable either to variation of the core component or perhaps indicating that this source should in reality have a more sharply peaked profile. \\
{\bf AS780} \\
2MASX J14592875-1810453, z=0.2344.  This object has previously been identified as a flat spectrum source \cite[e.g.][]{Healey07}.  The GHz range flat spectrum falls off above about 20~GHz.  The source is undetected in VLSS \cite[][]{Cohen07}, which suggests that the spectrum must turnover around 1~GHz.  The SED is fit with a GPS model,  that returns parameters 2.16$\pm$0.02, 0.34$\pm$0.02 and -0.35$\pm$0.01.  Note that the source is mildly confused in the WISH map \cite[][]{DeBreuck02}.  The flux used here is deblended BCG emission however there could still be wrongfully attributed flux contributing to this 352~MHz measurement.  Equally, this 352~MHz point lying above the curve could be indicative of a steeper spectrum component (which would push the GPS peak to higher frequency).  Future low frequency observations will determine this. \\
{\bf E1821+644} \\
HB89 1821+643, z=0.2970.  This source is one of the few radiatively efficient AGN known to reside within a cool core cluster \cite[see e.g.][and references therein]{Russell10}.  Nevertheless, the radio-properties appear to be consistent with its more inefficent brethren. Multiple high resolution VLA observations from the NRAO Image Archive allow the core to be isolated at multiple frequencies which are consistent with the spectral flattening seen above about 10~GHz.  The source shows significant variation.  Fit the core-only measurements with a GPS-like model, returning parameters of 0.83$\pm$0.10, 0.40$\pm$0.06 and -0.19$\pm$0.03.  This suggests a broad, flat profiled core.  Independently fit a steep power-law to the low frequency data, returning a fit with $\alpha$=1.06$\pm$0.13 and A=94.6$\pm$3.6. \\
{\bf Hydra-A} \\
This source is associated with the bright radio source 3C~218, z=0.0549.  This source is very well studied \cite[e.g.][and references therein]{McNamara00, Hamer14}.  VLBA observations at multiple frequencies \cite[e.g.][]{Taylor96, Araya10} allow the SED of the core only to be independently fit with a GPS model (returning parameters: 2.01$\pm$0.05, 0.95$\pm$0.06 and -0.59$\pm$0.06).  The integrated SED is dominated by flux from the extended emission and remains persistent to frequencies higher than 100~GHz.  Accounting for the minimal core contribution, fit the SED using a power-law with $\alpha$=1.00$\pm$0.02 and A=47214.3$\pm$43.0. \\
{\bf MACS0242-21} \\
PKS 0240-217, z=0.314.  This source appears to be a CSS or low-peaking GPS, that can be fit using a GPS model with parameters of A$_{0}$=3.14$\pm$0.02, A$_{1}$=-0.14$\pm$0.02 and A$_{2}$=-0.30$\pm$0.02.  VLBA observations of the source at S- and X-band \cite[][]{Beasley02} show that it is `wispy' on parsec scales, with the integrated flux consistent with the unresolved emission at VLA resolutions.  That the source has a low spectral peak, alongside it being very structured on parsec scales, is consistent with this source being in a late-stage of core activity. \\
{\bf MACS1931-26} \\
PMN J1931-2635, z=0.3520.  There are two radio-sources in close proximity within the cluster core; the BCG at 19:31:49.67 -26:34:33.4 and a Wide-Angled Tail (WAT) source, PMN 1931-2635, at 19:31:50.0 -26:35:16.4.  These sources are confused in many radio maps of the region.  The BCG core is isolated in ATCA-6km observations at 5.5 and 9.0~GHz along with CARMA and GISMO at higher frequencies.  The SED of only these points is well fit with a GPS model,  returning parameters 0.95$\pm$0.04, 0.42$\pm$0.03 and -0.21$\pm$0.01.  The sources are confused in lower resolution observations.  The sources cannot therefore be fully distinguished, however by fitting to the flux peaks coincident with the optical positions it appears as if PMN 1931-2635 has a steeper spectrum (as would be expected for a lobe-dominated WAT).  Further confounding the picture, high resolution VLA-A observations at L-band by \cite{Ehlert11} show that the BCG is surrounded by an amorphous halo \cite[see also][]{Mittal09} that is likely to be very steep spectrum.  This structure is most likely confused emission from previous activity cycles of the BCG \cite[][]{Hogan15}.  The integrated emission of this amorphous halo component does not account for all of the flux in the confused observations.  However, such high resolution observations at other frequencies to fully detangle this amorphous halo emission are not available currently.  Shown on the SED are the core only observations along with highlighted confused, amorphous halo and WAT data points. \\
{\bf RXJ0132-08} \\
PKS 0130-083, z=0.1489.  This source appears to be either a low-peaking GPS or a CSS.  The source does not appear above the 5$\sigma$ cut-off of the VLSS catalogue \cite[][]{Cohen07} but overlaying the VLSS map, the source is present at the optical position as a 3$\sigma$ detection.  The central engine is resolved in VLBI maps \cite[][[]{Bourda10,Bourda11,Petrov11}.  Fit a GPS model to the integrated emission (that returns parameters; 2.55$\pm$0.04, -0.24$\pm$0.03 and -0.11$\pm$0.03) and show the VLBI peaks on the SED (although note that these are not included in the fit). \\
{\bf RXJ0352+19} \\
2MASX J03525901+1940595, z=0.109.  The SED is relatively poorly sampled, consisting of only four detections and limits from the pointed GISMO observation (at 150~GHz) and the VLSS \cite[][]{Cohen07}. A single power-law fit returns $\alpha$=0.41$\pm$0.18 and A=20.4$\pm$5.8, which is consistent with the limits and suggests a flat spectrum source.  However, there could be a small GPS-like component with a tail but available data are not sufficient to confirm nor deny this, hence the power-law is perhaps the most robust fit possible with current data. \\
{\bf RXJ0439+05} \\
2MASX J04390223+0520443, z=0.208.  The SED can be well fit with a strong GPS component, that returns parameters 1.42$\pm$0.03, 1.67$\pm$0.02 and -0.72$\pm$0.01 and then a steep spectrum tail to lower frequency (less than 1~GHz) with $\alpha$=1.10$\pm$0.04 and A=60.2$\pm$5.5. \\
{\bf RXJ0747-19} \\
PKS 0745-191, z=0.1028. The SED potentially flattens below around 200~MHz but other than that presents as a steady power-law out to higher than 150~GHz, with $\alpha$=1.26$\pm$0.07 and A=3064.7$\pm$23.4. \\
{\bf RXJ1347-11} \\
GALEX J134730.7-114509, z=0.450.  The BCG is surrounded by a confirmed 500kpc radio mini-halo \cite[see][]{Gitti07}.  Using high resolution VLA observations from the NRAO Image Archive, an SED of only the BCG is created.  This suggests an active AGN, whose SED is well fit with the GPS model with parameters of A$_{0}$=1.35$\pm$0.05, A$_{1}$=0.30$\pm$0.01 and A$_{2}$=-0.35$\pm$0.01.  The mini-halo emission has a steep spectrum with $\alpha$=1.20$\pm$0.04 and A=47.2$\pm$4.1.  The similarity of the SED of this source to other observed spectra in systems without observed mini-halos, but with smaller amorphous haloes lends evidence to the belief that the seed populations for true mini-haloes are built from repeated AGN activity in the BCG \cite[][]{Hogan15}. \\
{\bf RXJ1350+09} \\
2MASX J13502209+0940109, z=0.1325.  This source appears to be highly variable, and has previously been identified as a BL-LAC \cite[e.g.][]{Massaro09, Richards11}. A non-detection in the VLSS \cite[][]{Cohen07} shows that this source has no steep component at low frequencies and hence the SED appears to be GPS-like. Fit the SED with a GPS model, which returns a sharp peaked spectrum with parameters A$_{0}$=2.42$\pm$0.04, A$_{1}$=0.70$\pm$0.03 and A$_{2}$=-0.65$\pm$0.03. \\
{\bf RXJ1504-02} \\
LCRS B150131.5-023636, z=0.2171.  This cluster contains a confirmed radio mini-halo \cite[][]{Giacintucci11}, 140kpc in extent.  This mini-halo manifests itself as a steep spectrum tail in the SED to low frequencies (below about 1~GHz).  There appears to be an active core within the central BCG that gives rise to a GPS-like component in the integrated SED, that can be fit using the GPS model with parameters A$_{0}$=1.67$\pm$0.01, A$_{1}$=-0.09$\pm$0.01 and A$_{2}$=-0.16$\pm$0.04.  The mini-halo emission can be isolated from that of the AGN at L-band \cite[][]{White97} and agrees well with lower frequency data (see SED), giving a power-law fit to this lower frequency component with $\alpha$=1.19$\pm$0.05 and A=31.4$\pm$3.9. \\
{\bf RXJ1558-14} \\
PKS 1555-140, z=0.0970.  This is an actively evolving system with good historical coverage.  See section \ref{R1558_section} for a more detailed description of this system.  Here we perform an average GPS model fit of the core over the previous 40~years, recovering parameters of 2.69$\pm$0.04, 0.31$\pm$0.10 and -0.45$\pm$0.06 and then fit a power-law for the steep spectrum tail which has $\alpha$=1.30$\pm$0.20 and A=71.6$\pm$4.0. \\
{\bf RXJ1715+57} \\
NGC6338, z=0.0282.  Considering the archival C-band measurements \cite[e.g.][]{Gregory91} suggests that this source may show some long-term variability in here.  Without the GISMO observation the SED would perhaps be interpreted as a GPS, however the 150~GHz detection shows that the spectrum remains flat up to high frequency.  The GISMO flux could be interpreted as evidence of `flickering' at high frequencies or variability.  Either way, the SED is most reliably fit using a power-law with $\alpha$=0.41$\pm$0.04 and A=53.1$\pm$3.1.  This shows the source to be flat spectrum and is consistent with the non-detection in VLSS \cite[][]{Cohen07}. \\
{\bf RXJ1832+68} \\
2MASX J18323551+6848059, z=0.205.  The SED is dominated by steep spectrum emission to about 10~GHz that is fit using a power-law with $\alpha$=0.92$\pm$0.03 and A=194.8$\pm$7.7.  Above this frequency the spectrum appears to `bump' outwards.  VLBA observations (Hogan \etal 2015c) at C-band show that there is an active core component in here \cite[see also][]{Laurent-Muehleisen97}.  By considering measurements of the core flux only, this active component can be well fit with a relatively high-peaking GPS component with returned parameters of 0.78$\pm$0.36, 1.31$\pm$0.53 and -0.56$\pm$0.20.  When combined with the power-law this interpretation appears to well explain the integrated spectral shape. \\
{\bf RXJ2341+00} \\
PKS 2338+000, z=0.2767.  This is an unresolved radio source (at few-arcsecond resolution) whose SED is steeper at higher frequency yet is flattening at less than a few GHz.  The 365~MHz flux from the TEXAS survey \cite[][]{Douglas96} shows that the SED does turnover.  Fit the SED with a low peaking GPS, with returned parameters of 2.62$\pm$0.01, -0.16$\pm$0.06 and -0.15$\pm$0.04.  This source is a non-detection at the 5$\sigma$ detection limit of the VLSS catalogue \cite[][]{Cohen07}, and the limit is consistent with the inferred turnover.  Looking at the VLSS map overlaid onto an optical image of the region, we see that there is a faint ($<$3$\sigma$) detection at the correct position.  This point was not included in the fit but is included on the SED to highlight that it is consistent with being a genuine detection. \\
{\bf Z235} \\
2MASX J00435213+2424213, z=0.083.  The SED of this source has a flat spectral index, that can be fit with a single power-law of $\alpha$=0.45$\pm$0.06 and A=61.7$\pm$7.3.  The returned index is consistent with the non-detection in VLSS \cite[][]{Cohen07}. \\
{\bf Z8193} \\
B3 1715+425, z=0.1754.  This source appears to contain a variable core component that is confirmed by VLBI measurements \cite[][see also Hogan \etal 2015c, in prep]{Bourda10, Bourda11}.  The SED is fit by a combination of a GPS model to account for the spectral flattening above about 1~GHz and a steep power-law tail to lower frequencies.  Note that the source is below the detection threshold of the VLSS catalogue \cite[][]{Cohen07} but appears as a $\approx$3$\sigma$ detection at the correct position in the map overlay.  The flux is therefore retrived using the AIPS task \textsc{IMFIT} from a VLSS cut-out.  The returned parameters for the GPS component are 1.82$\pm$0.01, 0.26$\pm$0.01 and -0.26$\pm$0.01.  Accounting for the core contribution, a power law with $\alpha$=0.82$\pm$0.04 and A=89.2$\pm$5.7 is fitted to the low frequency tail. \\
{\bf Z8276} \\
2MASX J17441450+3259292, z=0.075.  There is a distinct split in the spectral index of the SED at approximately 1~GHz.  There also appears to be large amplitude variability at higher frequencies.  Fit the steeper component using a power-law with $\alpha$=1.26$\pm$0.03 and A=80.0$\pm$6.0 and then fit the GPS model to the core only measurements, which returns GPS-model parameters of 1.57$\pm$0.50, 0.52$\pm$0.76 and -0.30$\pm$0.26. \\

\begin{figure*}    \label{SED_APPENDIX}
  \centering
    \subfigure[{\it 4C+55.16}]{\includegraphics[width=8cm]{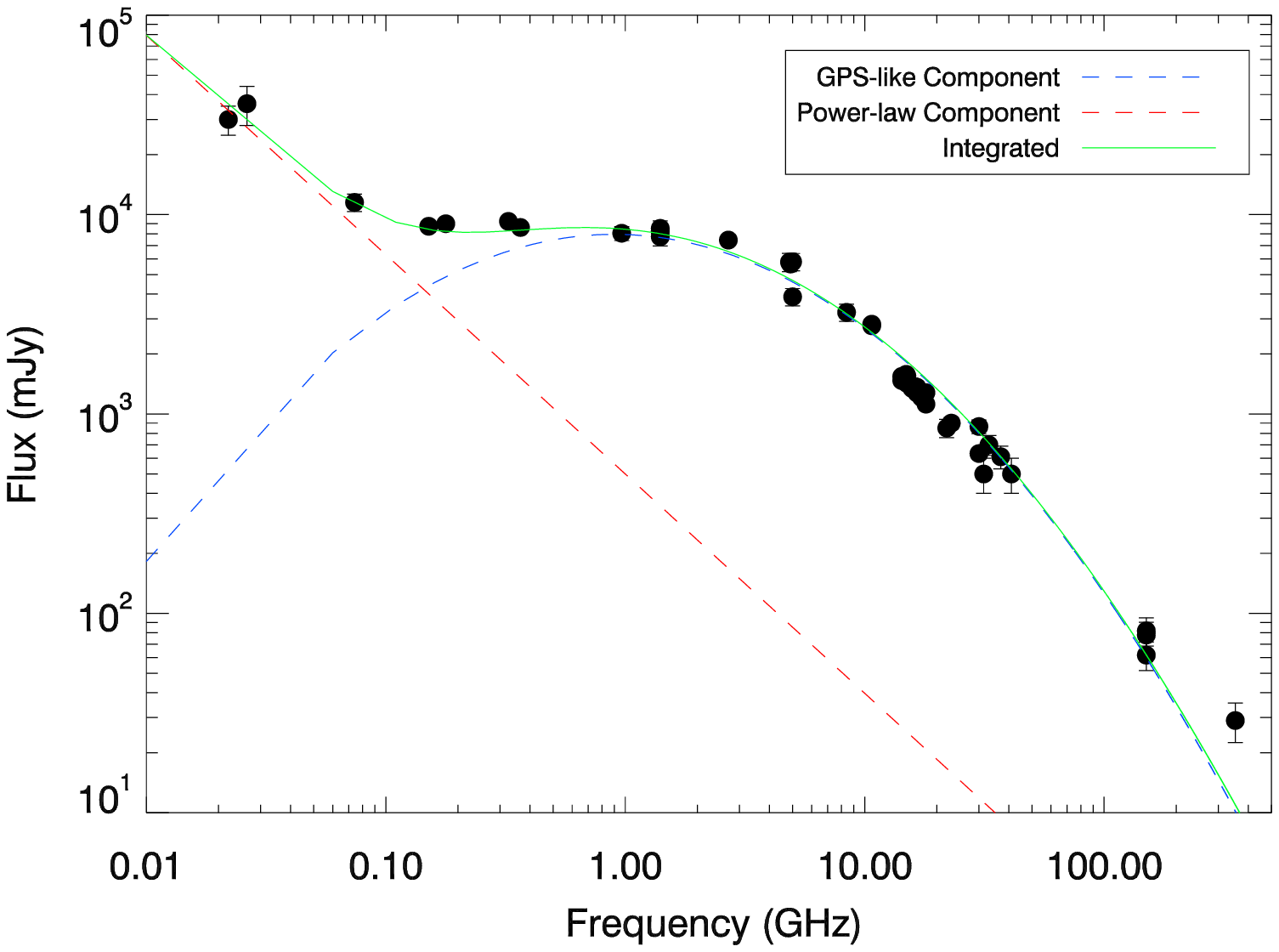}}
    \subfigure[{\it A1084}]{\includegraphics[width=8cm]{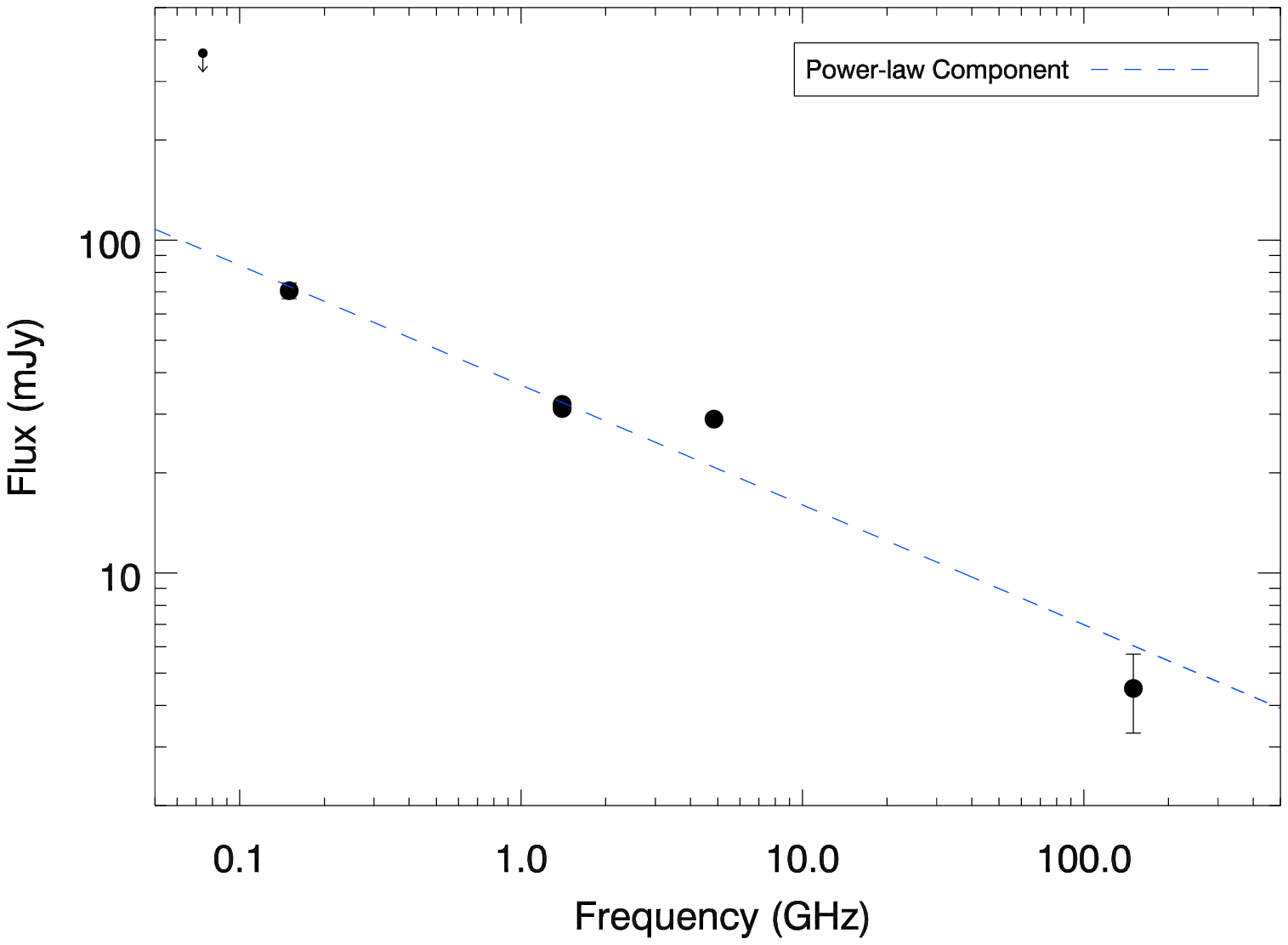}}
    \subfigure[{\it A1348}]{\includegraphics[width=8cm]{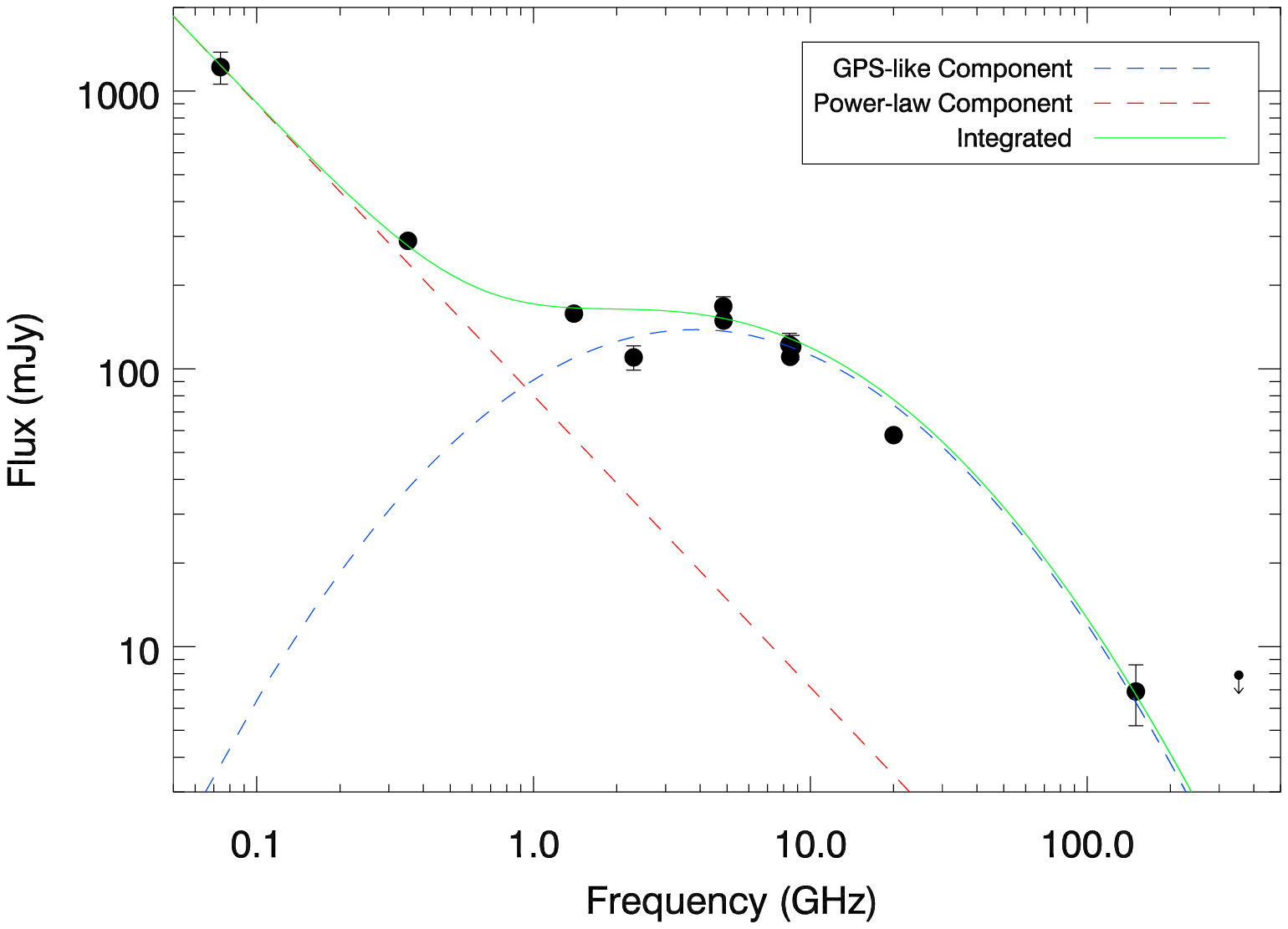}}
    \subfigure[{\it A1774}]{\includegraphics[width=8cm]{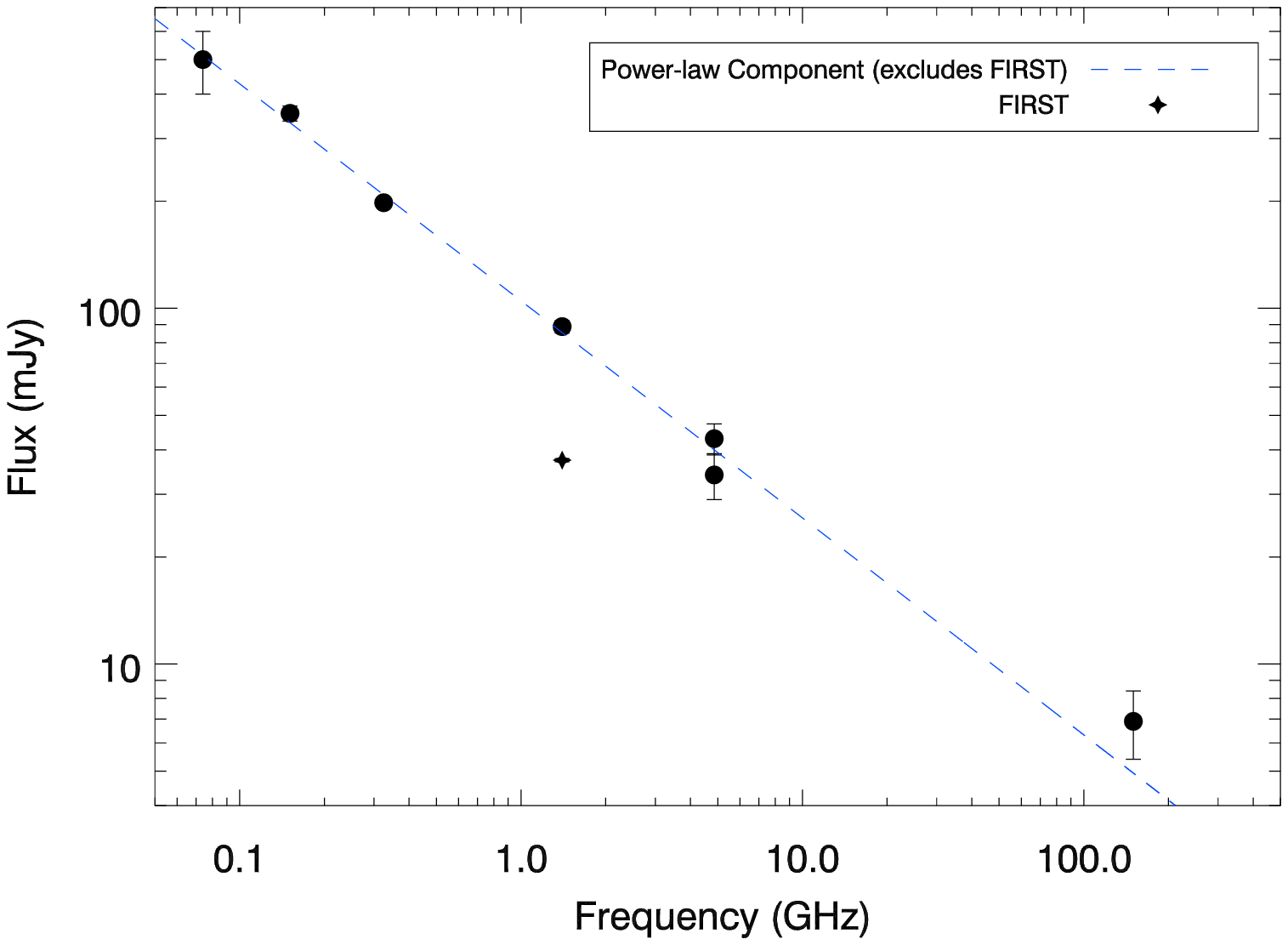}}
    \subfigure[{\it A1795}]{\includegraphics[width=8cm]{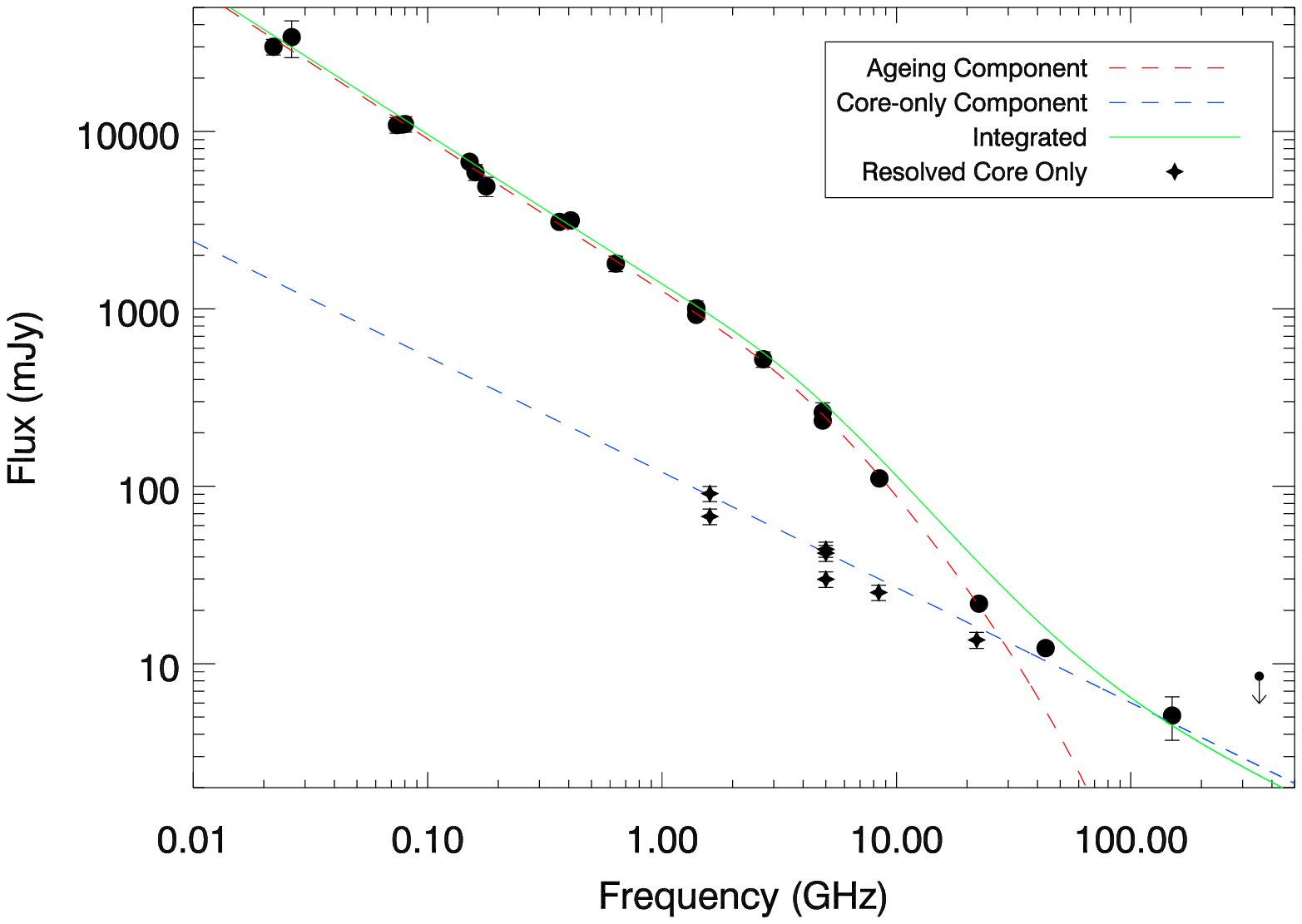}}
    \subfigure[{\it A1885}]{\includegraphics[width=8cm]{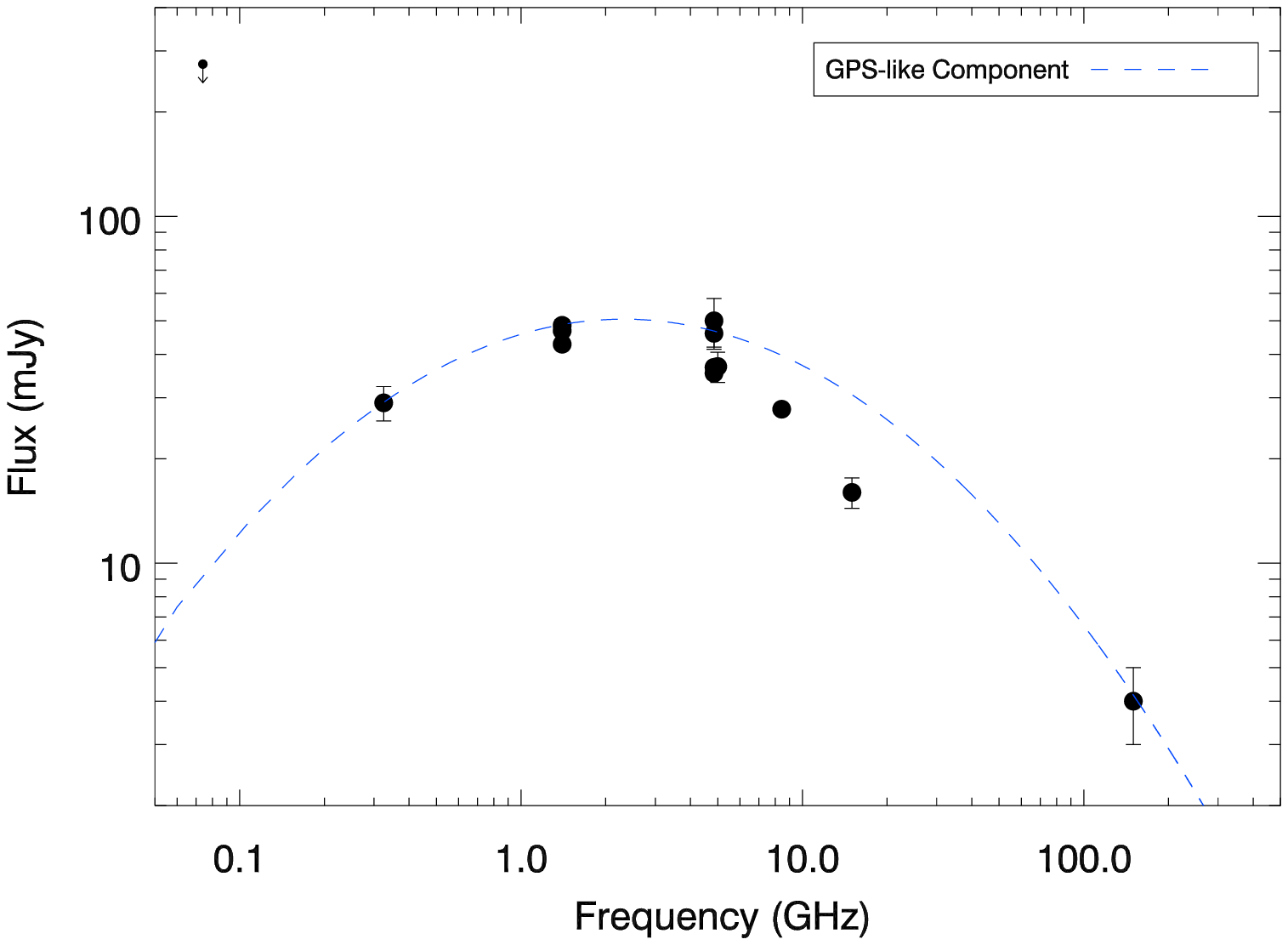}} 
  \caption{SEDs for sources withing our sample.  Typically, data-points represent the spatially-integrated flux density at the given frequency.  In many cases this is dominated by either the core, or the extended emission, as indicated by the fits.  Core only data-points have been highlighted for instances where a separate SED was fitted for cores that could be morphologically isolated, to differentiate these from instances where the integrated flux shows clear variability. Likewise, other `special' cases have been indicated on the plots.} 
 \label{SEDs}
\end{figure*}

\begin{figure*}    
  \centering
    \subfigure[{\it A2052}]{\includegraphics[width=8cm]{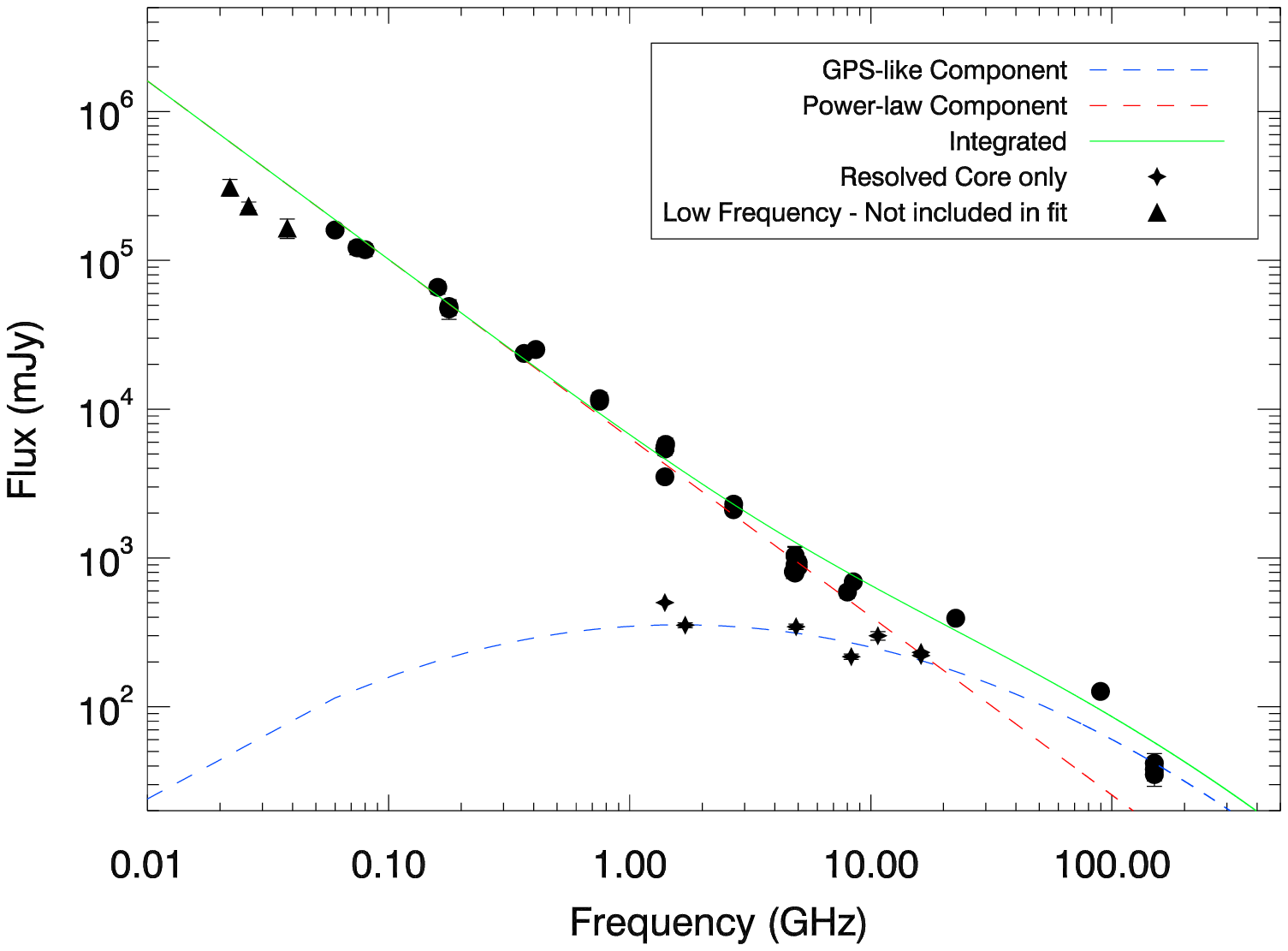}}
    \subfigure[{\it A2055}]{\includegraphics[width=8cm]{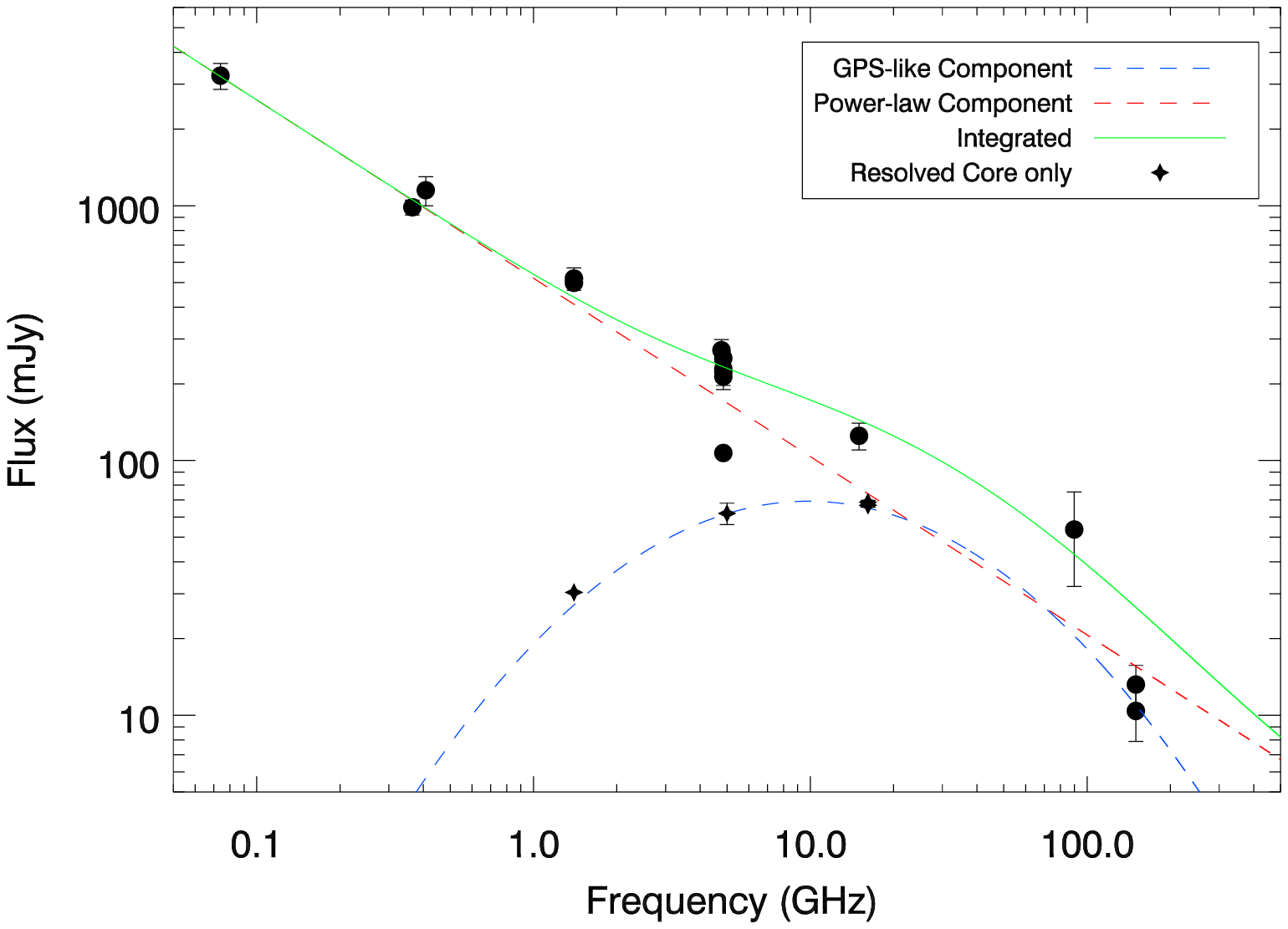}}
    \subfigure[{\it A2270}]{\includegraphics[width=8cm]{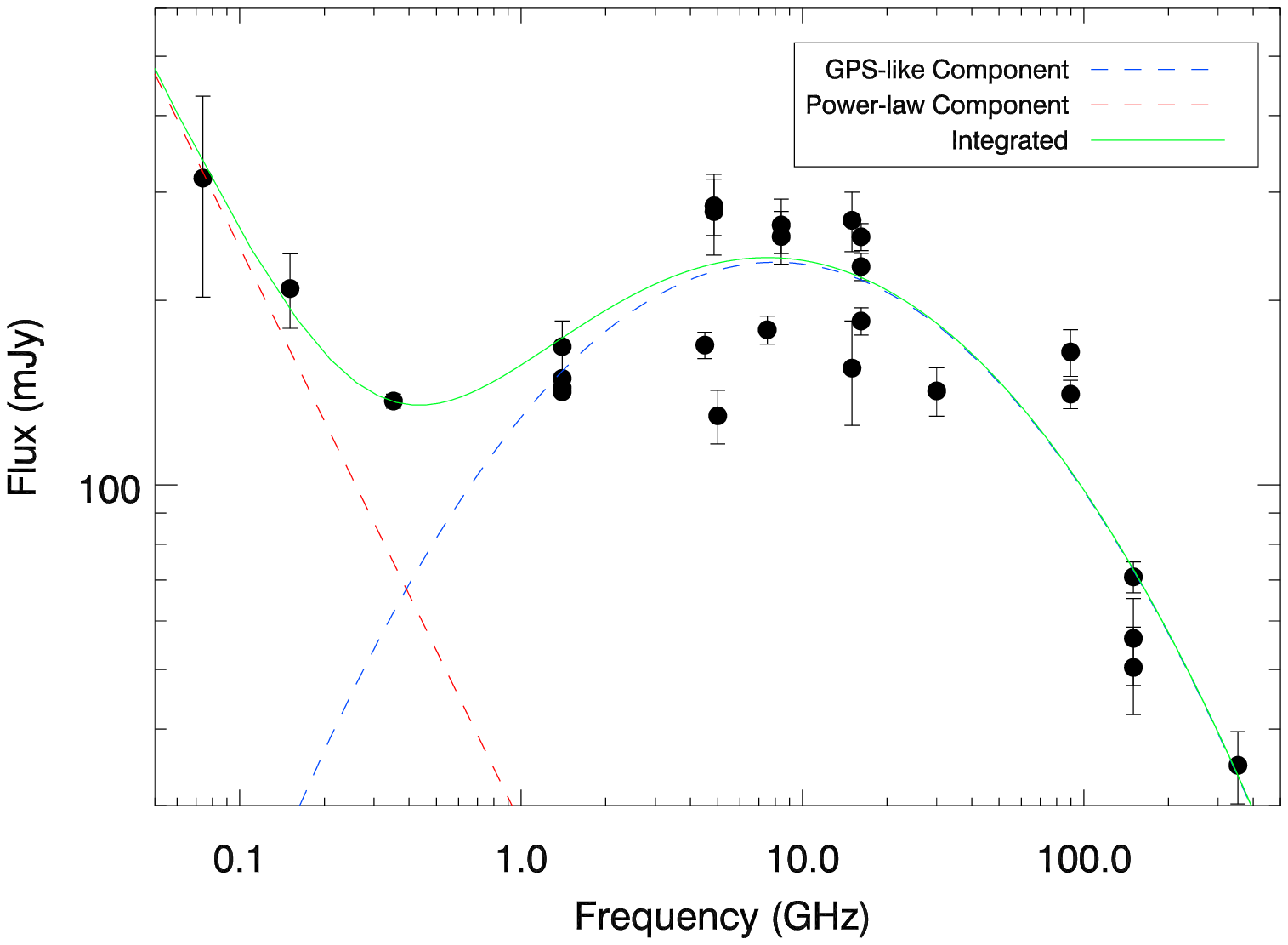}}
    \subfigure[{\it A2390}]{\includegraphics[width=8cm]{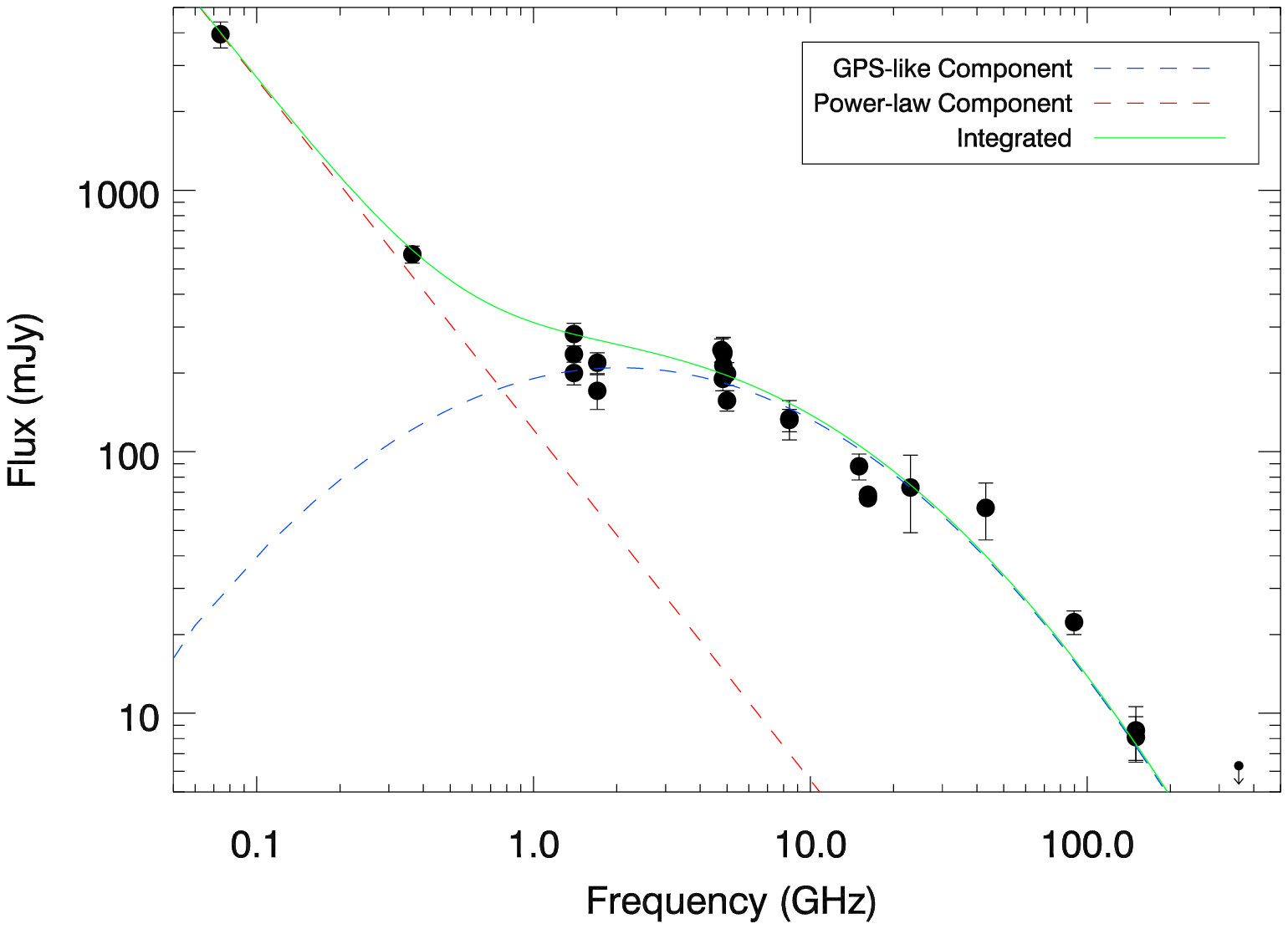}}
    \subfigure[{\it A2415}]{\includegraphics[width=8cm]{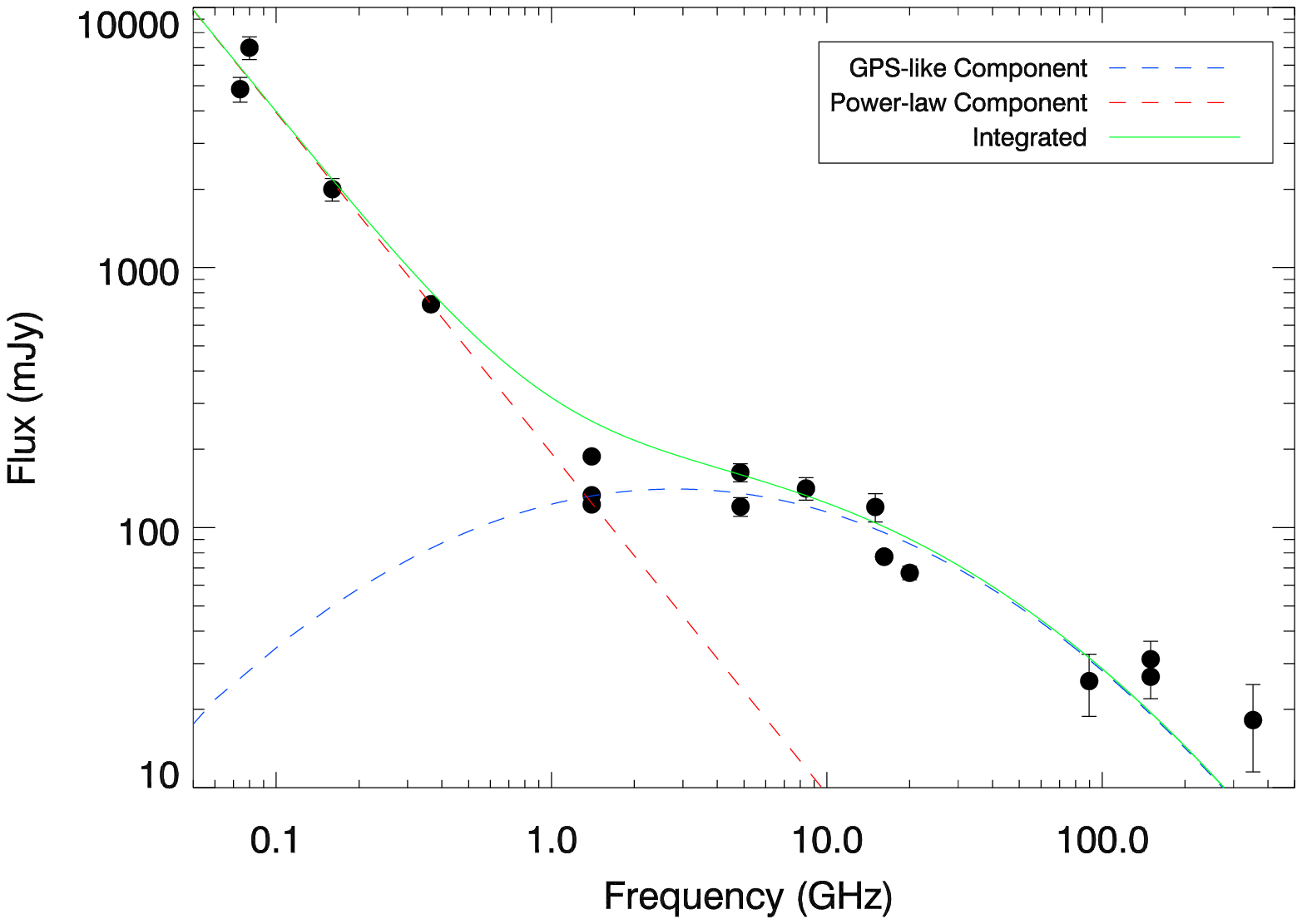}}
    \subfigure[{\it A2597}]{\includegraphics[width=8cm]{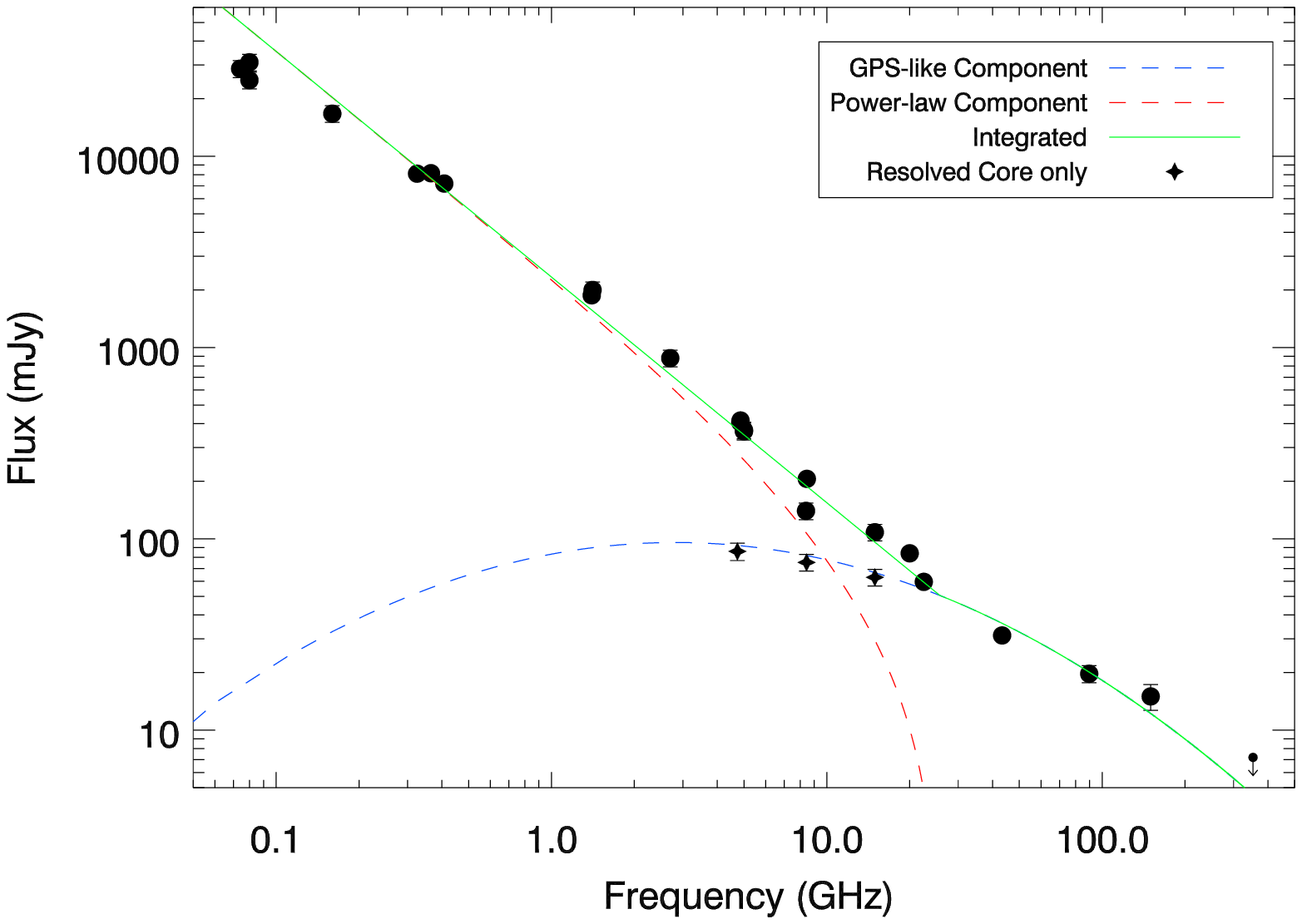}}
  \caption{SEDs, continuation of Fig. A1} 
\end{figure*}

\begin{figure*}    
  \centering
    \subfigure[{\it A2627}]{\includegraphics[width=8cm]{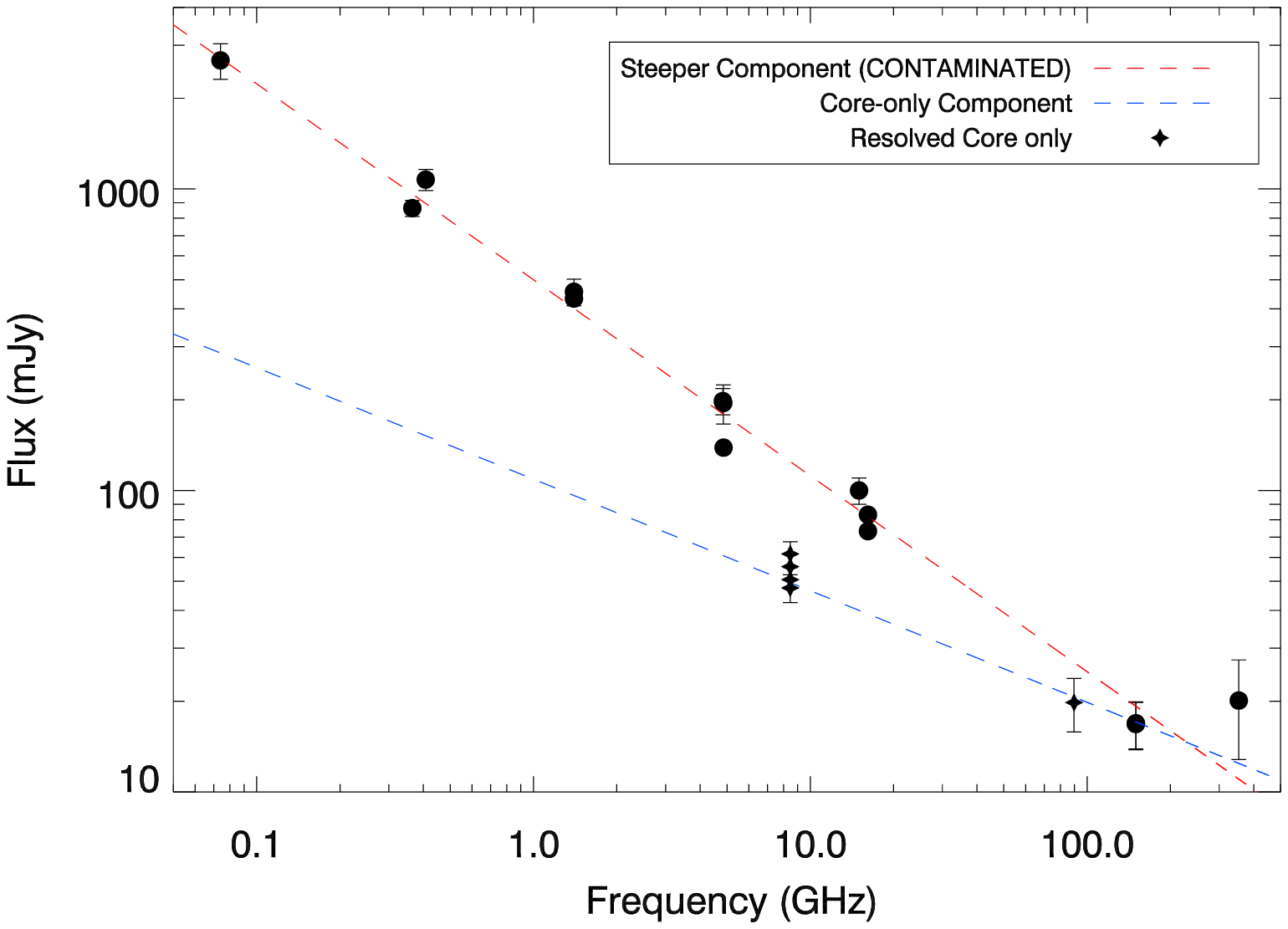}}
    \subfigure[{\it A3581}]{\includegraphics[width=8cm]{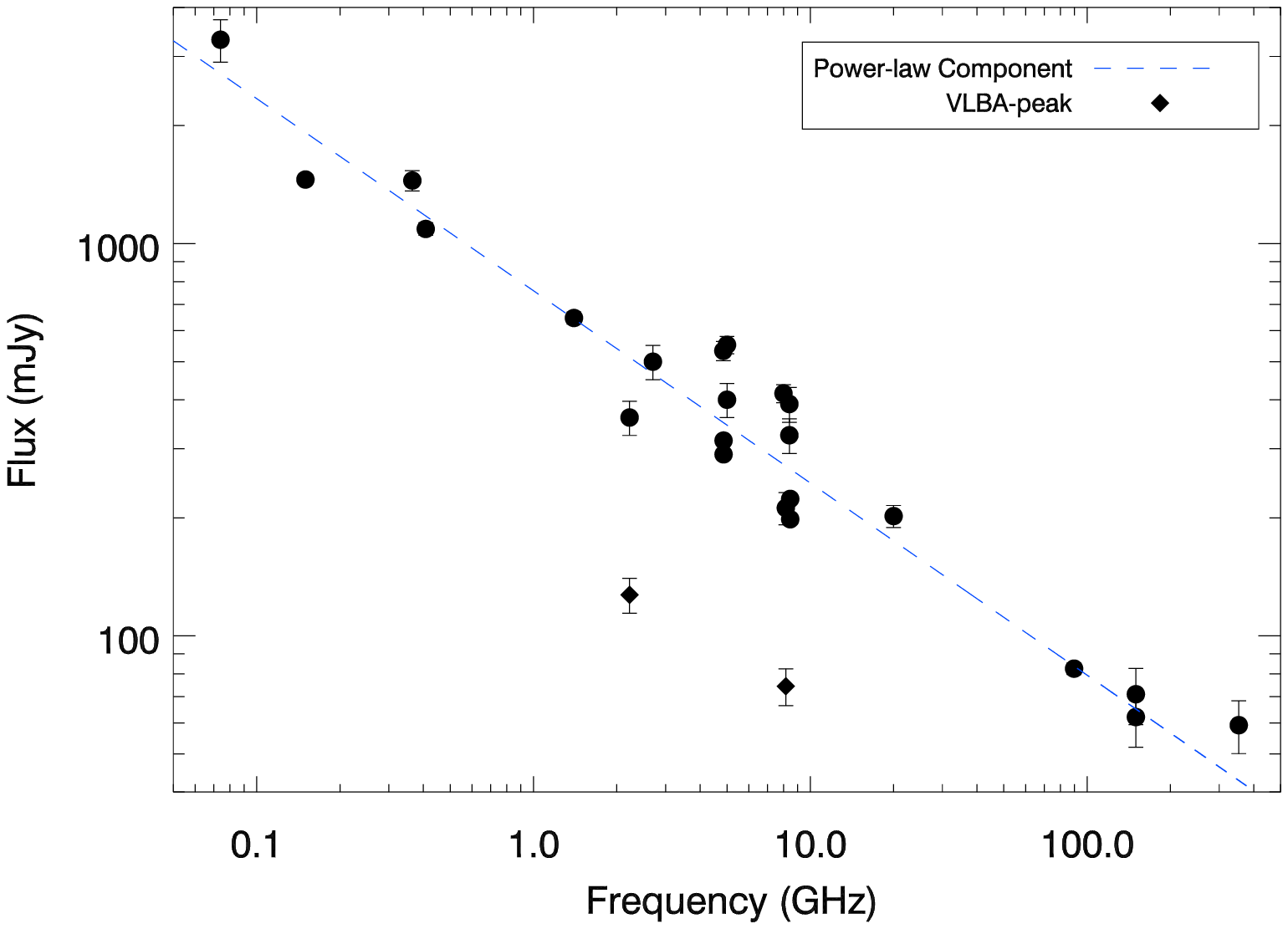}}
    \subfigure[{\it A496}]{\includegraphics[width=8cm]{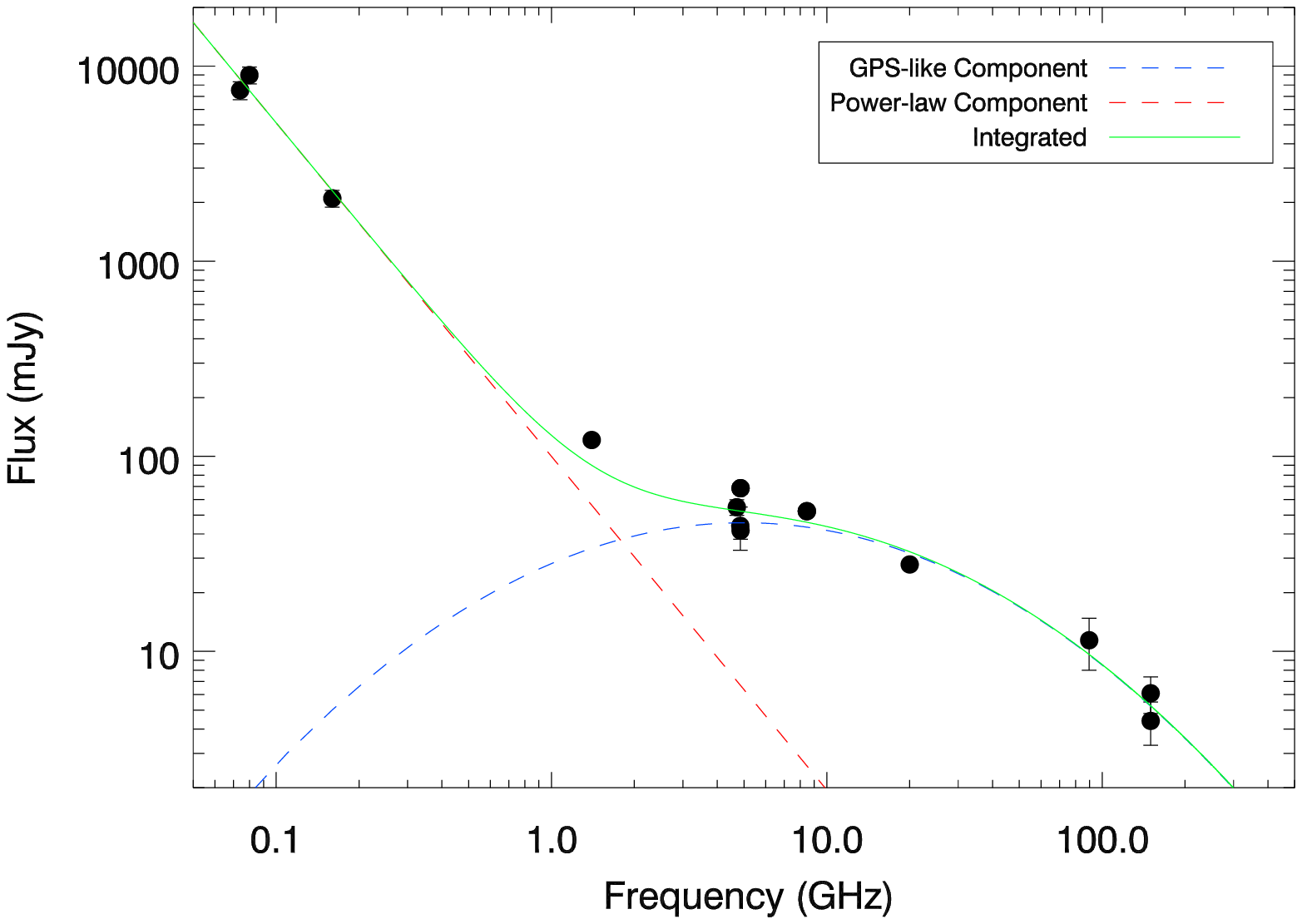}}
    \subfigure[{\it A646}]{\includegraphics[width=8cm]{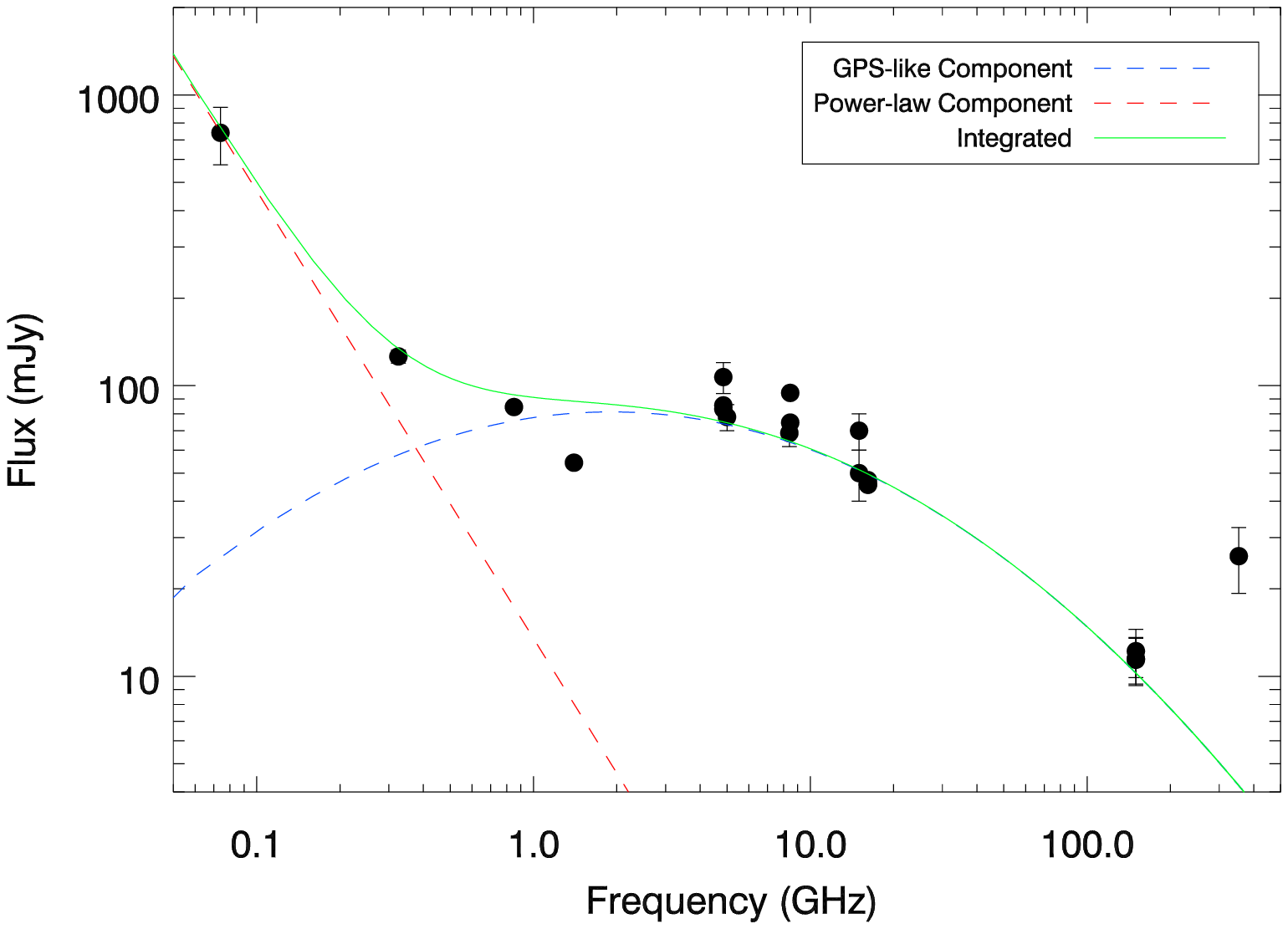}}
    \subfigure[{\it AS780}]{\includegraphics[width=8cm]{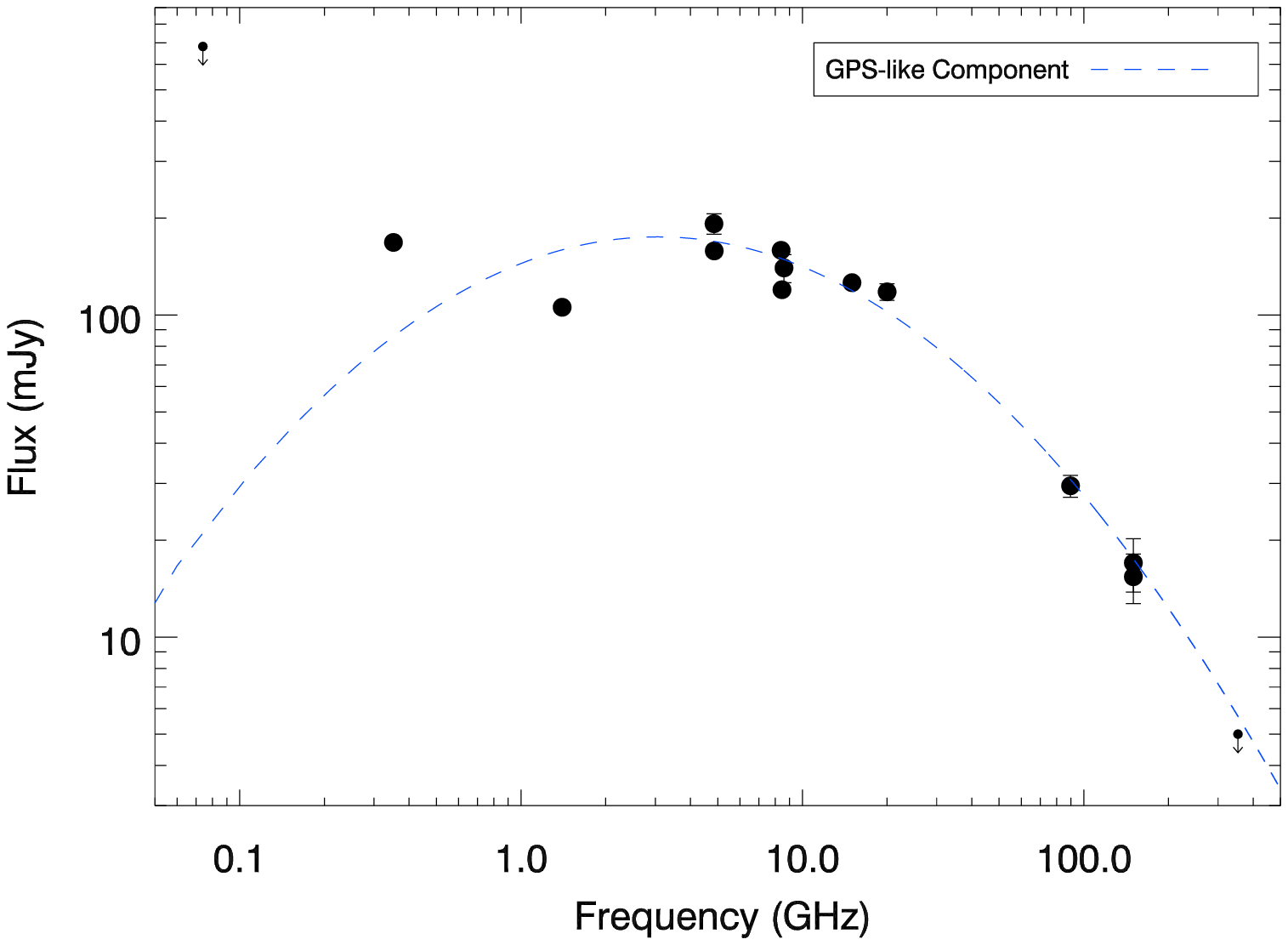}}
    \subfigure[{\it E1821+644}]{\includegraphics[width=8cm]{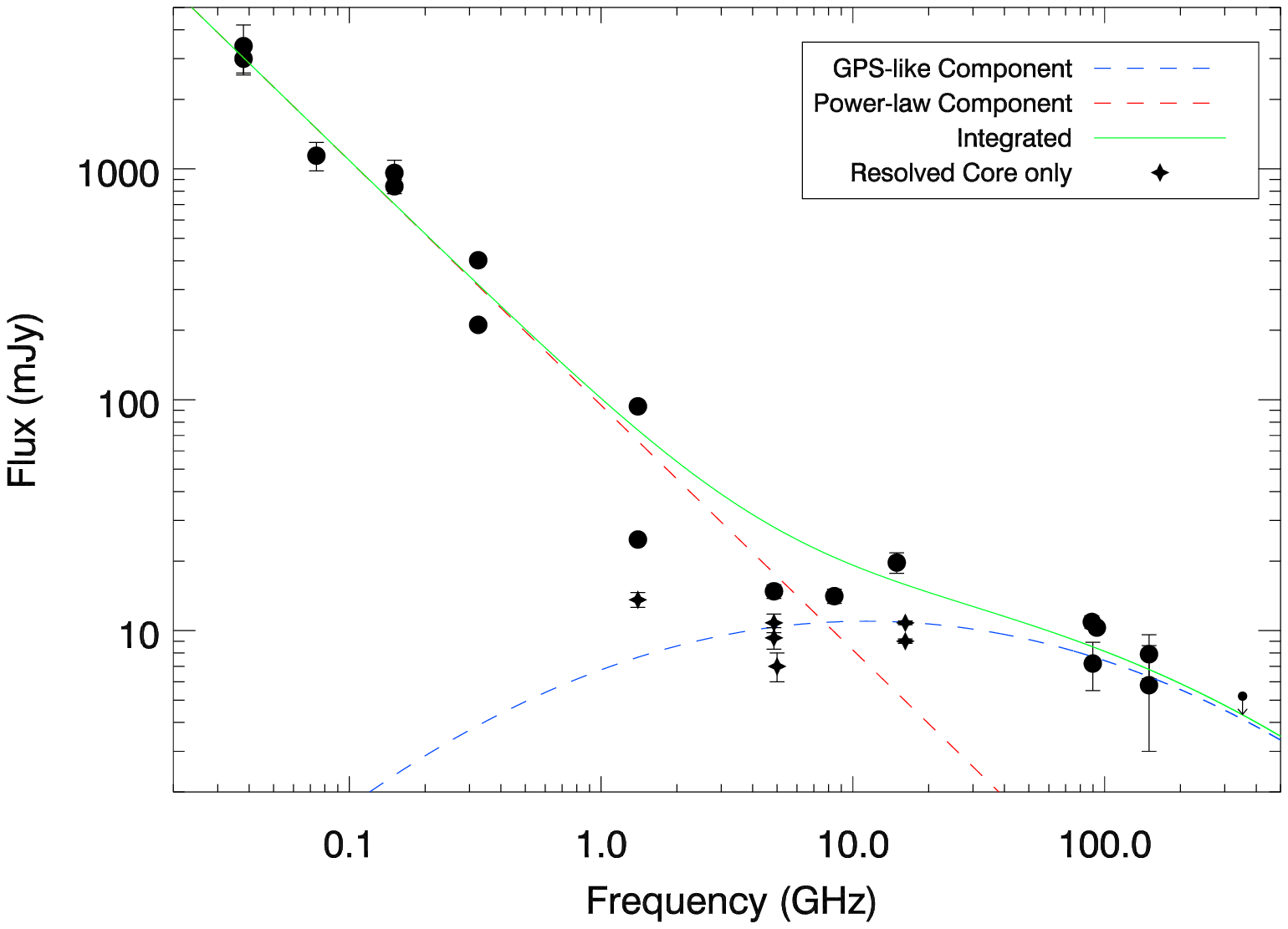}}
  \caption{SEDs, continuation of Fig. A1} 
\end{figure*}
\begin{figure*}    
  \centering
    \subfigure[{\it Hydra-a}]{\includegraphics[width=8cm]{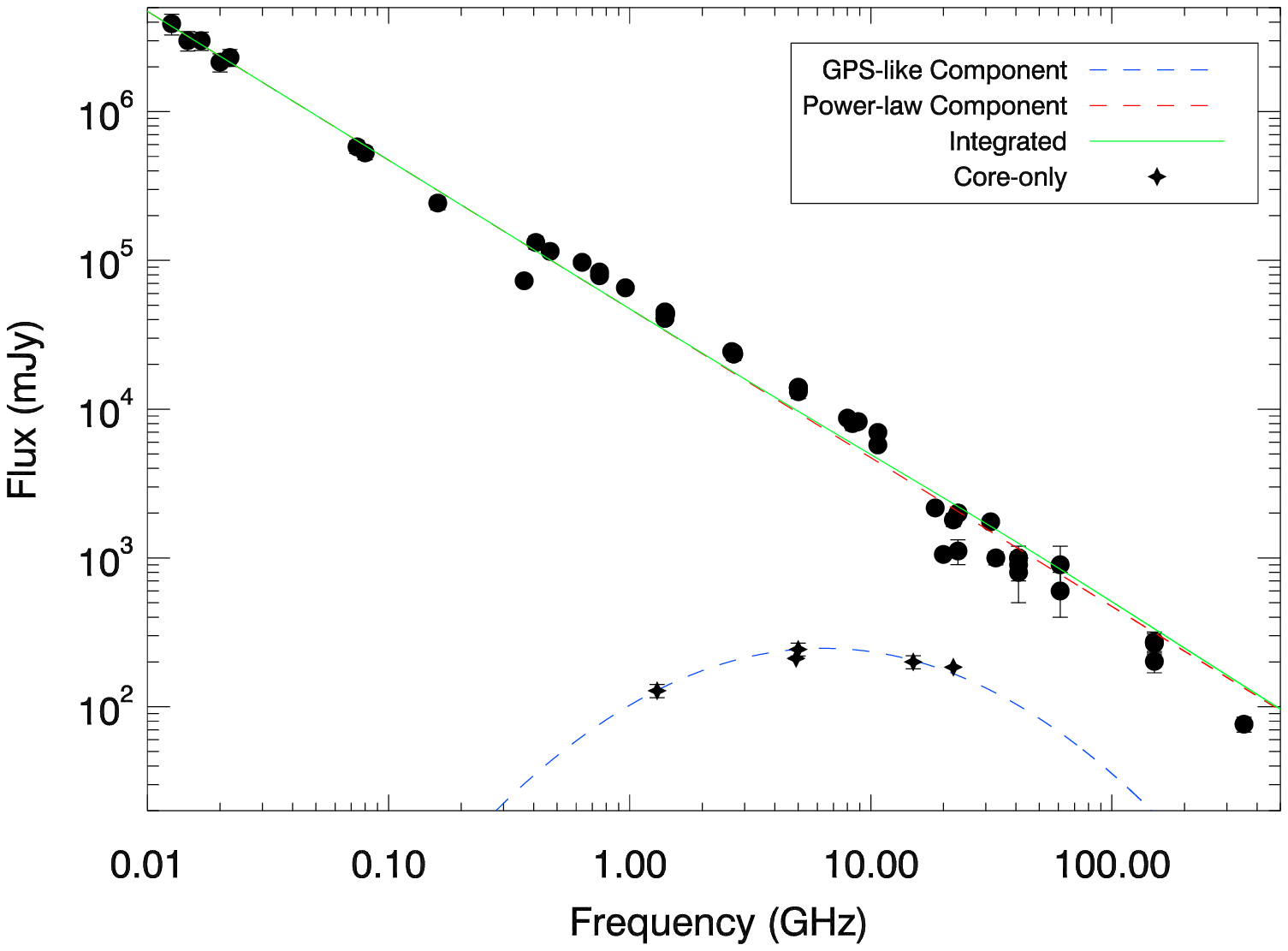}}
    \subfigure[{\it MACS0242-21}]{\includegraphics[width=8cm]{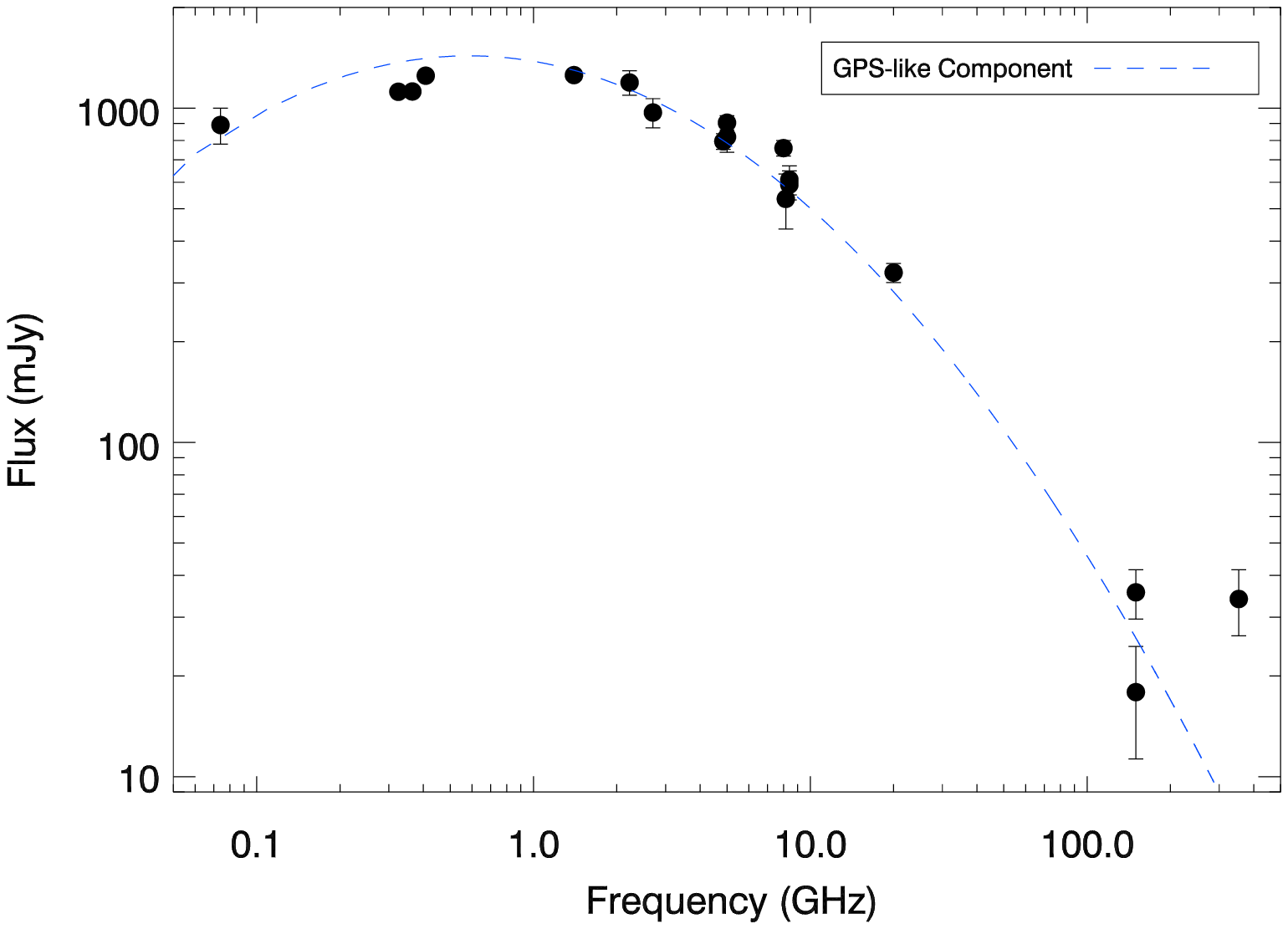}}
    \subfigure[{\it MACS1931-26}]{\includegraphics[width=8cm]{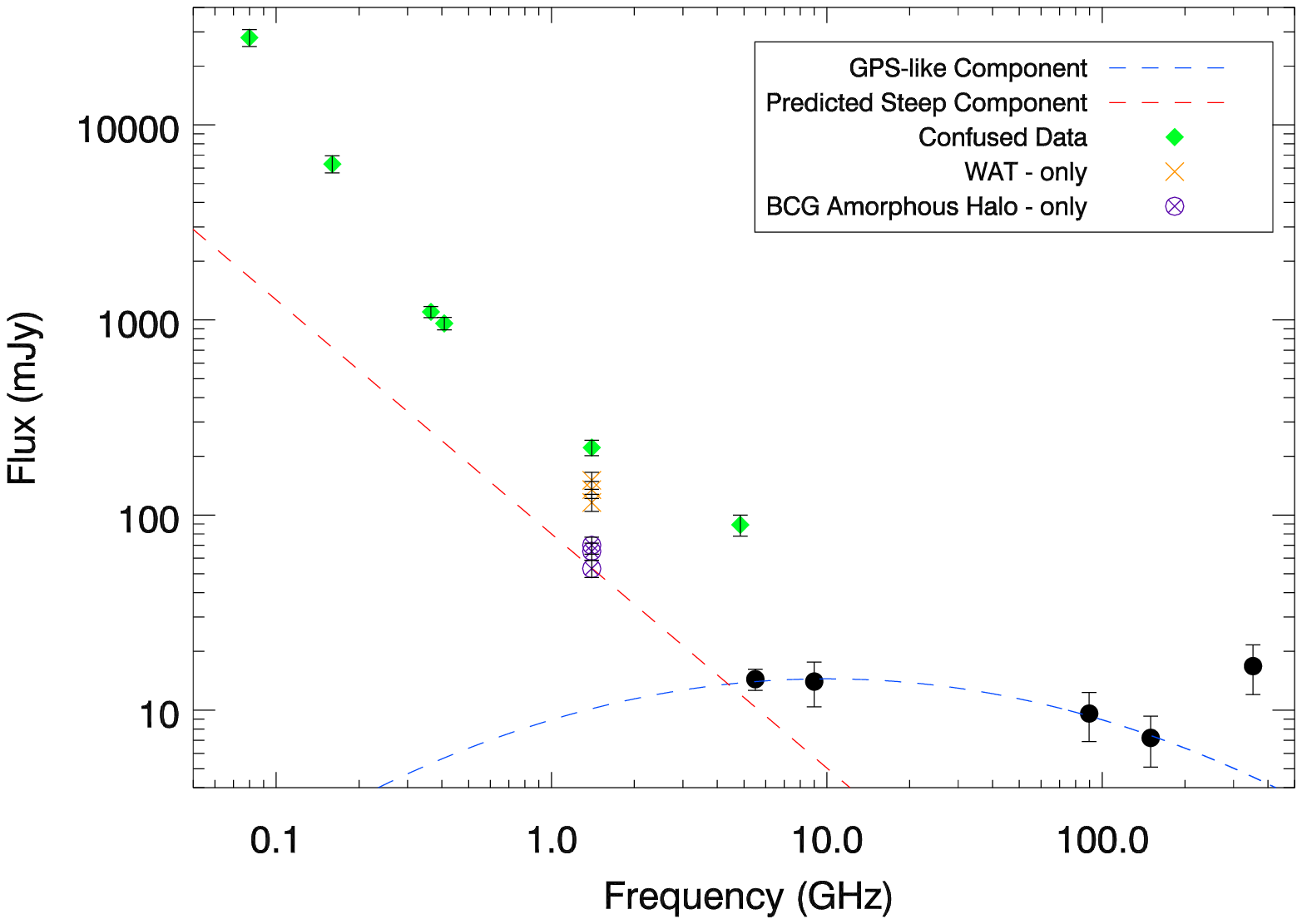}}
    \subfigure[{\it RXJ0132.6-0804}]{\includegraphics[width=8cm]{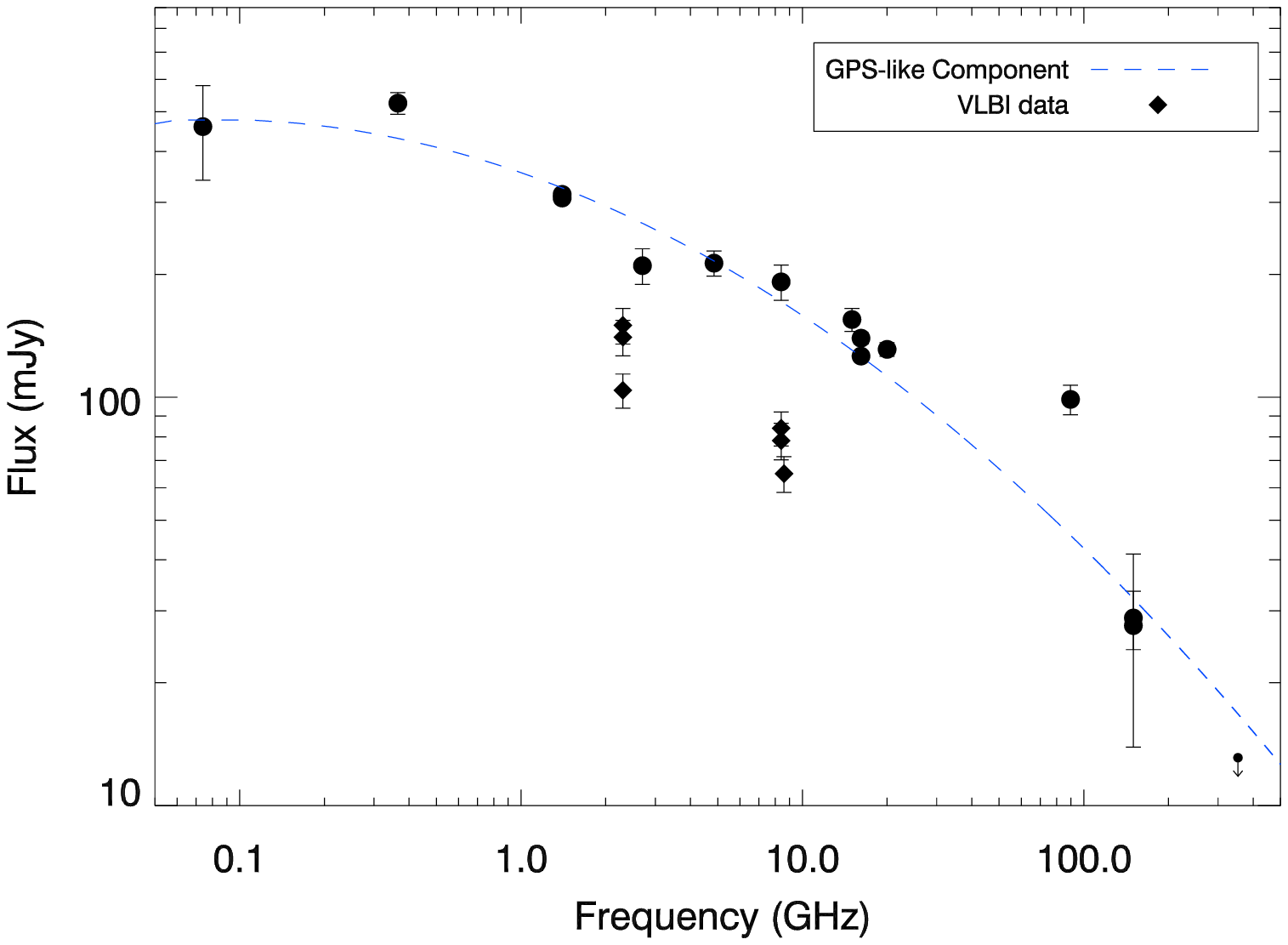}}
    \subfigure[{\it RXJ0352.9+1941}]{\includegraphics[width=8cm]{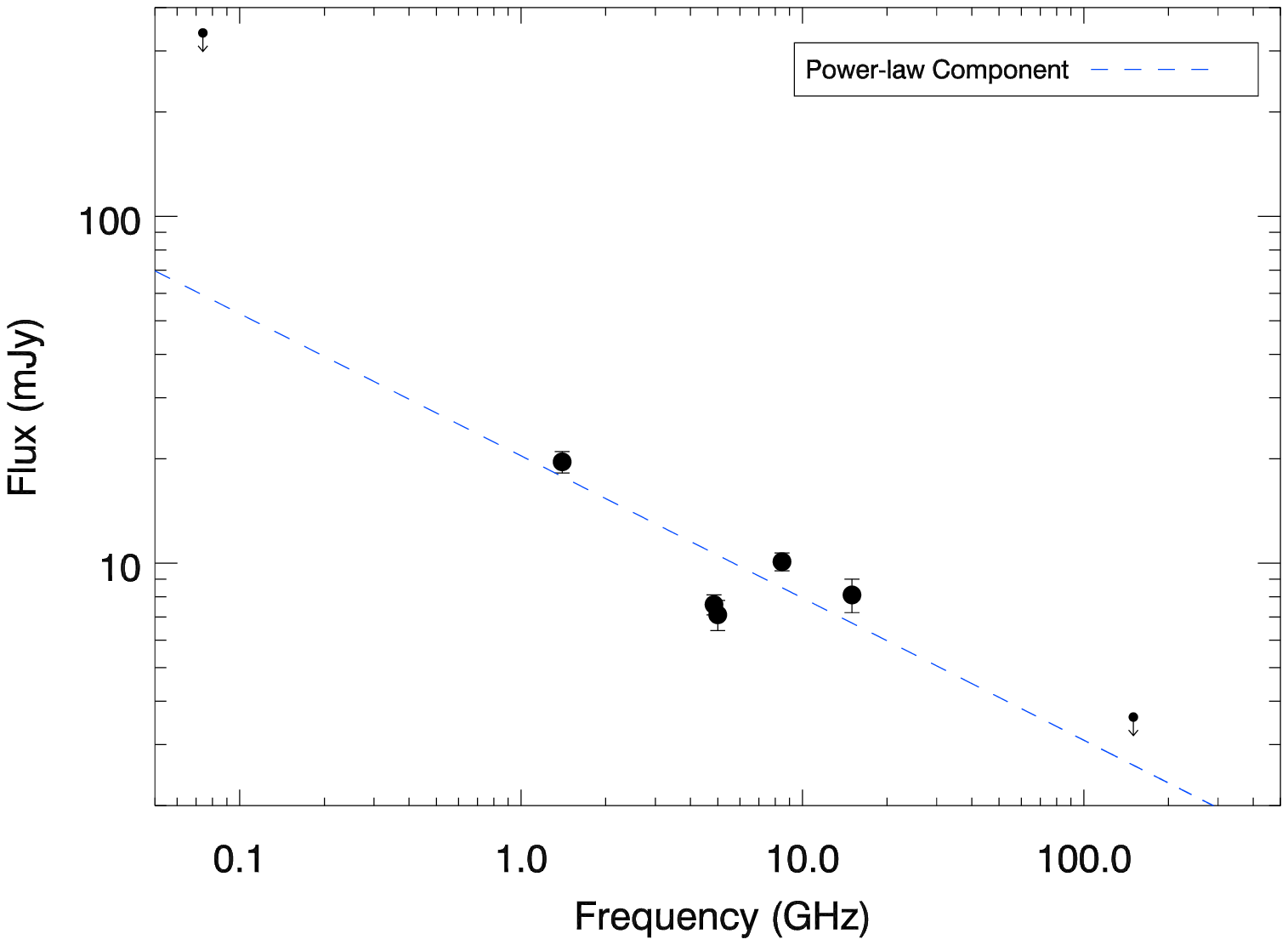}}
    \subfigure[{\it RXJ0439.0+0520}]{\includegraphics[width=8cm]{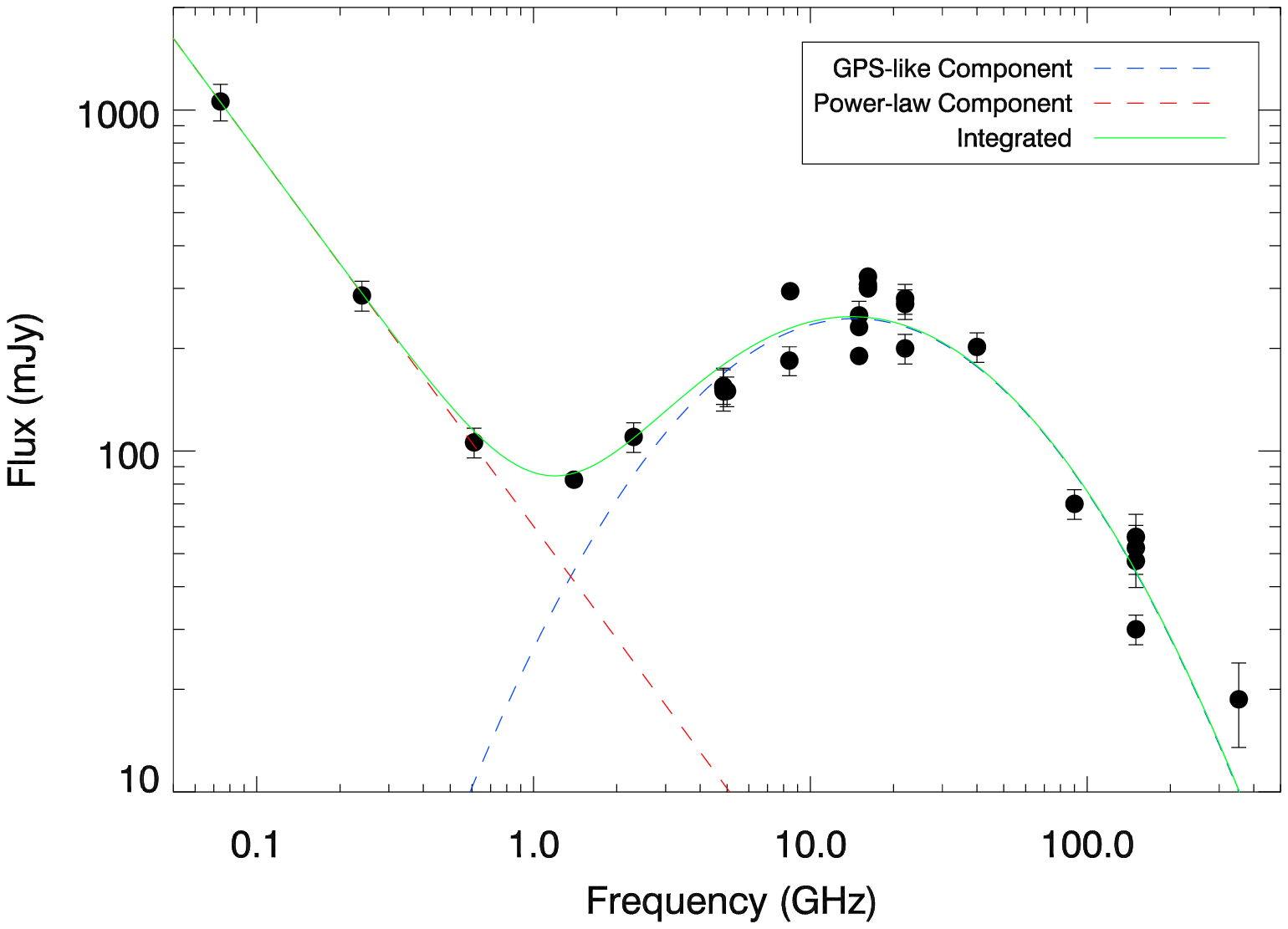}}
  \caption{SEDs, continuation of Fig. A1} 
\end{figure*}

\begin{figure*}    
  \centering
    \subfigure[{\it RXJ0747.5-1917}]{\includegraphics[width=8cm]{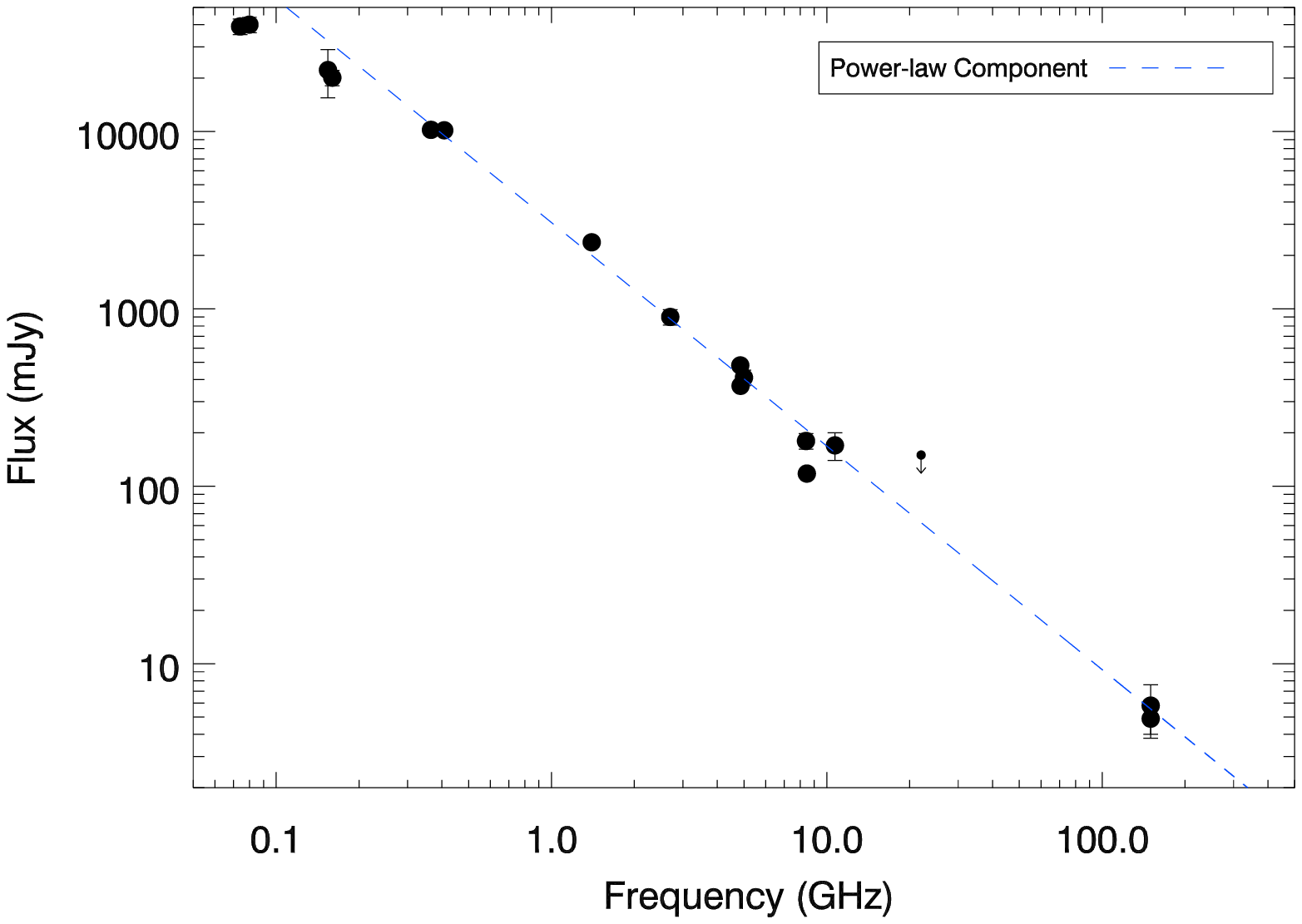}}
    \subfigure[{\it RXJ1347.5-1145}]{\includegraphics[width=8cm]{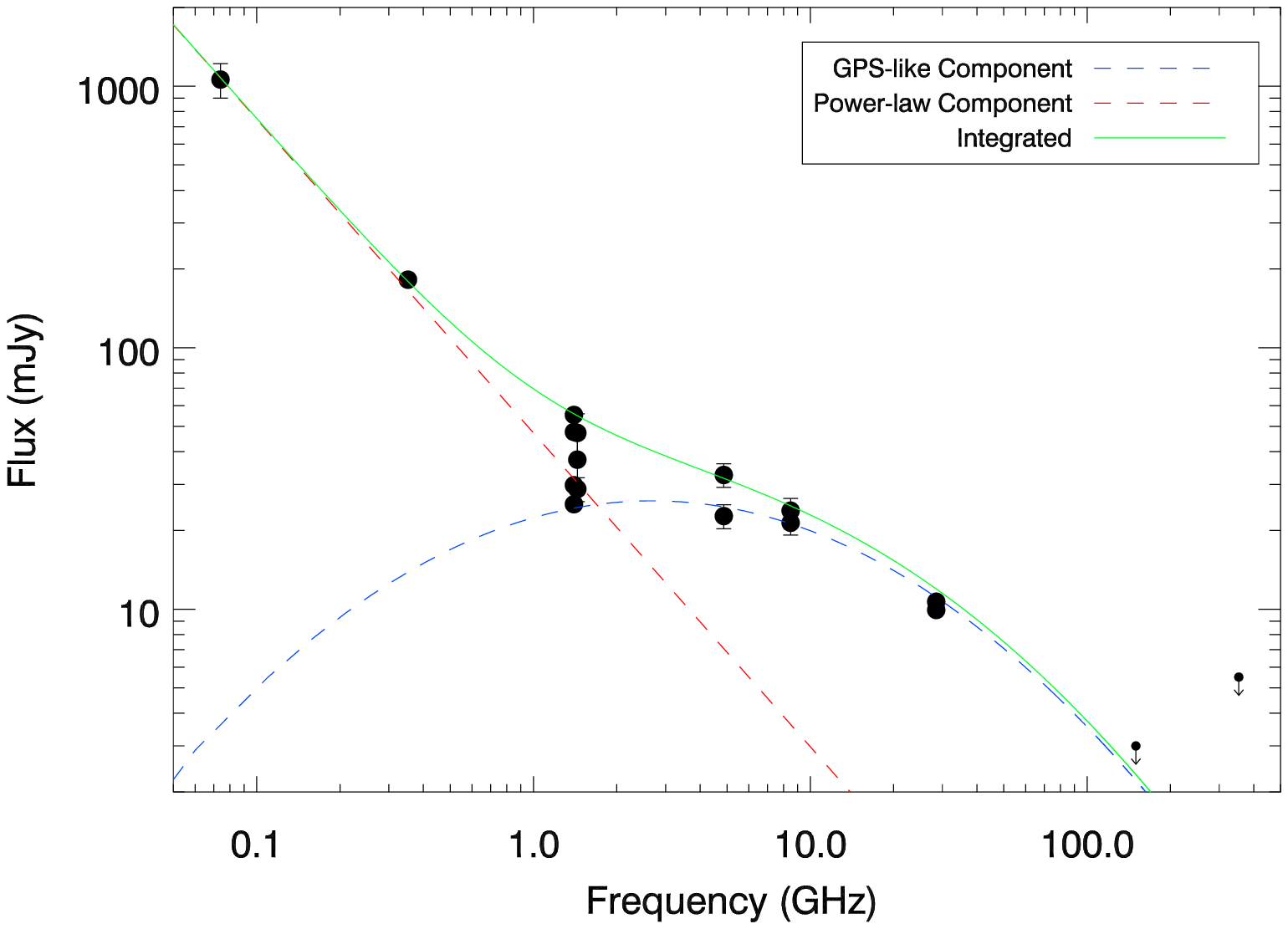}}
    \subfigure[{\it RXJ1350.3+0940}]{\includegraphics[width=8cm]{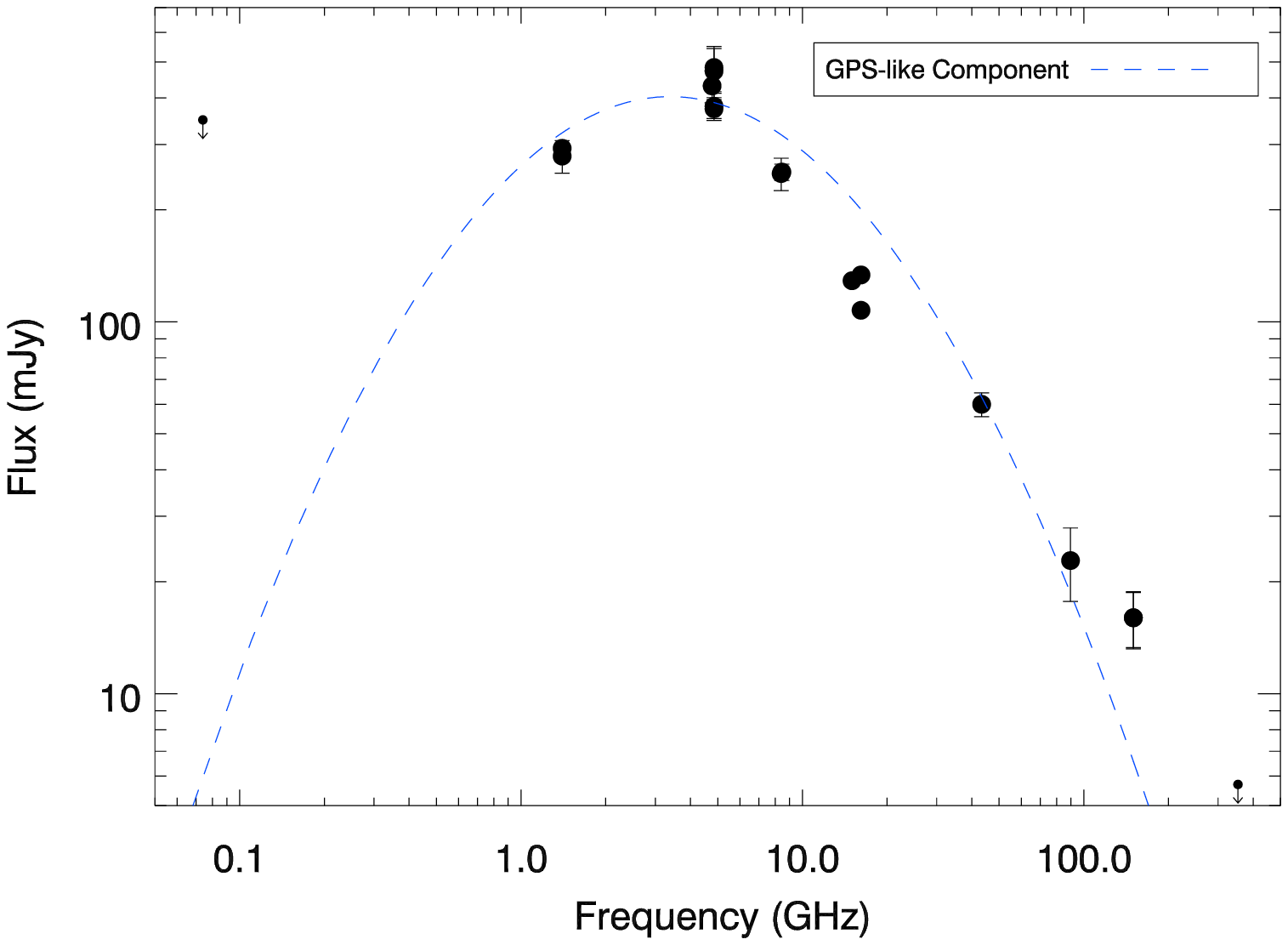}}
    \subfigure[{\it RXJ1504.1-0248}]{\includegraphics[width=8cm]{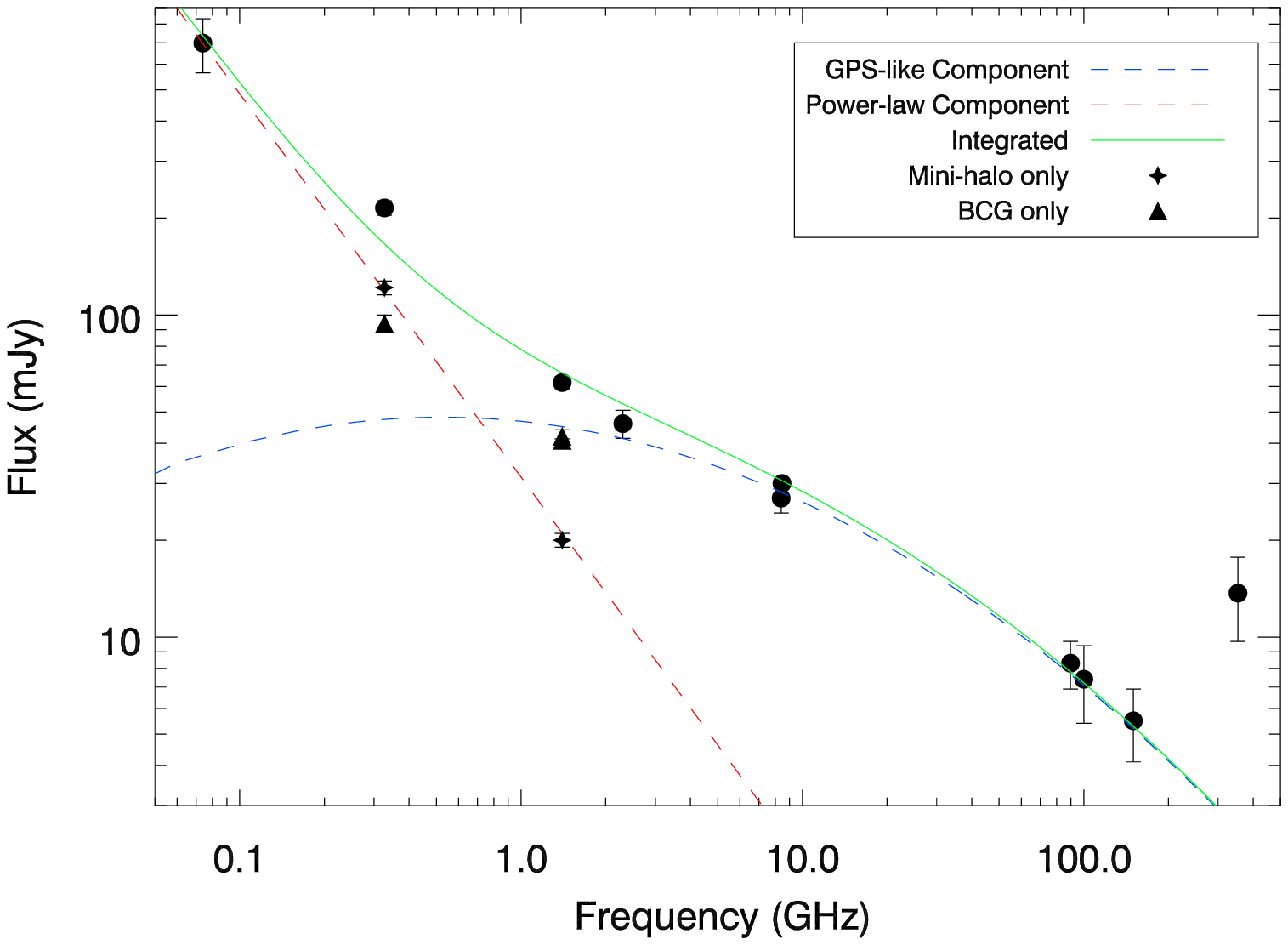}}
    \subfigure[{\it RXJ1558.4-1410}]{\includegraphics[width=8cm]{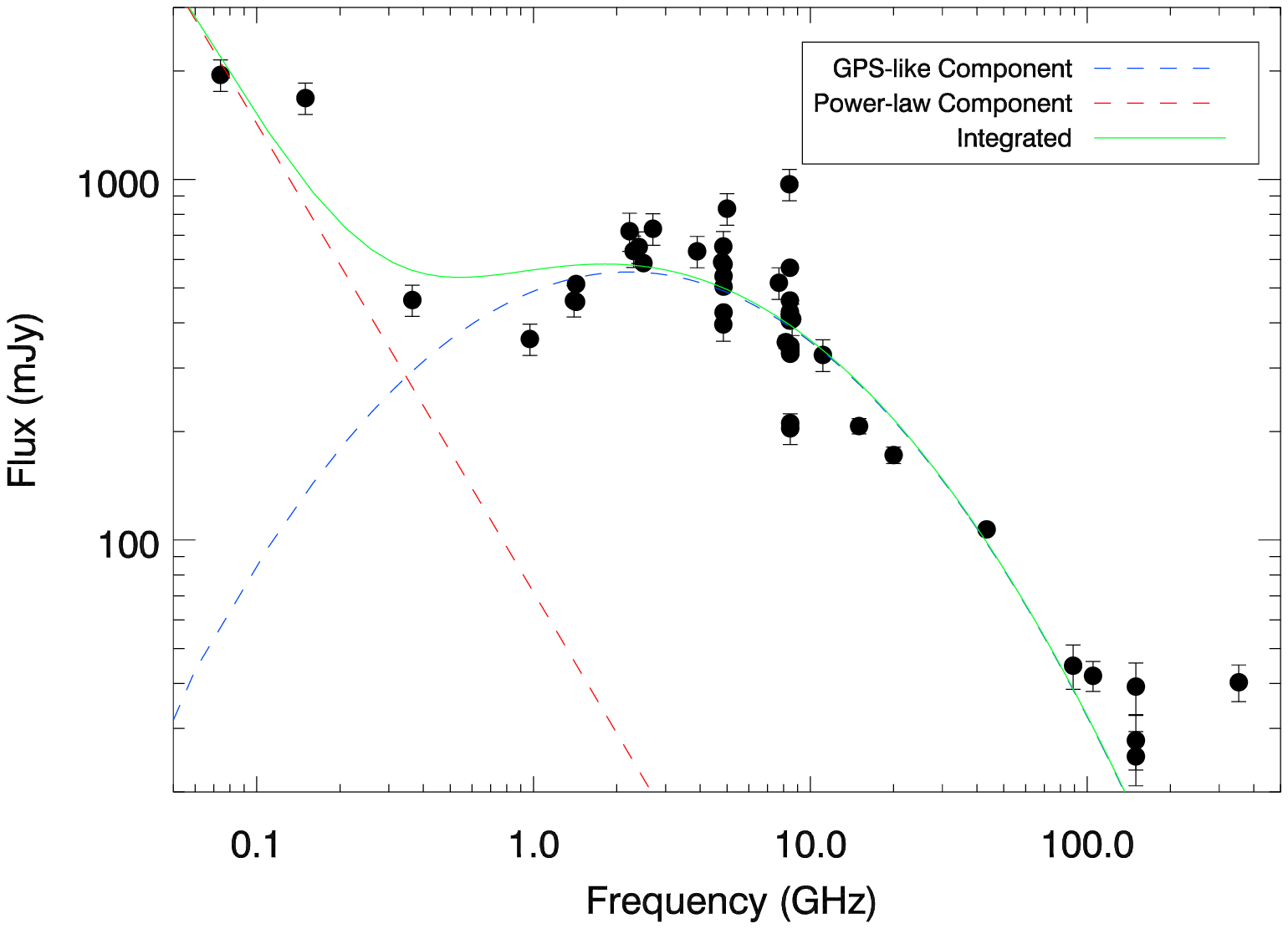}}
    \subfigure[{\it RXJ1715.3+5725 (NGC6338)}]{\includegraphics[width=8cm]{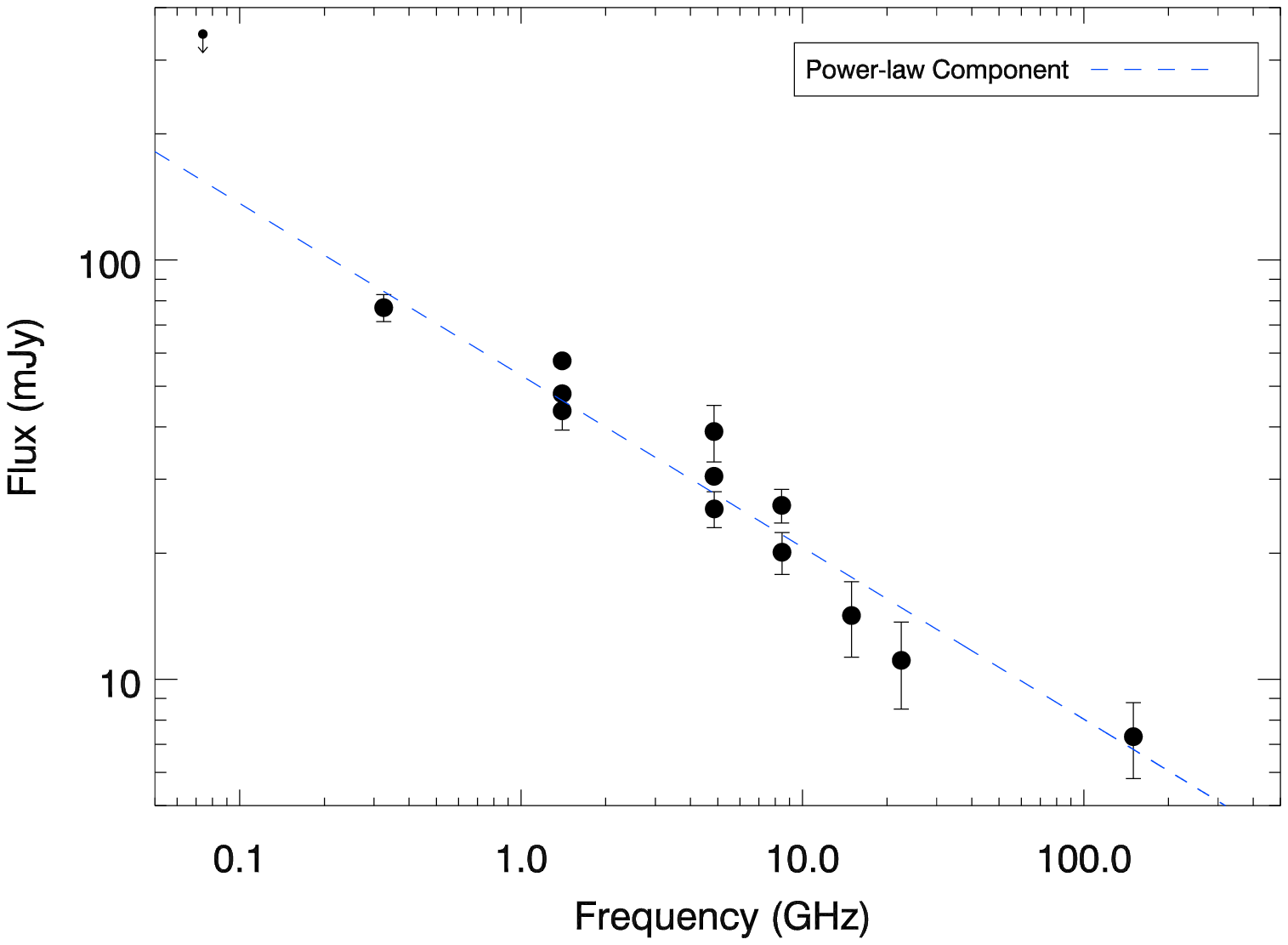}}
  \caption{SEDs, continuation of Fig. A1}
\end{figure*}

\begin{figure*}    
  \centering
    \subfigure[{\it RXJ1832.5+6848}]{\includegraphics[width=8cm]{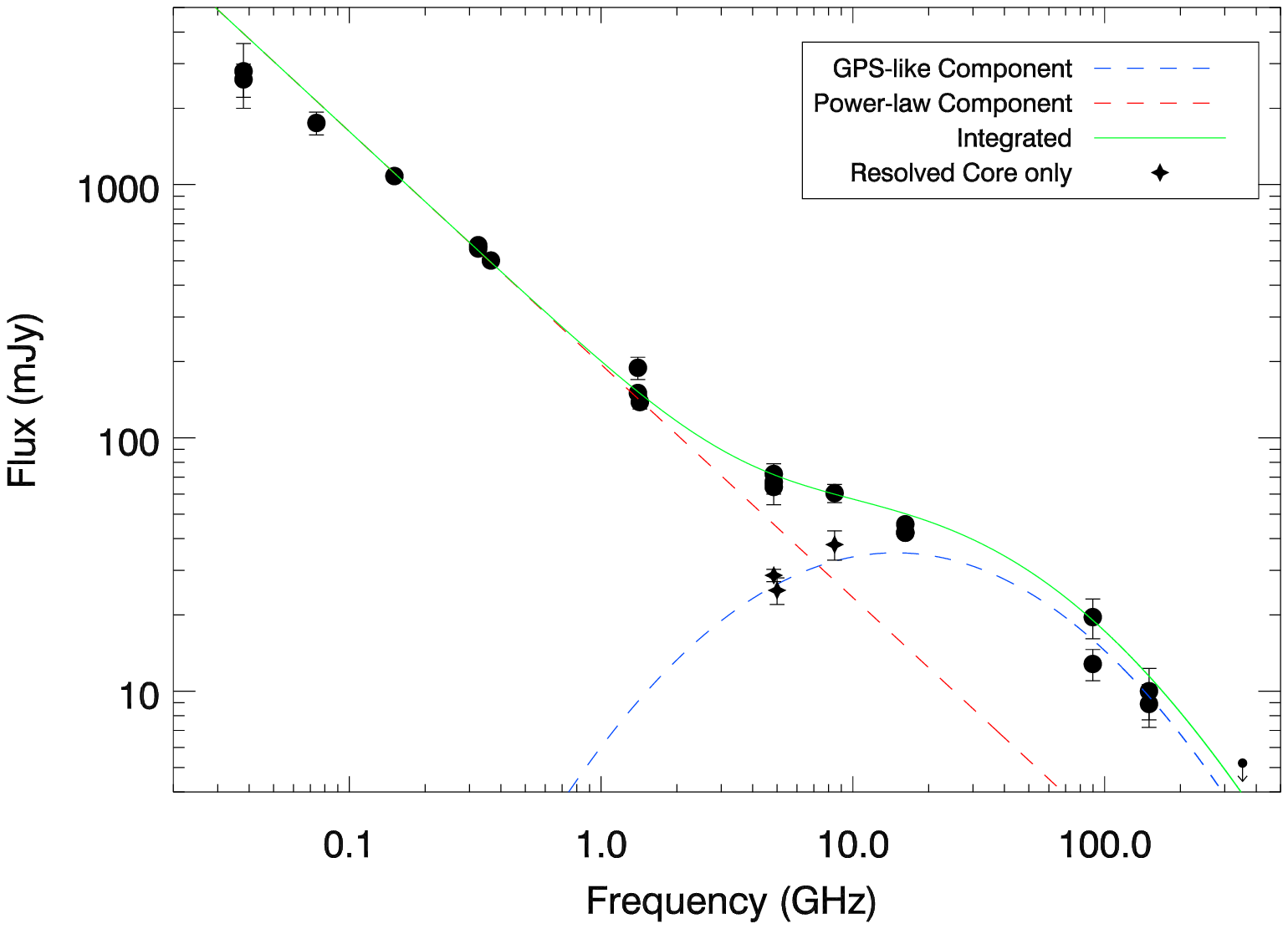}}
    \subfigure[{\it RXJ2341.1+0018}]{\includegraphics[width=8cm]{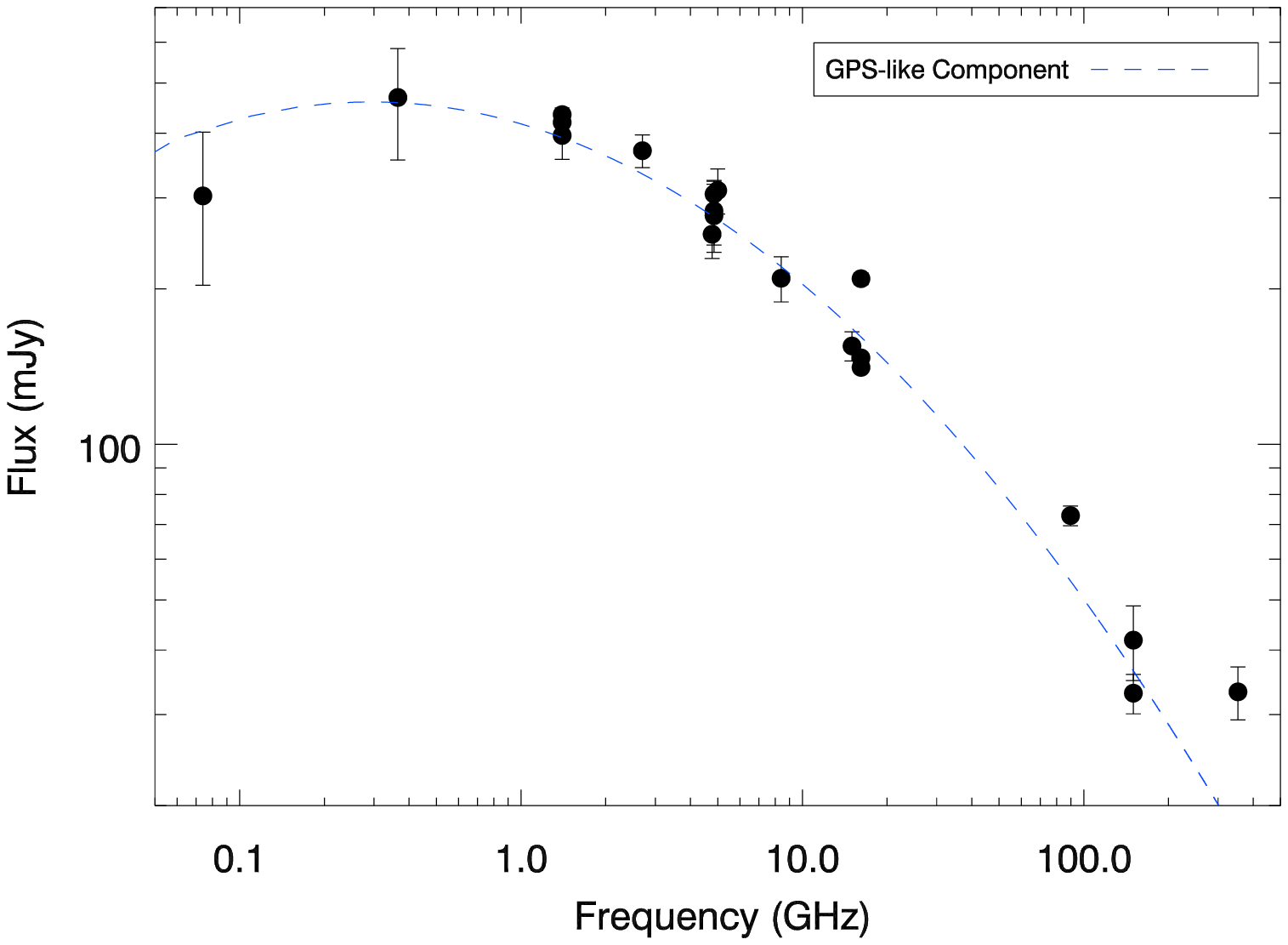}}
    \subfigure[{\it Z235}]{\includegraphics[width=8cm]{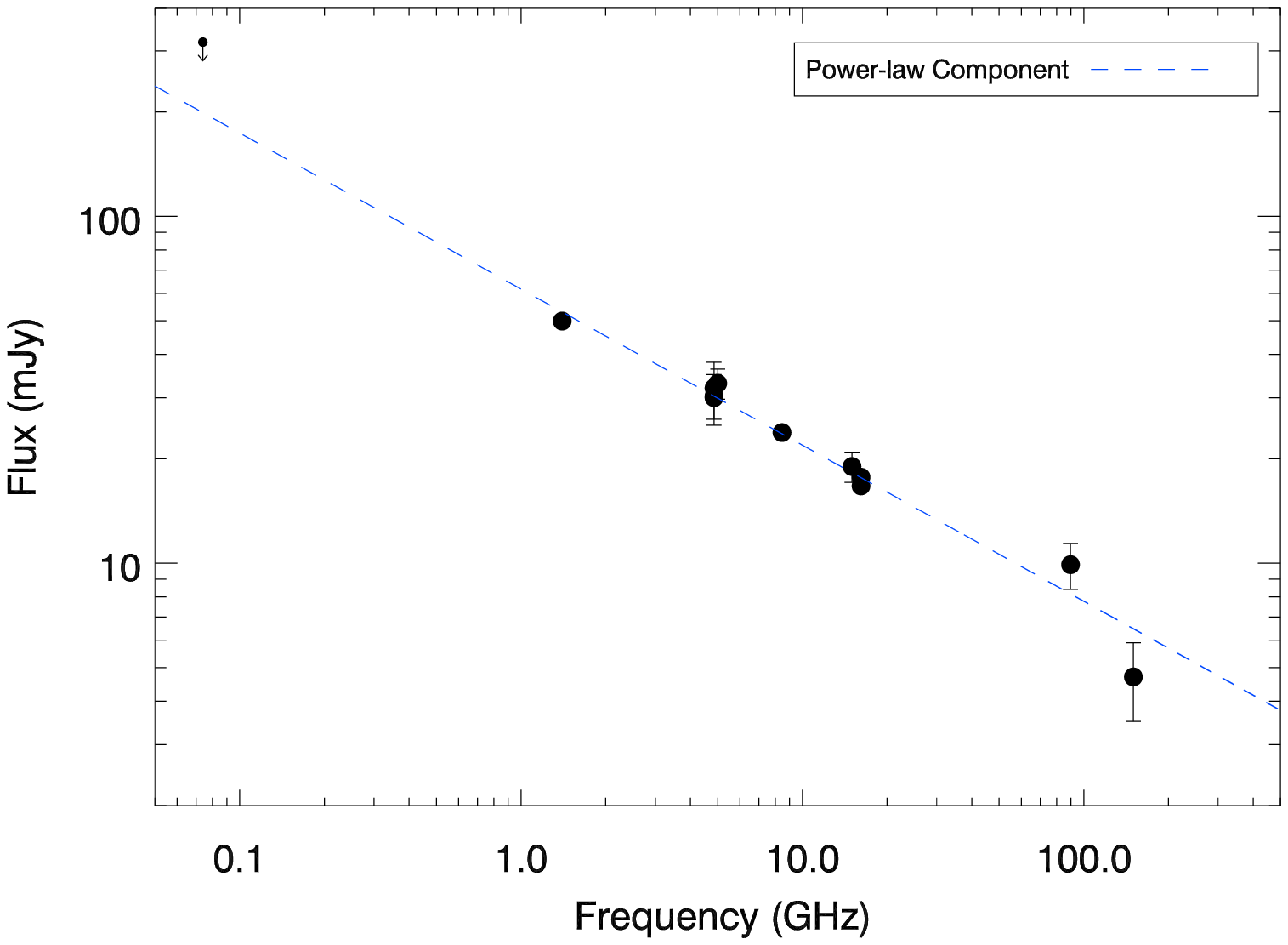}}
    \subfigure[{\it Z8193}]{\includegraphics[width=8cm]{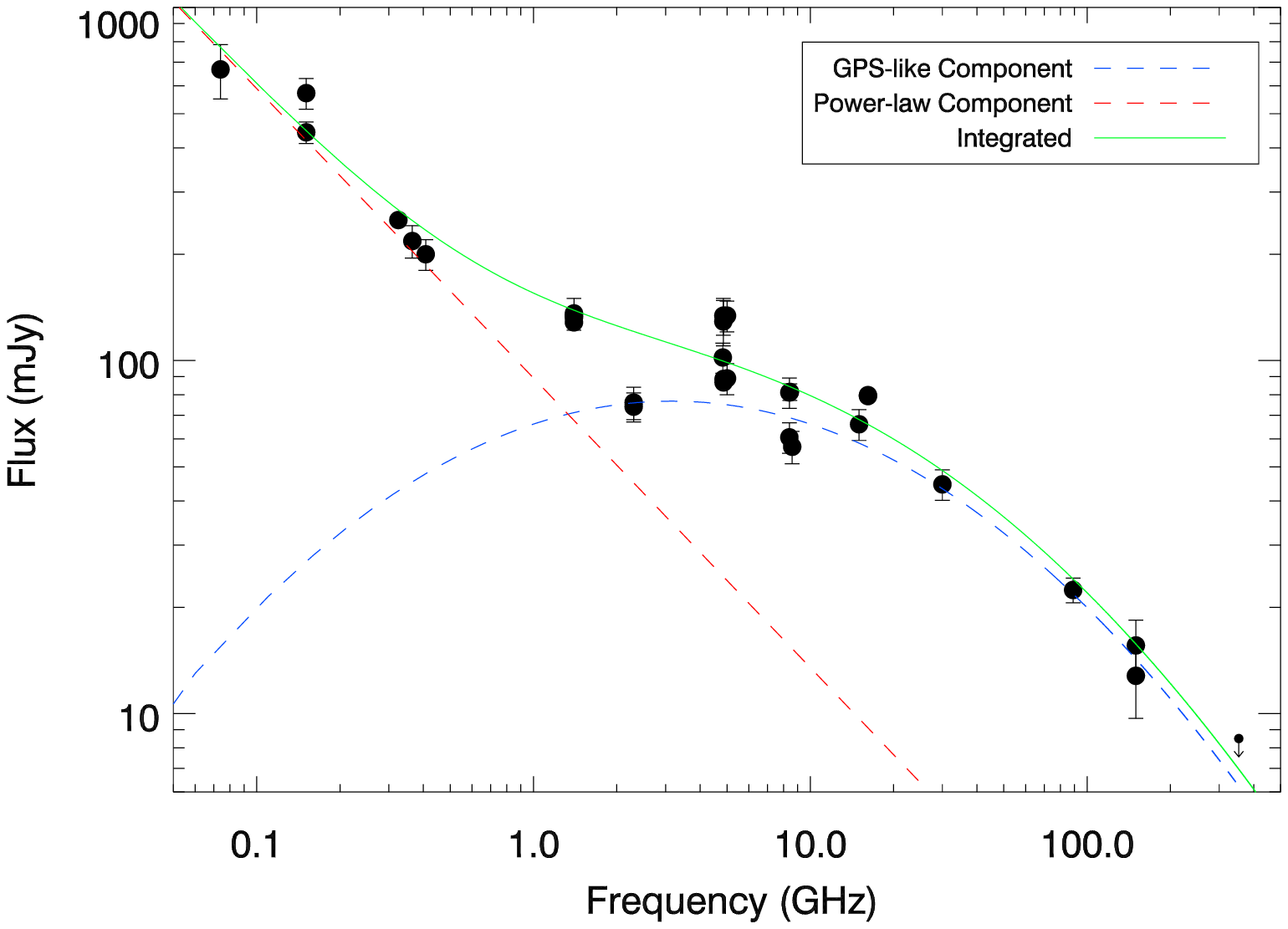}}
    \subfigure[{\it Z8276}]{\includegraphics[width=8cm]{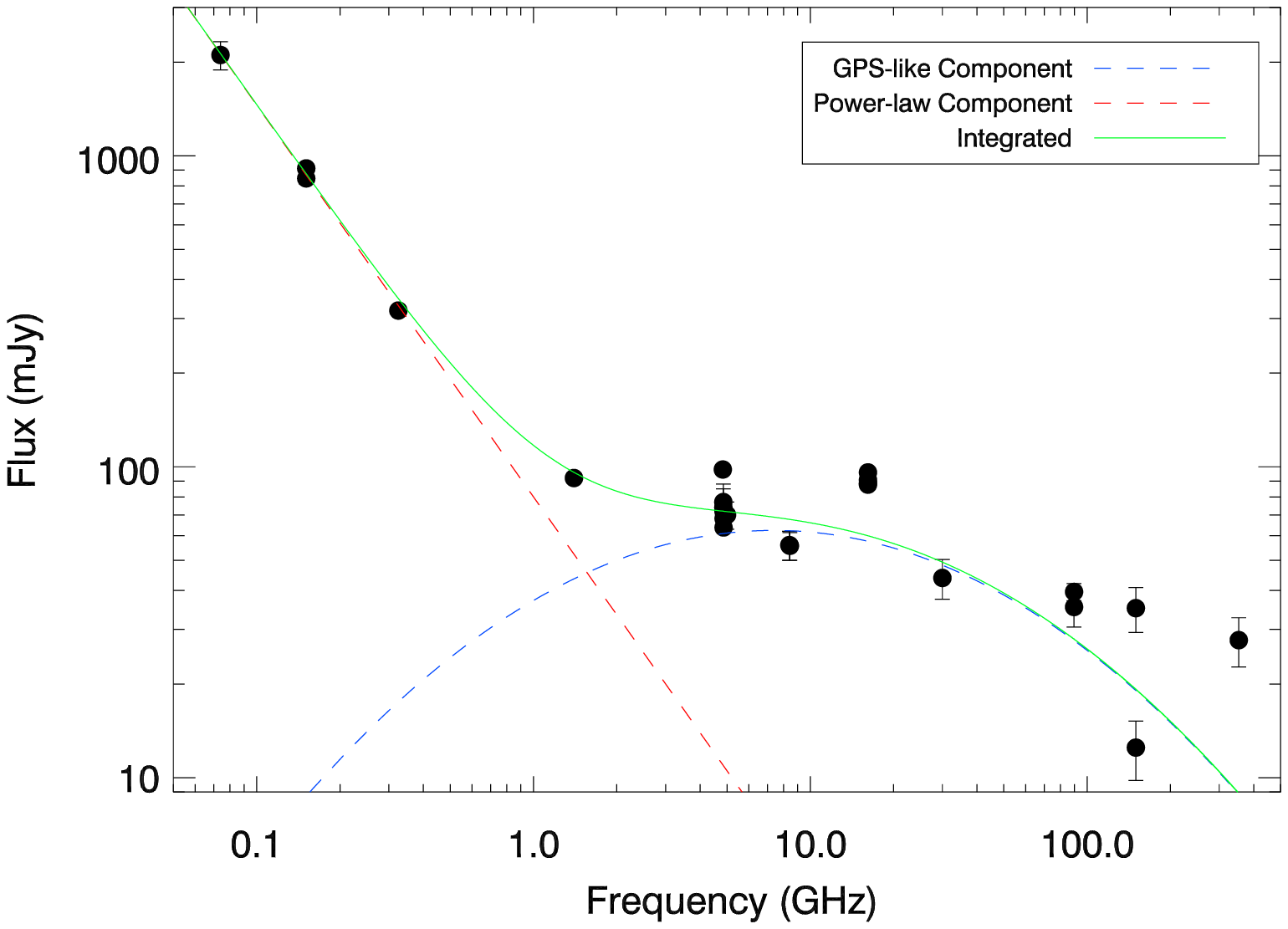}}
  \caption{SEDs, continuation of Fig. A1} 
\end{figure*}

\bsp

\label{lastpage}

\end{document}